\RequirePackage{fix-cm}

\documentclass[smallextended]{svjour3}   
\usepackage{todonotes}
\usepackage{amsmath,amssymb,amsfonts}
\usepackage{graphicx}
\usepackage{tikz}
\usepackage{siunitx}
\usepackage{booktabs}
\usepackage[table]{xcolor}
\usepackage{pgfplots}
\usepackage{pgfplotstable}
\usepgfplotslibrary{groupplots}
\usepackage{mdframed}
\usepackage{tcolorbox}
\usepackage{listings}
\usepackage{longtable}
\usepackage{textcomp}
\usepackage{url}
\usepackage{svg}
\usepackage{balance}
\usepackage{fontawesome5}
\usepackage{cite}
\usepackage{algorithmic}
\usepackage[normalem]{ulem}
\usepackage{float}
\usepackage{hyperref}
\usepackage{subcaption}  
\usepackage{rotating}
\usepackage{multirow}

\newcommand{\myColor}[1]{
    \color{black}{#1}
}

\lstdefinelanguage{Solidity}{
    keywords={contract, function, uint, address, require, emit},
    sensitive=true,
    morecomment=[l]{//},
    morecomment=[s]{/*}{*/},
    morestring=[b]",
}
\newcommand{\goal}[1]{
	\vspace{5pt}
	\noindent\colorbox{gray!05}{
		\parbox{0.97\linewidth}{
			\textbf{\faBullseye\ Our Goal.} #1
		}
	}
	\vspace{5pt}
}

\newcommand{\etal}{\textit{et al.}}

\newcommand{\major}[1]{\textcolor{black}{#1}}

\newcommand{\geribox}[1]{   
    \begin{mdframed}[hidealllines=true, backgroundcolor=gray!5]
        {\textit{#1}}
    \end{mdframed}
}

\lstdefinelanguage{Solidity}{
    keywords={modifier, require, contract, function, using, for, public, library, internal, pure, returns,external, payable, ifAdmin},
    ndkeywords={msg, sender, owner},
    ndkeywordstyle=\color{black},
    comment=[l]{//},
    commentstyle=\color{gray}\itshape,
    stringstyle=\color{red},
    basicstyle=\ttfamily,
    sensitive=true,
    morestring=[b]"
}

\smartqed 

\begin{document}

\title{How Do Solidity Versions Affect Vulnerability Detection Tools? An Empirical Study}

\titlerunning{How Do Solidity Versions Affect Vulnerability Detection Tools?}

\author{Gerardo Iuliano     \and
        Davide Corradini    \and
        Michele Pasqua      \and
        Matteo Cicalese     \and
        Mariano Ceccato     \and
        Dario Di Nucci
}
\institute{
        Gerardo Iuliano \at
        University of Salerno, Italy \\
        \email{geiuliano@unisa.it}
        \and
        Davide Corradini \at
        University of Luxembourg, Luxembourg \\
        \email{davide.corradini@uni.lu}
        \and
        Michele Pasqua \at
        University of Verona, Italy  \\
        \email{michele.pasqua@univr.it}
        \and
        Matteo Cicalese\at
        University of Salerno, Italy \\
        \email{mcicalese@unisa.it}
        \and
        Mariano Ceccato \at
        University of Verona, Italy \\
        \email{mariano.ceccato@univr.it}
        \and
        Dario Di Nucci \at
        University of Salerno, Italy \\
        \email{ddinucci@unisa.it}  
}

\authorrunning{Iuliano et al.}

\date{Received: date / Accepted: date}

\maketitle

\begin{abstract}
Smart contract vulnerabilities pose significant security risks for the Ethereum ecosystem, driving the development of automated tools for detection and mitigation. Smart contracts are written in Solidity, a rapidly evolving programming language that continues to add features and improvements to enhance smart contract security. New versions of Solidity change the compilation process, potentially affecting how tools interpret and analyze smart contract code.
We use Solidity as a case study to investigate a broader problem in software analysis: whether tool effectiveness remains stable as the underlying programming language and execution environment evolve.
In this continuously evolving landscape, we investigated the compatibility of detection tools with different Solidity versions. More specifically, we analyzed detection tools by empirically assessing (i) their crash resilience, (ii) their detection effectiveness, and (iii) their execution time across different versions of Solidity.
We collected 2,500 real-world smart contracts to construct a balanced wild dataset and created a separate injected-vulnerability dataset comprising 270 contracts.
We extended SmartBugs, a framework that enables the execution of vulnerability detection tools for Solidity, adding 11 new tools. Results indicate that most tools are crash-resilient to version 0.4. However, 20\% of the evaluated tools have been updated to remain compatible. Tool performance varies across Solidity versions and is affected by Solidity changes; execution time is higher in version 0.4.x, which was used for evaluation. We present a list of Solidity changes that affect tool performance and offer actionable takeaways for both researchers and practitioners.
\keywords{Blockchain \and Smart Contracts \and Software Evolution \and Software Vulnerabilities
}
\end{abstract}

\section{Introduction}
\label{intro}

\major{
Smart contracts enable programmable and autonomous interactions in block\-chain-based decentralized applications. However, their adoption has introduced significant security risks, as exploitable vulnerabilities can lead to severe and irreversible financial losses. Consequently, researchers and developers have devoted substantial effort to building automated vulnerability detection tools for the Ethereum ecosystem. More than 260 static and dynamic analysis tools have been proposed over the last decade~\cite{rameder2022review,iuliano_dinucci}, including widely used tools such as Mythril~\cite{mythril}, Slither~\cite{slither}, and Oyente~\cite{oyente}.
}

\major{
Software analysis tools, however, do not operate independently of the technological environments they target. Programming languages, compilers, libraries, and runtime environments continuously evolve, while analysis tools depend on assumptions about their syntax, semantics, and execution models. An evaluation conducted on a single technological configuration may therefore provide only temporary evidence and may lead to incomplete or misleading conclusions about a tool's general effectiveness. Solidity represents a concrete and security-relevant case study of this broader phenomenon because its language and compiler have undergone substantial changes since their inception.
}

\major{
Successive Solidity versions have introduced new language constructs and security mechanisms while deprecating or removing existing functionalities. For example, Solidity 0.5.x introduced stricter explicitness requirements, while Solidity 0.8.x enabled checked arithmetic operations by default. Although these changes improve the security and reliability of smart contracts, they can invalidate assumptions encoded in parsers, vulnerability rules, symbolic execution models, and test-input generators.
}

\major{
\autoref{fig:changesAcrossVersions} illustrates how the same arithmetic operation is handled under different Solidity configurations. Both implementations prevent the value from wrapping around, but they rely on different mechanisms. In Solidity 0.8.x, arithmetic operations are checked by default~\cite{solidity_08}, and the decrement automatically reverts. By contrast, contracts targeting versions earlier than Solidity 0.8.x commonly rely on libraries such as OpenZeppelin SafeMath to perform checked arithmetic. A tool whose detection rules assume unchecked arithmetic may incorrectly report the first implementation as an exploitable underflow, thereby generating a false positive. Conversely, a tool that recognizes only newer language patterns may fail to analyze or correctly classify vulnerabilities in older contracts.
}

\major{
Despite the large number of available vulnerability detection tools, to the best of our knowledge, no prior empirical study has systematically examined whether such tools remain compatible and effective across multiple Solidity versions. Existing evaluations predominantly rely on contracts written in Solidity 0.4.x and rarely explicitly report the compiler versions that the evaluated tools support. Consequently, it remains unclear whether changes in observed tool behavior result from limitations in vulnerability detection or from incompatibilities with the evolving language, compiler, and execution environment.
}

\major{
This limitation is visible in established evaluation practices. For example, Durieux et al.~\cite{emp_eval_tools} evaluated tools using the SmartBugs dataset, which predominantly contains contracts written in Solidity 0.4.x and only a limited number of contracts from Solidity 0.5.x. Some of the evaluated tools, such as Oyente~\cite{oyente}, were originally designed around even earlier Solidity and Ethereum configurations. Moreover, several recently proposed tools, including EVM-Shield~\cite{zhang2024evm}, Dfier~\cite{wang2024dfier}, sGuard+~\cite{gao2024sguard}, LENT-SSE~\cite{zheng2024lent}, ContractCheck~\cite{wang2024contractcheck}, and PrettySmart~\cite{zhong2024prettysmart}, continue to be evaluated using SmartBugs without explicitly declaring the range of Solidity versions they support. A preliminary analysis by Iuliano \etal~\cite{iuliano_dinucci} similarly shows that Solidity 0.4.x remains the version most frequently represented in tool evaluations.
}

\major{
In real-world auditing, incompatibility can manifest in several ways. A tool may explicitly fail by being unable to parse, compile, or analyze a contract, or it may crash during execution. More subtle incompatibilities may cause obsolete vulnerability rules to produce false positives or prevent the tool from recognizing vulnerabilities expressed through newer language constructs, thereby producing false negatives. False positives increase the manual effort required to inspect and validate tool reports, while false negatives may create a false sense of security. Moreover, a tool that completes execution without crashing may still provide unreliable vulnerability coverage if its internal assumptions do not align with the target Solidity version.
}

\major{
These consequences are particularly relevant because vulnerability detection tools are commonly used in practice. Chaliasos et al.~\cite{ChaliasosSurvey} report that 92\% of surveyed participants work in organizations that use open-source tools, while 59\% of developers and 48\% of auditors use at least one tool to detect logic-related vulnerabilities. The evolution of the OWASP Smart Contract Top 10 provides further evidence that technological changes can alter the relevance and manifestation of vulnerability classes: the 2025 edition removes or repositions several categories because changes in Solidity and the EVM have modified the conditions under which they occur or their practical impact~\cite{SoK}.
}

\major{
In practice, these tools still require manual validation: auditors must inspect and prioritize their findings before trusting them. This makes it essential to assess not just whether tools run correctly, but whether they stay reliable as Solidity evolves.
}

\major{
To address this gap, we use Solidity as a case study to investigate the broader need for technology-aware and version-aware evaluation of software analysis tools. Specifically, we examine (i) the crash resilience of vulnerability detection tools across different Solidity versions, (ii) their precision and recall on real-world and injected-vulnerability datasets, and (iii) their execution time. We further investigate which language changes affect tool behavior and whether these effects differ across static and dynamic analysis and between source-code and bytecode inputs.
}

\major{
We conducted an exploratory empirical study combining a structured assessment of the available tools with a controlled evaluation of their behavior. We built a real-world smart contract dataset covering multiple Solidity versions and an injected-vulnerability dataset covering different vulnerability types. We then used and extended the SmartBugs framework~\cite{smartbugs1.0,smartbugs2.0} to execute the selected tools under a common experimental infrastructure and measure their crash resilience, detection performance, and execution time.
}

\major{
We assessed 267 tools and integrated 26 reusable tools into SmartBugs. Our results show that tool behavior varies substantially across Solidity versions. Some configurations fail or crash on versions different from those used in their original evaluations, while others remain executable but experience reductions in precision or recall. This result demonstrates that successful execution alone does not guarantee effective vulnerability detection. Overall, static tools are generally more reliable than dynamic tools, while bytecode-based analysis is comparatively more resilient to Solidity evolution than source-code analysis, particularly in terms of recall. We also identify concrete language changes that explain several of the observed variations in tool behavior.
}

\begin{figure}[ht]
\centering
\begin{tcolorbox}[colback=gray!5!white, colframe=white, left=6pt, right=0pt, top=0pt, bottom=0pt]
\begin{lstlisting}[frame=none]
pragma solidity 0.8.0;
contract ExampleDC {
    function f() public pure {
        uint x = 0;
        x--; // Reverts on underflow
    }
}

pragma solidity >=0.4.0 <0.8.0;
import "@openzeppelin/contracts/SafeMath.sol";
contract ExampleSM {
    using SafeMath for uint;
    function f() public pure {
        uint x = 0;
        x = x.sub(1); // Reverts on underflow
    }
}
\end{lstlisting}
\end{tcolorbox}
\caption{Version-dependent handling of arithmetic underflow.}
\label{fig:changesAcrossVersions}
\end{figure}

In summary, our work provides the following contributions:

\begin{itemize}
\item We developed Docker images for a large set of vulnerability detection tools and extended SmartBugs with 26 reusable tools, providing a unified infrastructure for comparative and reproducible evaluations.

\item We created a real-world dataset covering five Solidity compiler versions and an injected-vulnerability dataset containing six vulnerability types across three compiler versions. These datasets support version-aware evaluations of future vulnerability detection tools.

\item We systematically measured three complementary dimensions of tool behavior, i.e., crash resilience, detection performance, and execution time, across multiple Solidity versions, revealing the limitations of evaluations restricted to a single technological configuration.

\item We identified concrete Solidity language changes that affect static and dynamic analyses and both source-code and bytecode inputs, providing empirical evidence of how technological evolution can alter the conclusions of tool evaluations.
\end{itemize}

The remainder of the paper is structured as follows. Section~\ref{bg} provides the necessary background and reviews related work. Section~\ref{rm} describes the research method, experimental setup, data collection, and procedures adopted to support transparency and reproducibility. Section~\ref{results} presents the results, which are discussed in Section~\ref{discussion}. Section~\ref{sec:implication} presents the implications of our findings. Section~\ref{threats} discusses the threats to validity and the measures adopted to mitigate them. Finally, Section~\ref{conclusion} concludes the paper.

\section{Background and Related Work}
\label{bg}

The rapid evolution of blockchain technology has brought significant advancements and challenges. However, vulnerabilities in Smart Contracts (SC) remain a persistent issue, exacerbated by the ongoing evolution of Solidity, the most common language for coding Ethereum smart contracts. Since its inception, Solidity has undergone numerous updates, introducing new features, deprecating keywords, and enhancing security. These changes address vulnerabilities but challenge developers and auditors to maintain compatibility and ensure robust security practices. Numerous tools leveraging various techniques have been developed to detect vulnerabilities. However, assessing the effectiveness and compatibility of these tools across different Solidity versions is unexplored. 

\subsection{Taxonomies of Vulnerabilities, Tools, and Benchmarks}

Rameder \etal~\cite{rameder2022review} analyzed the period from 2014 to January 2021 to conduct an SLR to investigate the vulnerabilities mentioned in the literature and how they are classified. They also focused on vulnerability detection tools, mapping them with a vulnerability taxonomy. Finally, they compiled a list of benchmarks for evaluating tools. They obtained a taxonomy that included 54 vulnerabilities, 140 tools, and 33 benchmarks.

Iuliano and Di Nucci~\cite{iuliano_dinucci} updated the work of Rameder \etal~\cite{rameder2022review}, analyzing the period from 2021 to 2025. 
The work extends the previous taxonomy, condensing the main classification in the literature by organizing vulnerabilities into 13 hierarchical categories and resolving nomenclature issues. The result is a taxonomy that includes 192 vulnerabilities. We also compiled a list of 219 tools and analyzed their functionality, code transformation techniques, and methods. We also mapped tools to the taxonomy and finally collected a list of 133 benchmarks used for tool evaluation or other experiments.
Our previous study shows that Solidity 0.4 is predominant in the benchmarks used for tool evaluation. Moreover, tools focus on a narrow set of vulnerabilities, covering 41\% of the taxonomy.  

Vidal \etal~\cite{vidal} proposed OpenSCV, an open hierarchical taxonomy for smart contract vulnerabilities designed to accommodate community contributions. They analyzed previous vulnerability classifications, community-main\-tained classification schemes, and smart contract vulnerability detection methods. Finally, they validated the taxonomy through an expert-based analysis by surveying researchers in the field of smart contract security.

\subsection{Related Work}
\label{rw}

Durieux \etal~\cite{emp_eval_tools} compared nine automated analysis tools for Ethereum smart contracts using one dataset with 69 annotated vulnerable contracts and another with 47,518 Etherscan contracts. Using SmartBugs, an execution framework for evaluating tools, they ran 428,337 analyses over 564 days. Only 42\% of vulnerabilities in the annotated dataset were detected by all tools, with Mythril achieving the highest accuracy (27\%). Most tools flagged 97\% of contracts as vulnerable, suggesting many false positives and limited overlap among detected vulnerabilities.

Ren \etal~\cite{emp_eval_settings} propose a unified standard to evaluate smart contract testing tools to reduce evaluation bias. They compile 46,186 source-available contracts from diverse organizations, creating a benchmark suite that includes real-world, artificially generated, and confirmed-vulnerable contracts. They introduce a four-step evaluation process that highlights differences from previous methods. Experiments with nine tools reveal that experimental settings significantly affect performance, leading to misleading results and underscoring the need for principled evaluation approaches in vulnerability detection.

Kado \etal~\cite{Kado} investigate the prevalence of vulnerabilities in Ethereum smart contracts across different compiler versions. They analyzed 503,572 contracts with publicly available Solidity source code on the Ethereum blockchain. Focusing on three high-severity vulnerabilities, Locked Money, Using tx.origin, and Unchecked Call, they observed a decline in the appearance rates of these vulnerabilities following significant updates to the Solidity compiler.

Salzano \etal~\cite{salzano2026empirical} present an empirical evaluation of automated vulnerability analysis tools for Solidity smart contracts, leveraging SmartBugs. In addition to comparing traditional tools, they evaluate an LLM-based detection method on two popular datasets and obtain inconsistent results, suggesting unreliable detection when analyzing real-world smart contracts. Their study demonstrates the advantages of combining multiple detection methods, achieving up to 76.78\% of vulnerabilities found while running in under one minute on average.

The main differences between our study, Durieux \etal~\cite{emp_eval_tools}, Ren \etal~\cite{emp_eval_settings}, and Salzano \etal~\cite{salzano2026empirical} lie in the scope of analysis and in the focus on compatibility with the Solidity version. Kado \etal~\cite{Kado} focused on Solidity versions and their impact on vulnerability rates, but did not consider tool compatibility or how changes in Solidity versions affect tool performance. While previous work evaluated the performance of tools on large datasets of Ethereum smart contracts, we examined how detection tools perform and how compatible they are across different Solidity versions. Unlike Durieux \textit{et al.} and Ren \etal, Salzano \etal~\cite{salzano2026empirical} evaluate tool effectiveness at the vulnerability-line level and explore combinations of tools to maximize detection, but they do not examine how tool performance varies with the evolution of Solidity versions, which remains the central focus of our study. Unlike these previous studies, our work evaluated a broader range of tools, including their practical usability, to understand how easily they can be integrated into real-world workflows.

\section{Research Method}
\label{rm}

The following section presents the study's details, highlighting the main goal and research questions. 

\subsection{Preliminary Dataset Analysis}
\label{sec:preliminary_analysis}
First, we performed a preliminary analysis on the datasets collected by Iuliano \etal~\cite{iuliano_dinucci} to understand which Solidity versions (identified by inspecting pragma directives) tool evaluations are based on. Starting from 102 datasets, we filtered them to include only those that were public and available and contained the Solidity source code of the smart contracts and their bytecode. This first phase reduced the number of datasets to 37. We extracted 130,974 smart contracts. To remove duplicates, we removed all spaces, tabs, and comments from the source code and applied a hash function. We identified 111,223 unique smart contracts and 19,751 duplicates. From the 111,223 smart contracts, we removed those that lacked the pragma directive. We obtained 99,894 smart contracts that meet all selection criteria. Finally, we grouped the smart contracts into pragma versions.

\begin{table}[ht]
    \centering
    \caption{Number of smart contracts for each pragma version}
    \label{tab:pragma_freq}
    \begin{tabular}{l|r|r|r|r|r|r}
        \toprule
        \textbf{Pragma} & 0.3.x & 0.4.x & 0.5.x & 0.6.x & 0.7.x & 0.8.x\\ \midrule
        \textbf{Number of SC} & 8 & 83,631 & 10,338 & 5,580 & 325 & 12 \\ \bottomrule
    \end{tabular}
\end{table}

\autoref{tab:pragma_freq} shows the number of smart contracts for each pragma version from 0.3.x to 0.8.x. The most common Solidity version is 0.4.x, followed by 0.5.x, suggesting that most tools have been evaluated on older versions.

Afterward, we assessed the presence of curated smart contracts, i.e., labeled or manually injected vulnerabilities. Eleven of the 37 datasets contain curated smart contracts. We collected 21,821 contracts, of which 6,319 were duplicates. Of the remaining 15,502 smart contracts, we removed those without the pragma directive, leaving 11,693. Finally, we mapped each contract and its vulnerability using the taxonomy created by Iuliano \etal~\cite{iuliano_dinucci} to make a uniform nomenclature. \autoref{tab:sc_labeled} shows the occurrences of labeled smart contracts for each type of vulnerability.

\begin{table}[!ht]
    \centering
    \rowcolors{2}{gray!10}{white}
    \caption{Number of labeled Smart Contracts for each pragma}
    \label{tab:sc_labeled}
    \begin{tabular}{l|rrrrr|r}
    
        \toprule
        \textbf{Vulnerability} & \textbf{0.4.x} & \textbf{0.5.x} & \textbf{0.6.x} & \textbf{0.7.x} & \textbf{0.8.x} & \textbf{Total}  \\ \midrule
        1A1 & 1,390 & 83 & 17 & 0 & 1 & 1,491  \\ 
        1B & 288 & 0 & 1 & 0 & 0 & 289  \\ 
        1B1 & 0 & 30 & 13 & 0 & 0 & 43  \\ 
        1B2 & 0 & 33 & 15 & 0 & 0 & 48  \\ 
        1D & 74 & 5 & 3 & 3 & 2 & 87  \\ 
        2A1 & 1,586 & 37 & 0 & 0 & 0 & 1,623  \\ 
        2A2 & 367 & 0 & 0 & 0 & 0 & 367  \\ 
        2B & 10 & 74 & 13 & 0 & 0 & 97  \\ 
        2B1 & 4 & 32 & 0 & 0 & 0 & 36  \\ 
        2C & 272 & 1 & 0 & 0 & 0 & 273  \\ 
        3A & 1,162 & 112 & 35 & 0 & 1 & 1,310  \\ 
        3B1 & 8 & 2 & 0 & 0 & 0 & 10  \\ 
        3C & 20 & 0 & 0 & 0 & 0 & 20  \\ 
        4A & 75 & 48 & 10 & 2 & 0 & 135  \\ 
        4C & 6 & 82 & 0 & 0 & 0 & 88  \\ 
        4C1 & 8 & 0 & 0 & 0 & 0 & 8  \\ 
        5D & 0 & 33 & 13 & 0 & 0 & 46  \\ 
        6A1 & 22 & 74 & 13 & 0 & 0 & 109  \\ 
        6B & 0 & 34 & 12 & 0 & 0 & 46  \\ 
        6B1 & 0 & 3 & 0 & 0 & 0 & 3  \\ 
        6C & 19 & 32 & 0 & 0 & 0 & 51  \\ 
        6C1 & 101 & 25 & 27 & 3 & 0 & 156  \\ 
        6C2 & 46 & 0 & 0 & 0 & 0 & 46  \\ 
        6D3 & 0 & 26 & 2 & 0 & 0 & 28  \\ 
        7A & 5,139 & 64 & 15 & 0 & 0 & 5,218  \\ 
        8J & 13 & 9 & 0 & 0 & 0 & 22  \\ 
        9A & 1 & 35 & 3 & 0 & 0 & 39  \\  
        10C & 4 & 0 & 0 & 0 & 0 & 4 \\ 
        \midrule
        \textbf{Total} & \textbf{10,615} & \textbf{874} & \textbf{192} & \textbf{8} & \textbf{4} & \textbf{11,693}\\
        \bottomrule
    \end{tabular}
\end{table}

The preliminary analysis results evidence a low number of smart contracts from recent Solidity versions, highlighting the need for updated datasets. We created a large, real-world dataset of smart contracts deployed to the Ethereum blockchain to mitigate this issue. We extracted smart contracts from the last block of the chain to obtain the most recent ones. We used Etherscan, a widely used blockchain explorer and analytics platform for the Ethereum network, to search for and explore transactions and blocks. 
The dataset we created is balanced; it has the same number of contracts for each version of Solidity (from 0.4.x to 0.8.x).

\subsection{Dataset Extraction}
\major{
As a first step, we implemented EthereumMiner\footnote{\url{https://github.com/GerardoIuliano/EthereumMiner}}, a tool that retrieves Ethere\-um smart contracts by traversing the blockchain backward from a specified starting block to an ending block. Preliminary analyses indicate that more than 4.5 million contracts have been deployed on Ethereum since 2018, while fewer than 4\%, approximately 180,000, have publicly available source code~\cite{huang2024sword}. Because we did not have direct access to the complete population of contracts associated with each Solidity version, the dataset cannot be sampled randomly. Instead, we adopted an exhaustive backward-mining procedure: starting from block 22,573,538, the latest block available at the time of the experiment on May 27, 2025, we sequentially inspected deployed contracts until we collected 500 eligible contracts for each of the five Solidity versions. This procedure resulted in a balanced dataset of 2,500 smart contracts.
}

\major{
The balanced design was intentional and followed the objective of our study. We did not aim to estimate tool performance based on the natural distribution of Solidity versions on the blockchain, but rather to compare tool behavior across Solidity versions under comparable conditions. Therefore, each version contributed the same number of contracts. Using a naturally distributed dataset would have made the comparison less informative, because Solidity versions are highly imbalanced in practice and less frequent versions would contribute too few observations. In our dataset, all contracts are still real-world contracts mined from Ethereum; only their selection was controlled to avoid confounding version effects with differences in data availability.
}

We collected smart contracts (i) that do not import an external library, (ii) whose source code is available, and (iii) that are written in Solidity.
We excluded smart contracts that import external libraries to avoid additional threats to validity. Libraries can use different Solidity versions than the contract that imports them. 
Bytecode could be significant in the long term, as it can remain unchanged even when new syntactic sugar is introduced in Solidity. Syntactic sugar is introduced to existing programming languages to improve their readability and brevity. It is particularly interesting to examine how bytecode-based analysis tools behave across different Solidity versions.
To facilitate this investigation, we collected the runtime bytecode and the creation bytecode of smart contracts deployed to the blockchain, along with the publicly available source code.
This approach eliminates bias and ensures a fair comparison between source code and bytecode analysis tools. Access to the source code also enabled us to manually validate the results, thereby improving the reliability of our evaluation.  

\major{
To inject vulnerabilities, we relied on existing vulnerability-injection tools whose original implementations and injected instances had been evaluated by their authors. Although this does not eliminate potential injection errors, we consider these strategies a reasonable starting point given the large-scale nature of our empirical study, which does not aim to provide new vulnerability injection strategies. Nevertheless, these tools, much like vulnerability detection tools, are not fully compatible with all Solidity versions.
}

\major{
MuSe~\cite{muse} relies on pattern matching to inject vulnerabilities. Specifically, it traverses the AST to find a suitable path to inject a vulnerability. MuSe was implemented exclusively for version 0.4.x and can therefore only traverse ASTs produced for that version. Later versions, once compiled, produce a different AST that lacks the patterns required for the mutation that introduces the vulnerability. Adapting it would have required an in-depth analysis of the AST nodes visited when injecting each vulnerability and of how these nodes change across versions, involving substantial technical challenges.
}

\major{
SCAnoGenerator~\cite{SCAnoGenerator} is likewise based on pattern matching, although compatible with versions 0.5.x to 0.7.x. Nevertheless, it does not guarantee that such patterns are always present in the code, which limits its usage. If the required patterns are not present in the contract, the tool cannot inject the vulnerability, as also highlighted by its original authors, who noted that, on average, 21,603 contracts must be examined to find one containing a pattern suitable for injecting the SWC-127 vulnerability. Indeed, across our pool of 500 contracts per version, SCAnoGenerator injected only two vulnerabilities in two contracts, emphasizing its limited effectiveness in our dataset.
}

\major{
SolidiFI~\cite{solidifi} does not rely on pattern matching but instead injects a vulnerable code snippet into the contract, requiring no preconditions for the injection to succeed. Nevertheless, it was not designed to be compatible with multiple versions, and its fixed set of vulnerable snippets is compatible only with version 0.4.x. Unlike previous tools, SolidiFI allows selection of the solc compiler version to compile the contract and perform the necessary checks to identify a suitable injection point, verify that the injection was successful, and ensure the contract remains syntactically valid. The ability to freely choose the compiler version allowed us to run SolidiFI across all versions. This approach also required modifying the set of vulnerable snippets injected by SolidiFI. We created five sets of vulnerable snippets, one per version, so that the contract would still compile after injection. These new snippets are simply the original snippets, adapted to remain compilable across the other versions. In particular, we adapted the keywords used in the snippet for each target version. For example, if the snippet contains the keyword \textsc{now}, we substituted it with \textsc{block.timestamp} in version 0.7.x, since \textsc{now} is no longer supported and has been replaced by \textsc{block.timestamp}. We considered starting with an already validated solution and adapting the vulnerable snippets, so they compile across other versions, making this the most scientifically reliable choice. Unfortunately, SolidiFI exhibited anomalies in version 0.5.x, where, for reasons unknown to us, its executions entered an infinite loop. Furthermore, in version 0.8.x, although execution completed successfully, the contracts generated after injection did not actually contain the vulnerable snippets, even though the output log indicated successful execution. Therefore, we applied it across the three remaining Solidity versions: 0.4.x, 0.6.x, and 0.7.x.
}

\major{
We injected six vulnerabilities (SWC-101, SWC-104, SWC-107, SWC-114, SWC-115, SWC-116) based on SolidiFI's injection capabilities and to maximize the number of tools able to detect them. Expanding the set of injected vulnerability types would not have provided a proportional benefit for our comparison, as the vulnerability classes we did not inject were mostly detected by only one of the considered tools; therefore, increasing the complexity and cost of the injection process would not meaningfully broaden cross-tool comparison. Finally, implementing new injection strategies would have required validating the method itself, which falls outside the scope of our work.
}

\major{
Overall, the dataset comprises vulnerable smart contracts for Solidity  0.4.x, 0.6.x, and 0.7.x. 
For each of these versions, we generated 15 vulnerable contracts for each of the six selected vulnerability classes, resulting in $6 \times 15 = 90$ vulnerable contracts per version and 270 contracts overall. We injected vulnerabilities in 18\% of our initial pool of 1,500 smart contracts, as done by Demissie \etal~\cite{demissie2020security}, who injected vulnerabilities in 15\% of their initial sample. 
These vulnerabilities are among the most relevant in the literature, ranking among the top eight most extensively analyzed vulnerabilities~\cite{iuliano_dinucci}. Considering the limitations of existing injectors, we deem this set of vulnerabilities constitutes an optimal trade-off.
}

\subsection{Goal and Research Questions}
The goal of this study is to analyze the crash resilience of state-of-the-art vulnerability detection tools across Solidity versions and compare their performance.

\goal{

\textbf{Purpose:} Explore

\textbf{Issue:} the crash resilience and the performance

\textbf{Object:} of vulnerability detection tools for Ethereum SCs

\textbf{Viewpoint:} from the points of view of smart contract auditors, practitioners, and researchers.
}

To address this goal, we formulated the following RQs.

\geribox{\faSearch \hspace{0.2mm} \textbf{RQ\textsubscript{1}.} How many state-of-the-art tools are exercisable in practice?}
Considering that the literature provides numerous tools to detect software vulnerabilities in smart contracts, \emph{\textbf{RQ}$_1$} examines how many of them can be built and executed.

\geribox{\faSearch \hspace{0.2mm} \textbf{RQ\textsubscript{2}.}
How crash-resilient are detection tools across different versions of Solidity?}

\emph{\textbf{RQ}$_2$} is motivated by the rapid evolution of Solidity, with new features and security improvements. Many analysis tools, evaluated on older versions, may struggle with newer pragma directives. As developers adopt newer Solidity versions, assessing how well these tools adapt is crucial to ensure they remain relevant. This crash resilience evaluation helps identify gaps and improve the tools currently in use. 

\begin{figure}[ht]
    \centering
    \includegraphics[width=0.75\textwidth]{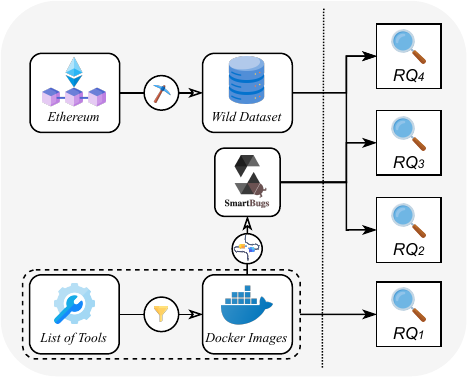}
    \caption{Overview of the Research Method.}
    \label{fig:method}
\end{figure}

\geribox{\faSearch \hspace{0.2mm} \textbf{RQ\textsubscript{3}.}
What is the performance of the detection tools across different versions of Solidity?}
\emph{\textbf{RQ}$_3$} examines the performance of the tools using a standardized benchmark. This experiment ensures a fair evaluation of all tools, enabling an understanding of their relative strengths and weaknesses based on Solidity versions, functionality, and effectiveness. \emph{\textbf{RQ}$_3$} allows us to determine whether Solidity upgrades have introduced new challenges that detection tools ignore, potentially affecting their performance.

\geribox{\faSearch \hspace{0.2mm} \textbf{RQ\textsubscript{4}.}
What is the impact of Solidity versions on the execution time of smart contract detection tools?}

\emph{\textbf{RQ}$_4$} is critical for understanding the practical usability and scalability of tools in real-world settings, where execution time becomes a key performance metric. High execution times affect scalability and delay development or deployment. 
Rather than identifying the best tool, we aim to determine if language changes impact execution time and help users choose the most suitable option based on their needs and resources. This understanding is crucial for maximizing vulnerability detection while minimizing execution time and resource consumption across different language versions. It helps organizations to allocate resources, balancing accuracy with computational costs.

\begin{table*}[ht]
\centering
\caption{Artifact evaluation and SmartBugs integration checklist}
\label{tab:artifacts-checklist}
\begin{tabular}{p{0.4cm}p{3.6cm}p{4.4cm}p{1.3cm}}
\toprule
\multicolumn{4}{c}{\textbf{Artifact evaluation checklist}}   \\
\midrule
\textbf{Step} & \textbf{Question} & \textbf{Description} & \textbf{Badge} \\
\midrule
1 & Are the artifacts available in a public, permanent online repository? &
Artifacts must be made available via an archival repository, such as Software Heritage, FigShare, Zenodo, or GitHub. & Available \\
2 & Does the artifact provide sufficient documentation to enable its exercise? &
Artifact documentation must explain how to obtain, unpack, install, configure, and use the artifact in detail. & Functional \\
3 & Do the artifacts associated with the paper demonstrate a level of quality that significantly exceeds minimal functionality? &
Artifacts are carefully documented and well-structured, facilitating reuse and repurposing. & Reusable \\
\midrule
\multicolumn{4}{c}{\textbf{SmartBugs integration checklist}}   \\
\midrule
& \textbf{Question} & \textbf{Description} & \textbf{Resource} \\
\midrule
4 & Does the artifact include a Dockerfile, a pre-built Docker image, or sufficient instructions to enable practitioners to create and use a Docker image?	& The Docker image must be uploaded to DockerHub. A configuration file (config.yaml) is required and must include, at a minimum, the image field and at least one of the following sections: Solidity, bytecode, or runtime, depending on whether the tool analyzes source code, bytecode, or runtime code. & Docker image\\

5 & Is the Command Line Interface of the artifact sufficiently complete and documented to allow users to run the tool on their own inputs? & A Bash script is required to execute the tool within the Docker image. & script.sh\\
6 & Is the output produced by the artifact machine-readable? & An output parser is required to standardize the output of each tool. & parser.py\\
\bottomrule
\end{tabular}
\end{table*}

\subsection{Data Analysis}
\autoref{fig:method} depicts the method we adopted in the study.
To answer \textbf{RQ}$_1$, we selected tools from the list provided by Iuliano \etal~\cite{iuliano_dinucci} and Rameder \etal~\cite{rameder2022review} because they offer a comprehensive list of tools and provide information like code transformation techniques, methods, and functionality.
We selected all open-source and publicly available tools that support execution via a Command-Line Interface (CLI). If available, we followed the tool's official guidelines, i.e., those provided in the README file, to correctly use it. Tools that fail to build despite following the provided instructions or that systematically crash during execution were excluded. We excluded all tools whose licenses prohibited their usage and commercial tools that required fees to use.
We applied guidelines inspired by those applied by top conferences, such as ICSME and ICSE, for the Artifact Evaluation Track. The guidelines allow us to determine the most appropriate badge for the tool evaluation, i.e., Available, Functional, or Reusable.
\autoref{tab:artifacts-checklist} shows the criteria to apply a badge.
After assessing the number of functional and reusable tools, we integrated them into \emph{SmartBugs}~\cite{smartbugs1.0, smartbugs2.0}, a framework that supports adding new tools by providing a Docker image and a configuration file. We followed the official \emph{SmartBugs} documentation\footnote{\url{https://github.com/smartbugs/smartbugs/wiki}} to integrate tools.
\autoref{tab:artifacts-checklist} lists the resources required to integrate a new tool into the SmartBugs framework.

To answer \textbf{RQ}$_2$, we analyzed the tool logs for each Solidity version. We considered Solidity versions from 0.4.x up to the most recent 0.8.x. 
We adopted our extended \emph{SmartBugs} version to answer \textbf{RQ}$_1$ and check for compilation errors, generic errors, and exceptions. 
The \emph{SmartBugs} infrastructure allows tools not strictly connected to a specific compiler version to choose the compiler version based on the contract they analyze. The framework enabled us to observe the behavior of the tools in different Solidity versions. In addition, we defined a parser to standardize the tools' output into a standardized format. 
\major{
To evaluate the crash resilience of the tools across different Solidity versions, we analyzed the results produced by each tool on the wild dataset described in Section \ref{sec:preliminary_analysis}, comprising 2,500 smart contracts, 500 per version. We use the full wild dataset for this analysis because crash resilience concerns whether a tool can successfully process contracts written for a given Solidity version, independently of whether those contracts contain a specific vulnerability. Using the largest available set of contracts, therefore, exposes tools to a broader range of real-world syntax constructs and dependency patterns, offering a realistic representation of real-world scenarios and yielding more generalizable results.} 
First, we excluded results that terminated due to timeouts and produced no findings, errors, or failures because it is not possible to assess the tool’s behavior without any output.
After removing empty and timeout-terminated results, we analyzed the errors and failures generated by the tools across versions. We defined the tool’s crash resilience as the proportion of successful executions among all executions.
For each tool, we calculated its crash resilience across Solidity versions, depending on the input type analyzed. Since some tools can analyze both source code and bytecode, we computed crash resilience separately for each input type.

To answer \textbf{RQ$_3$}, we calculated the precision and recall of tools across different Solidity versions. 
For the precision analysis, we distinguish between creation bytecode and runtime bytecode. The former is the bytecode generated by the Solidity compiler and executed only once during contract deployment. The latter is the bytecode ultimately stored on the blockchain and invoked during contract execution. The two representations differ primarily in the presence of deployment-specific instructions, such as constructor logic and initialization code, that are stripped from the runtime bytecode. 

\major{
To compute precision, we manually validated a convenience sample~\cite{baltes2022sampling, etikan2016comparison}. In particular, to ensure manual validation covers as many vulnerable instances as possible, we extracted a convenience sample from each set of 500-contract-per-version corpus of the wild dataset, particularly 133 smart contracts for version 0.4.x, 145 for 0.7.x, and 135 for 0.8.x, for a total of 413 smart contracts. We aimed to minimize the number of contracts while maximizing the number of detection reports produced by the tools. Therefore, we considered only smart contracts on which at least two tools have produced a finding. As a result, the actual number of validations is 4,191, representing potential vulnerability instances reported by the tools across 413 smart contracts (1,887 instances for version 0.4, 1,137 for version 0.7, and 1,167 for version 0.8). Each finding was labeled as either a true positive (TP) or a false positive (FP) to compute precision.
}
Since our study focuses on the compatibility of vulnerability detection tools across Solidity versions, we first filtered the reported warnings to retain only vulnerability-related findings.
Several tools (e.g., Slither, Solhint, and Semgrep) also report style issues, best-practice violations, and non-security recommendations, which are not vulnerability instances. Because precision is defined with respect to actual vulnerabilities, we excluded these categories from manual validation.
Manual validation of the filtered results enabled us to identify false positives and true positives and to understand how changes introduced by Solidity version upgrades affect the behavior and performance of the analyzed tools. 
\major{
For manual validation, we prioritized three versions, i.e., 0.4.x, 0.7.x, and 0.8.x, for the following reasons. Version 0.4.x is the oldest version analyzed in our study and the most frequently used for tool evaluation in the state of the art, i.e., the version with the highest occurrence across existing benchmarks. The majority of the tools we surveyed were evaluated on this version; therefore, we considered it essential to include it in our validation. The other two versions, 0.7.x and 0.8.x, are the most recent ones considered in our study and inherit many of the changes introduced in versions 0.5.x and 0.6.x. Moreover, since subsequent changes to Solidity have been largely security-oriented, these two versions best represent the most recent evolution of the language from a security perspective. A notable example is that, starting from version 0.8.x, the compiler natively handles arithmetic overflow and underflow checks. This change has a strong impact on security and, transversally, also reduces the occurrence of other vulnerabilities, such as reentrancy. Finally, we believe that selecting the oldest but most widely used version and comparing it with the two most recent versions, introducing substantial security-related changes, is a reasonable approach for providing useful, up-to-date results. This choice is impactful not only for researchers aiming to develop new detection solutions, but also for auditors who rely on these tools in practice, as well as for developers and practitioners working with Solidity.
}
The observed changes and their implications are discussed in Section \ref{discussion}.

\begin{table}[!ht]
    \centering
    \rowcolors{2}{gray!8}{white}
    \caption{Injected vulnerabilities detected by tools.}
    \label{tab:tool_x_swc}
    \resizebox{\textwidth}{!}{
    \begin{tabular}{lcccccc}
    \toprule
        Tool & SWC-101 & SWC-104 & SWC-107 & SWC-114 & SWC-115 & SWC-116 \\ 
    \midrule
        ConFuzzius & \checkmark & \checkmark & \checkmark & \checkmark & ~ & ~ \\
        ConFuzziusBI & \checkmark & \checkmark & \checkmark & \checkmark & ~ & ~ \\
        Conkas & \checkmark & \checkmark & \checkmark & \checkmark & ~ & \checkmark \\
        Detecti & \checkmark & \checkmark & \checkmark & ~ & ~ & ~ \\
        EtherSolve & ~ & ~ & \checkmark & ~ & \checkmark & ~ \\
        EtherTrust & ~ & ~ & \checkmark & ~ & ~ & ~ \\
        ItyFuzz & ~ & ~ & \checkmark & ~ & ~ & \checkmark \\
        Manticore  & \checkmark & ~ & \checkmark & ~ & ~ & ~ \\
        Mythril & \checkmark & \checkmark & \checkmark & ~ & \checkmark & \checkmark \\
        Osiris & \checkmark & ~ & \checkmark & \checkmark & ~ & \checkmark \\
        Oyente & \checkmark & ~ & \checkmark & \checkmark & ~ & \checkmark \\
        Securify & ~ & \checkmark & \checkmark & \checkmark & ~ & ~ \\
        Semgrep & \checkmark & ~ & \checkmark & ~ & ~ & ~ \\
        sFuzz & \checkmark & \checkmark & \checkmark & ~ & ~ & \checkmark \\
        Slither & ~ & \checkmark & \checkmark & ~ & \checkmark & \checkmark \\
        SmartCheck & \checkmark & \checkmark & ~ & ~ & \checkmark & \checkmark \\
        SolidityCheck & ~ & ~ & \checkmark & ~ & ~ & ~ \\
        Solhint & ~ & ~ & \checkmark & ~ & \checkmark & \checkmark \\
        Vandal & ~ & \checkmark & \checkmark & ~ & \checkmark & ~ \\
        \bottomrule
    \end{tabular}}
\end{table}

To calculate recall, we used the previously described injected dataset and computed recall for 19 tools, considering only the injected vulnerabilities each tool could detect, as shown in \autoref{tab:tool_x_swc}. We analyzed whether detection tools fail to detect injected vulnerabilities, resulting in False Negatives, or correctly detect them, resulting in True Positives.

To simplify performance metric calculations, we used the SWC Registry taxonomy to standardize vulnerability names. 
Manual analysis, alongside a deeper investigation into the tools' behavior, allowed us to understand the reasons behind detection failures and whether newer versions of Solidity have introduced challenges that existing tools do not address. We provided a comprehensive analysis of the limitations and strengths of existing tools that can be helpful for follow-up work on improving analysis tools.

To conduct the statistical tests, we first analyzed all tools together, then analyzed static and dynamic tools separately. As shown by Duriex \etal~\cite{emp_eval_tools}, static analysis tends to produce false positives, whereas Jiang \etal~\cite{ContractFuzzer} show that dynamic analysis reduces them.

We hypothesized that different Solidity versions may affect performance detection and made the following null and alternative hypotheses:

\geribox{\textbf{H}$_{0,3a}$: \textit{There is no dependence between Solidity versions and the detection performance of tools.}\\
\textbf{H}$_{a,3a}$: \textit{There is a dependence between Solidity versions and the detection performance of tools.}}

\geribox{\textbf{H}$_{0,3b}$: \textit{There is no dependence between Solidity versions and the detection performance of static analysis tools.}\\
\textbf{H}$_{a,3b}$: \textit{There is a dependence between Solidity versions and the detection performance of static analysis tools.}}

\geribox{\textbf{H}$_{0,3c}$: \textit{There is no dependence between Solidity versions and the detection performance of dynamic analysis tools.}\\
\textbf{H}$_{a,3c}$: \textit{There is a dependence between Solidity versions and the detection performance of dynamic analysis tools.}}

The wild dataset has also been used to address \textbf{RQ}$_4$. 
We measured the execution times of the tools across different Solidity versions using the SmartBugs framework. \emph{SmartBugs} records the start and duration of the analysis, enabling precise computation for each tool execution. This approach ensures an accurate and systematic evaluation of their efficiency in real-world conditions, contributing to a comprehensive understanding of their performance. We set a 20-minute timeout (1,200 seconds). 
The tools have different implementations and employ different code-transformation techniques. 
They also differ in the type of analysis employed, which significantly affects execution time. Given this, we distinguished between static and dynamic analysis tools and, for each group, constructed five execution-time distributions across Solidity versions from 0.4.x to 0.8.x.
The static analysis tool set comprises 19 tools, whereas the dynamic analysis tool set comprises 7.
For both groups of distributions, we first assessed the assumptions required to apply parametric statistical tests. Since the execution-time distributions did not follow a normal distribution, we employed the Kruskal–Wallis test, a nonparametric test that does not assume normality, to compare the distributions across Solidity versions.

We used statistical tests to determine the extent to which Solidity versions affect the performance of such tools. We hypothesized that different versions may differ in execution time. We make the following null and alternative hypotheses: 

\geribox{\textbf{H}$_{0,4a}$: \textit{There is no statistically significant difference between the execution time of static analysis tools run on different Solidity versions.}\\
\textbf{H}$_{a,4a}$: \textit{
There is a statistically significant difference between the execution time of static analysis tools run on different Solidity versions.}}

\geribox{\textbf{H}$_{0,4b}$: \textit{There is no statistically significant difference between the execution time of dynamic analysis tools run on different Solidity versions.}\\
\textbf{H}$_{a,4b}$: \textit{
There is a statistically significant difference between the execution time of dynamic analysis tools run on different Solidity versions.}}

For both \textbf{RQ}$_3$ and \textbf{RQ}$_4$, we checked the assumptions about the normality of the data and, consequently, chose the appropriate statistical test, i.e., parametric or non-parametric, to verify whether the performance differences are significant. 

\major{
When answering \textbf{RQ}$_3$, we recognized that the underlying outcomes were more naturally represented as categorical counts. Specifically, each manually validated warning was classified as either a true positive (TP) or a false positive (FP) for the precision analysis, while each detectable injected vulnerability was classified as either a true positive (TP) or a false negative (FN) for the recall analysis. Exact tests for contingency tables therefore provide a closer match to the domain and structure of our data than traditional statistical tests, e.g., the Kruskal–Wallis test and the One-Way ANOVA test, which analyze continuous distributions rather than discrete ones, and would instead compare ranks derived from aggregated per-tool precision and recall values. Accordingly, we used the Fisher–Freeman–Halton Exact test~\cite{freeman1951note}, an extension of Fisher’s Exact Test~\cite{fisher} to ($r \times c$) contingency tables, to evaluate the global association between Solidity versions and detection outcomes across versions. We then used Fisher’s Exact Test on ($2 \times 2$) contingency tables to investigate pairwise differences between Solidity versions. Pairwise (p)-values were adjusted using the Holm–Bonferroni method~\cite{holm}. Effect sizes were quantified using Cramér’s (V)~\cite{cramer} for the global analyses and the ($\phi$) coefficient for the pairwise comparisons, allowing us to assess both statistical significance and the practical magnitude of the observed associations.
The choice of the statistical test was motivated by the closer correspondence between exact contingency-table tests and the categorical nature of the underlying TP, FP, and FN outcomes. 
}

\major{
Finally, when answering \textbf{RQ}$_4$, since execution time distributions are continuous, we verified the assumptions for applying the One-Way ANOVA test~\cite{freund2003statistical, zimmerman1993relative}, a robust statistical method for testing differences between continuous distributions when data are normally distributed and variances are equal.
Nevertheless, data were not normally distributed, so we relied on the Kruskal-Wallis test~\cite{freund2003statistical, zimmerman1993relative}, which does not require normality or homogeneity of variances. We applied appropriate post-hoc tests to identify which group differs, e.g., the Dunn post-hoc test with Bonferroni correction for the Kruskal-Wallis test~\cite{freund2003statistical, zimmerman1993relative}, and Cliff's $\delta$~\cite{macbeth2011cliff} for the effect size.
}
The results for \textbf{RQ}$_3$ and \textbf{RQ}$_4$ were statistically significant at $\alpha=0.05$. 

All experiments were conducted on a high-performance workstation with an Intel i9-14900K CPU (6 GHz), 128 GB RAM, and Windows 11.

\subsection{Public Data Availability}
To support the reproducibility of our study and enable research on future Solidity versions or other languages, we provided all materials, including scripts and datasets, in an online appendix hosted on permanent storage platforms.\footnote{\url{https://figshare.com/s/ec72394075625a122300}}

\section{Results}
\label{results}

\subsection{\textbf{RQ\textsubscript{1}.} How many state-of-the-art tools are exercisable in practice?}

\begin{figure}[ht]
    \centering
    \includegraphics[width=0.99\textwidth]{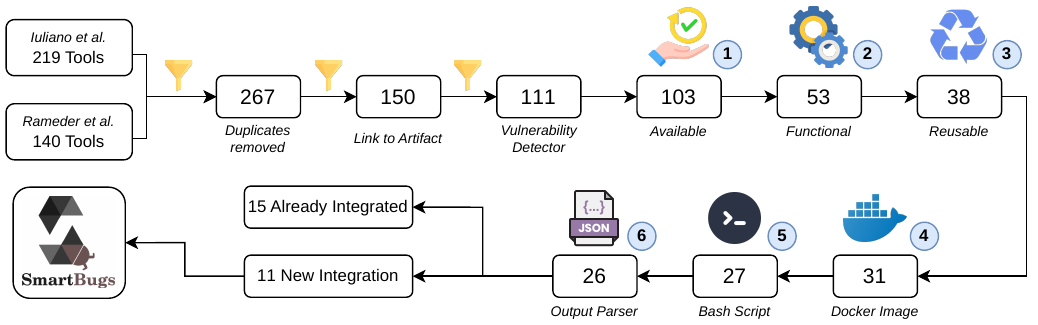}
    \caption{Overview of the artifact evaluation.}
    \label{fig:artifact_overview}
\end{figure}

\major{
We merged the lists of tools compiled by Iuliano et al.~\cite{iuliano_dinucci} and Rameder et al.~\cite{rameder2022review}, resulting in 267 tools after removing duplicates, of which 150 have a link to a code repository or another publicly accessible resource. We removed tools that are not vulnerability detectors, i.e., tools designed for mutation testing, code repair, metric extraction, attack detection, and exploit generation, reducing the final number of tools to 111. Each tool was evaluated based on its availability; if available, we assessed its functionality, and, if it worked properly, its reusability.
}

\begin{table}[!ht]
    \centering
    \caption{Summary of the analyzed tools.}
    \label{tab:analyzed_tools}
    \rowcolors{2}{gray!8}{white}
    \resizebox{\textwidth}{!}{
    \begin{tabular}{lcccccc}
        \toprule
        Tool & Link & Static & Dynamic & Bytecode & Source Code & New Integrations\\ 
        \midrule        
        ConFuzzius~\cite{ConFuzzius} & \href{https://github.com/christoftorres/ConFuzzius}{Link} & ~ & \checkmark & ~ & \checkmark &  \\ 
        Conkas~\cite{Conkas} & \href{https://github.com/nveloso/conkas}{Link} & \checkmark & ~ & \checkmark & \checkmark &  \\ 
        HoneyBadger~\cite{HoneyBadger} & \href{https://github.com/christoftorres/HoneyBadger}{Link} & \checkmark & ~ & ~ & \checkmark &  \\
        MAIAN~\cite{Maian} & \href{https://github.com/ivicanikolicsg/MAIAN}{Link} & \checkmark & ~ & \checkmark & ~ &  \\ 
        Manticore-0.3.7~\cite{Manticore} & \href{https://github.com/trailofbits/manticore/}{Link} & ~ & \checkmark & \checkmark & ~ &  \\ 
        Mythril-0.24.7~\cite{mythril} & \href{https://github.com/wflk/mythril}{Link} & ~ & \checkmark & \checkmark & \checkmark &  \\ 
        Osiris~\cite{Osiris} & \href{https://github.com/christoftorres/Osiris}{Link} & \checkmark & ~ & \checkmark & \checkmark &  \\ 
        Oyente~\cite{oyente} & \href{https://github.com/enzymefinance/oyente}{Link} & \checkmark & ~ & \checkmark & \checkmark &  \\ 
        Securify~\cite{Securify} & \href{https://github.com/eth-sri/securify}{Link} & \checkmark & ~ & \checkmark & \checkmark &  \\ 
        Semgrep~\cite{semgrep} & \href{https://github.com/semgrep/semgrep}{Link} & \checkmark & ~ & ~ & \checkmark  &  \\
        sFuzz~\cite{sFuzz} & \href{https://github.com/duytai/sFuzz}{Link} & ~ & \checkmark & \checkmark & \checkmark &  \\ 
        Slither-0.10.4~\cite{slither} & \href{https://github.com/crytic/slither}{Link} & \checkmark & ~ & ~ & \checkmark &  \\ 
        SmartCheck~\cite{SmartCheck} & \href{https://github.com/smartdec/smartcheck}{Link} & \checkmark & ~ & ~ & \checkmark &  \\ 
        Solhint-3.3.8~\cite{solhint} & \href{https://github.com/protofire/solhint/tree/develop}{Link}  & \checkmark & ~ & ~ & \checkmark &  \\ 
        Vandal~\cite{Vandal} & \href{https://github.com/usyd-blockchain/vandal}{Link} & \checkmark & ~ & \checkmark & ~ &  \\ \hline 
        ItyFuzz~\cite{ItyFuzz} & \href{https://github.com/fuzzland/ityfuzz}{Link} & ~ & \checkmark & ~ & \checkmark & \checkmark \\ 
        RNVulDet~\cite{RNVulDet} & \href{https://github.com/Messi-Q/RNVulDet}{Link} & ~ & \checkmark & \checkmark & ~ & \checkmark \\ 
        eTainter~\cite{eTainter} & \href{https://github.com/DependableSystemsLab/eTainter}{Link} & \checkmark & ~ & \checkmark & ~ & \checkmark \\ 
        aChecker~\cite{aChecker} & \href{https://github.com/DependableSystemsLab/Achecker}{Link} & \checkmark & ~ & \checkmark & ~ & \checkmark \\ 
        SIGUARD~\cite{Siguard} & \href{https://github.com/Jiashuo-Zhang/Siguard}{Link} & \checkmark & ~ & \checkmark & ~ & \checkmark \\ 
        EtherSolve2.0~\cite{EtherSolve} & \href{https://github.com/SeUniVr/EtherSolve/releases/tags/V1.1}{Link} & \checkmark & ~ & \checkmark & ~ & \checkmark \\ 
        ConFuzzius-BI~\cite{ConFuzzius-BI} & \href{https://github.com/SmartcontractTest/ConFuzzius-BI/tree/main}{Link} & ~ & \checkmark & ~ & \checkmark & \checkmark \\ 
        EtherTrust~\cite{EtherTrust} & \href{https://www.netidee.at/ethertrust}{Link} & \checkmark & ~ & \checkmark & ~ & \checkmark \\ 
        SCAnalyzer~\cite{SCAnalyzer} & \href{https://github.com/sjmini/icse2020-Solidity}{Link} & \checkmark & ~ & ~ & \checkmark & \checkmark \\ 
        SolidityCheck & \href{https://github.com/xf97/SolidityCheck}{Link} & \checkmark & ~ & ~ & \checkmark & \checkmark \\ 
        Detecti~\cite{Detecti} & \href{https://github.com/Oss28/Detecti}{Link} & \checkmark & ~ & ~ & \checkmark & \checkmark \\
        \bottomrule
    \end{tabular}}
\end{table}

\major{
A small number of tools are currently unavailable, primarily because the links provided by the original authors are no longer accessible~\cite{cbgru,contractarmor,smartscan}. Out of 111 tools, 103 are available (step 1).
Although several tools were publicly accessible, some could not be executed due to technical or documentation-related issues~\cite{esaf,etherprov,smartcondetect,irfuzz,safesmartcontracts,vuldeesmartcontract,mar}.
}
\major{Specifically, ESAF~\cite{esaf} provides only pseudocode without an implementation; EtherProve~\cite{etherprov} lacks a README or usage instructions; the SmartConDetect~\cite{smartcondetect} repository contains only a README pointing to a dataset, with no accompanying source code; IR-Fuzz~\cite{irfuzz} fails to compile; SafeSmartContracts~\cite{safesmartcontracts} does not provide sufficient information to replicate the study or run the tool; VulDeeSmartContract~\cite{vuldeesmartcontract} executes but encounters runtime errors that prevent correct execution; and M-A-R~\cite{mar} relies on hardcoded file paths that result in missing-file errors. The number of working tools is 53 (step 2).}
\major{Finally, we evaluated whether the remaining tools' artifacts exhibit quality that significantly exceeds minimal functionality. Specifically, we verified that each tool satisfies all Functional-level criteria and, in addition, is carefully documented and well-structured, facilitating reuse and repurposing via the CLI. However, we found that many tools are released as compiled artifacts (i.e., Java jars or Docker images) without source code and do not provide detailed documentation for reuse on personalized configurations. 
As reported in \autoref{fig:artifact_overview}, out of the 53 open-source vulnerability detectors considered available and functional, 38 are reusable (step 3).}

\major{
We evaluated the reusable tools to verify whether they meet the SmartBugs integration checklist reported in \autoref{tab:artifacts-checklist}. First, we checked whether it is possible to create a Dockerfile for tools that did not have one. During this phase, we found that seven tools are not suitable for execution in a Docker image because one is an extension for the VS Code IDE, and the others have source code that hardcodes absolute paths incompatible with containerized environments. After the Docker images (step 4), we checked the possibility of implementing a bash script to run the Docker image following the requirements of the SmartBugs documentation. 
For some tools, it is not possible to specify input or output paths through the command-line interface; instead, they rely on hard-coded directories to provide inputs or store results. Although SmartBugs mounts a temporary working directory inside the container, its standard execution model does not support tool-specific volume mounts to arbitrary hard-coded locations. Therefore, when these directories cannot be populated or accessed through the wrapper script, the tool cannot be integrated without modifying the SmartBugs execution infrastructure (step 5). Some tools take in input a Hardhat or Truffle project, which is not compatible with the SmartBugs architecture. Finally, some tools do not produce parsable output: instead of reporting the type and location of detected vulnerabilities, they provide only bytecode-level symbols. As a result, it is impossible to identify either the nature or the location of the vulnerability, preventing the implementation of a reliable result parser, as there is insufficient information to reverse-engineer the findings (step 6).
As shown in \autoref{fig:artifact_overview}, 26 tools meet the SmartBugs integration checklist, of which 15 are already integrated and functioning in SmartBugs. 
In addition to these tools, we integrated 11 new vulnerability detectors. \autoref{tab:analyzed_tools} lists all 26 tools employed in our study, providing information on the types of analysis they perform (static or dynamic) and the inputs they analyze (source code or bytecode). We released our extended SmartBugs as open source~\footnote{\url{https://github.com/GerardoIuliano/SmartBugs-3.0}}.
}

\geribox{\hspace{0.2mm} \textbf{RQ\textsubscript{1} Summary.} We derived a list of tools from two SLRs, with the latter serving as an update of the former. Together, the two SLRs cover research published between 2014 and 2025. Across this time span, we identified 53 tools that are effectively exercisable, of which 38 are reusable. We integrated 11 of them into SmartBugs.}

{
\rowcolors{4}{gray!8}{white}
\begin{longtable}{llrrrrr}
    \caption{Crash resilience of tools across different Solidity versions.}
    \label{tab:rq2_compatibility}    \\
    \toprule
    Tool & Input Type & 0.4.x & 0.5.x & 0.6.x & 0.7.x & 0.8.x \\
    \midrule
    \endfirsthead
    \rowcolor{white}
    \multicolumn{7}{r}{{\footnotesize Continued from previous page}} \\
    \toprule
    \rowcolor{white}
    Tool & Input Type & 0.4.x & 0.5.x & 0.6.x & 0.7.x & 0.8.x \\
    \midrule
    \endhead
    \midrule
    \rowcolor{white}
    \multicolumn{7}{r}{{\footnotesize Continued on next page}} \\
    \midrule
    \endfoot
    \bottomrule
    \endlastfoot
        aChecker & Creation & 1.00 & 1.00 & 0.99 & 1.00 & 1.00 \\ 
        aChecker & Runtime & 0.99 & 0.97 & 0.98 & 1.00 & 1.00 \\ 
        ConFuzzius & Sourcecode & 0.72 & 0.28 & 0.16 & 0.15 & 0.62 \\ 
        ConFuzziusbi & Sourcecode & 0.73 & 0.33 & 0.19 & 0.16 & 0.62 \\ 
        Conkas & Sourcecode & 0.87 & 0.66 & 0.00 & 0.00 & 0.00 \\ 
        Detecti & Sourcecode & 0.96 & 0.90 & 0.97 & 0.97 & 1.00 \\ 
        eTainter & Creation & 1.00 & 1.00 & 0.99 & 1.00 & 0.99 \\ 
        eTainter & Runtime & 1.00 & 0.97 & 0.98 & 1.00 & 1.00 \\ 
        EtherSolve & Creation & 0.88 & 0.94 & 0.93 & 0.90 & 0.92 \\ 
        EtherSolve & Runtime & 0.01 & 0.04 & 0.01 & 0.00 & 0.00 \\ 
        EtherTrust & Creation & 0.96 & 0.84 & 0.88 & 0.98 & 0.99 \\ 
        EtherTrust & Runtime & 0.96 & 0.84 & 0.87 & 0.97 & 0.99 \\ 
        HoneyBadger & Sourcecode & 0.87 & 0.01 & 0.00 & 0.00 & 0.00 \\ 
        ItyFuzz & Sourcecode & 1.00 & 1.00 & 1.00 & 1.00 & 1.00 \\ 
        Maian & Creation & 0.91 & 0.47 & 0.68 & 0.81 & 0.95 \\ 
        Maian & Runtime & 0.96 & 0.95 & 0.97 & 0.96 & 0.96 \\ 
        Maian & Sourcecode & 0.83 & 0.45 & 0.64 & 0.17 & 0.93 \\ 
        Manticore & Sourcecode & 0.93 & 0.90 & 0.93 & 0.99 & 1.00 \\ 
        Mythril & Creation & 0.99 & 1.00 & 0.99 & 1.00 & 0.99 \\ 
        Mythril & Runtime & 1.00 & 0.99 & 1.00 & 1.00 & 0.99 \\ 
        Mythril & Sourcecode & 0.91 & 0.78 & 0.72 & 0.97 & 0.96 \\ 
        Osiris & Sourcecode & 0.95 & 0.00 & 0.00 & 0.00 & 0.00 \\ 
        Oyente & Sourcecode & 0.93 & 0.01 & 0.00 & 0.00 & 0.00 \\ 
        RNVulDet & Creation & 0.98 & 0.96 & 0.97 & 0.98 & 0.99 \\ 
        RNVulDet & Runtime & 1.00 & 1.00 & 1.00 & 1.00 & 1.00 \\ 
        SCAnalyzer & Sourcecode & 0.51 & 0.19 & 0.03 & 0.02 & 0.22 \\ 
        Securify & Sourcecode & 0.94 & 0.92 & 0.93 & 1.00 & 1.00 \\ 
        Semgrep & Sourcecode & 1.00 & 1.00 & 1.00 & 1.00 & 1.00 \\ 
        sFuzz & Sourcecode & 0.92 & 0.00 & 0.00 & 0.00 & 0.00 \\ 
        Siguard & Creation & 0.99 & 0.95 & 0.99 & 1.00 & 1.00 \\ 
        Siguard & Runtime & 0.99 & 0.98 & 1.00 & 1.00 & 1.00 \\ 
        Siguard & Sourcecode & 0.95 & 0.91 & 0.94 & 1.00 & 1.00 \\ 
        Slither & Sourcecode & 0.95 & 0.93 & 0.94 & 1.00 & 1.00 \\ 
        SmartCheck & Sourcecode & 1.00 & 1.00 & 1.00 & 1.00 & 1.00 \\ 
        Solhint & Sourcecode & 1.00 & 1.00 & 1.00 & 1.00 & 1.00 \\ 
        SolidityCheck & Sourcecode & 1.00 & 0.99 & 1.00 & 1.00 & 1.00 \\ 
        Vandal & Runtime & 1.00 & 1.00 & 1.00 & 1.00 & 0.98 \\
        Vandal & Creation & 0.96 & 0.99 & 1.00 & 1.00 & 0.98 \\
        \midrule
        \multicolumn{2}{l}{Average} & 0.91 & 0.74 & 0.72 & 0.73 & 0.79\\    
\end{longtable}}

\begin{figure*}[!ht]
\centering
\resizebox{\textwidth}{!}{
\begin{tikzpicture}
\begin{groupplot}[
    group style={
        group size=5 by 8,
        horizontal sep=0.4cm,
        vertical sep=0.8cm
    },
    width=0.25\textwidth,
    height=0.25\textwidth,
    ymin=0, ymax=1,
    ytick={0,1},
    xtick=data,
    xticklabels={0.4,0.5,0.6,0.7,0.8},
    tick label style={font=\tiny},
    title style={font=\scriptsize},
    ymajorgrids=true,
    axis background/.style={fill=gray!10},
]
\nextgroupplot[x=0.4cm, title style={font=\scriptsize, yshift=-2mm}, title={\texttt{osiris} - src}]
\addplot[ybar, bar width=6pt] coordinates {
    (0,0.95)
    (1,0.0)
    (2,0.0)
    (3,0.0)
    (4,0.0)
};

\nextgroupplot[x=0.4cm, title style={font=\scriptsize, yshift=-2mm}, title={\texttt{oyente} - src}]
\addplot[ybar, bar width=6pt] coordinates {
    (0,0.93)
    (1,0.01)
    (2,0.0)
    (3,0.0)
    (4,0.0)
};

\nextgroupplot[x=0.4cm, title style={font=\scriptsize, yshift=-2mm}, title={\texttt{honeybadger} - src}]
\addplot[ybar, bar width=6pt] coordinates {
    (0,0.87)
    (1,0.01)
    (2,0.0)
    (3,0.0)
    (4,0.0)
};

\nextgroupplot[x=0.4cm, title style={font=\scriptsize, yshift=-2mm}, title={\texttt{sfuzz} - src}]
\addplot[ybar, bar width=6pt] coordinates {
    (0,0.92)
    (1,0.0)
    (2,0.0)
    (3,0.0)
    (4,0.0)
};

\nextgroupplot[x=0.4cm, title style={font=\scriptsize, yshift=-2mm}, title={\texttt{conkas} - src}]
\addplot[ybar, bar width=6pt] coordinates {
    (0,0.87)
    (1,0.66)
    (2,0.0)
    (3,0.0)
    (4,0.0)
};

\nextgroupplot[x=0.4cm, title style={font=\scriptsize, yshift=-2mm}, title={\texttt{ethersolve} - cb}]
\addplot[ybar, bar width=6pt] coordinates {
    (0,0.88)
    (1,0.94)
    (2,0.93)
    (3,0.9)
    (4,0.92)
};

\nextgroupplot[x=0.4cm, title style={font=\scriptsize, yshift=-2mm}, title={\texttt{ethersolve} - rb}]
\addplot[ybar, bar width=6pt] coordinates {
    (0,0.01)
    (1,0.04)
    (2,0.01)
    (3,0.0)
    (4,0.0)
};

\nextgroupplot[x=0.4cm, title style={font=\scriptsize, yshift=-2mm}, title={\texttt{maian} - cb}]
\addplot[ybar, bar width=6pt] coordinates {
    (0,0.91)
    (1,0.47)
    (2,0.68)
    (3,0.81)
    (4,0.95)
};

\nextgroupplot[x=0.4cm, title style={font=\scriptsize, yshift=-2mm}, title={\texttt{confuzzius} - src}]
\addplot[ybar, bar width=6pt] coordinates {
    (0,0.72)
    (1,0.28)
    (2,0.16)
    (3,0.15)
    (4,0.62)
};

\nextgroupplot[x=0.4cm, title style={font=\scriptsize, yshift=-2mm}, title={\texttt{confuzziusbi} - src}]
\addplot[ybar, bar width=6pt] coordinates {
    (0,0.73)
    (1,0.33)
    (2,0.19)
    (3,0.16)
    (4,0.62)
};

\nextgroupplot[x=0.4cm, title style={font=\scriptsize, yshift=-2mm}, title={\texttt{scanalyzer} - src}]
\addplot[ybar, bar width=6pt] coordinates {
    (0,0.51)
    (1,0.19)
    (2,0.03)
    (3,0.02)
    (4,0.22)
};

\nextgroupplot[x=0.4cm, title style={font=\scriptsize, yshift=-2mm}, title={\texttt{ethertrust} - cb}]
\addplot[ybar, bar width=6pt] coordinates {
    (0,0.96)
    (1,0.84)
    (2,0.88)
    (3,0.98)
    (4,0.99)
};

\nextgroupplot[x=0.4cm, title style={font=\scriptsize, yshift=-2mm}, title={\texttt{ethertrust} - rb}]
\addplot[ybar, bar width=6pt] coordinates {
    (0,0.96)
    (1,0.84)
    (2,0.87)
    (3,0.97)
    (4,0.99)
};

\nextgroupplot[x=0.4cm, title style={font=\scriptsize, yshift=-2mm}, title={\texttt{maian} - rb}]
\addplot[ybar, bar width=6pt] coordinates {
    (0,0.96)
    (1,0.95)
    (2,0.97)
    (3,0.96)
    (4,0.96)
};

\nextgroupplot[x=0.4cm, title style={font=\scriptsize, yshift=-2mm}, title={\texttt{maian} - src}]
\addplot[ybar, bar width=6pt] coordinates {
    (0,0.83)
    (1,0.45)
    (2,0.64)
    (3,0.17)
    (4,0.93)
};

\nextgroupplot[x=0.4cm, title style={font=\scriptsize, yshift=-2mm}, title={\texttt{manticore} - src}]
\addplot[ybar, bar width=6pt] coordinates {
    (0,0.93)
    (1,0.9)
    (2,0.93)
    (3,0.99)
    (4,1.0)
};

\nextgroupplot[x=0.4cm, title style={font=\scriptsize, yshift=-2mm}, title={\texttt{achecker} - cb}]
\addplot[ybar, bar width=6pt] coordinates {
    (0,1.0)
    (1,1.0)
    (2,0.99)
    (3,1.0)
    (4,1.0)
};

\nextgroupplot[x=0.4cm, title style={font=\scriptsize, yshift=-2mm}, title={\texttt{achecker} - rb}]
\addplot[ybar, bar width=6pt] coordinates {
    (0,0.99)
    (1,0.97)
    (2,0.98)
    (3,1.0)
    (4,1.0)
};

\nextgroupplot[x=0.4cm, title style={font=\scriptsize, yshift=-2mm},  title={\texttt{detecti} - src}]
\addplot[ybar, bar width=6pt] coordinates {
    (0,0.96)
    (1,0.9)
    (2,0.97)
    (3,0.97)
    (4,1.0)
};

\nextgroupplot[x=0.4cm, title style={font=\scriptsize, yshift=-2mm}, title={\texttt{etainter} - cb}]
\addplot[ybar, bar width=6pt] coordinates {
    (0,1.0)
    (1,1.0)
    (2,0.99)
    (3,1.0)
    (4,0.99)
};

\nextgroupplot[x=0.4cm, title style={font=\scriptsize, yshift=-2mm}, title={\texttt{etainter} - rb}]
\addplot[ybar, bar width=6pt] coordinates {
    (0,1.0)
    (1,0.97)
    (2,0.98)
    (3,1.0)
    (4,1.0)
};

\nextgroupplot[x=0.4cm, title style={font=\scriptsize, yshift=-2mm}, title={\texttt{mythril} - cb}]
\addplot[ybar, bar width=6pt] coordinates {
    (0,0.99)
    (1,1.0)
    (2,0.99)
    (3,1.0)
    (4,0.99)
};

\nextgroupplot[x=0.4cm, title style={font=\scriptsize, yshift=-2mm}, title={\texttt{mythril} - rb}]
\addplot[ybar, bar width=6pt] coordinates {
    (0,1.0)
    (1,0.99)
    (2,1.0)
    (3,1.0)
    (4,0.99)
};

\nextgroupplot[x=0.4cm, title style={font=\scriptsize, yshift=-2mm}, title={\texttt{mythril} - src}]
\addplot[ybar, bar width=6pt] coordinates {
    (0,0.91)
    (1,0.78)
    (2,0.72)
    (3,0.97)
    (4,0.96)
};

\nextgroupplot[x=0.4cm, title style={font=\scriptsize, yshift=-2mm}, title={\texttt{rnvuldet} - cb}]
\addplot[ybar, bar width=6pt] coordinates {
    (0,0.98)
    (1,0.96)
    (2,0.97)
    (3,0.98)
    (4,0.99)
};

\nextgroupplot[x=0.4cm, title style={font=\scriptsize, yshift=-2mm}, title={\texttt{rnvuldet} - rb}]
\addplot[ybar, bar width=6pt] coordinates {
    (0,1.0)
    (1,1.0)
    (2,1.0)
    (3,1.0)
    (4,1.0)
};

\nextgroupplot[x=0.4cm, title style={font=\scriptsize, yshift=-2mm}, title={\texttt{securify} - src}]
\addplot[ybar, bar width=6pt] coordinates {
    (0,0.94)
    (1,0.92)
    (2,0.93)
    (3,1.0)
    (4,1.0)
};

\nextgroupplot[x=0.4cm, title style={font=\scriptsize, yshift=-2mm}, title={\texttt{semgrep} - src}]
\addplot[ybar, bar width=6pt] coordinates {
    (0,1.0)
    (1,1.0)
    (2,1.0)
    (3,1.0)
    (4,1.0)
};

\nextgroupplot[x=0.4cm, title style={font=\scriptsize, yshift=-2mm}, title={\texttt{siguard} - cb}]
\addplot[ybar, bar width=6pt] coordinates {
    (0,0.99)
    (1,0.95)
    (2,0.99)
    (3,1.0)
    (4,1.0)
};

\nextgroupplot[x=0.4cm, title style={font=\scriptsize, yshift=-2mm}, title={\texttt{siguard} - rb}]
\addplot[ybar, bar width=6pt] coordinates {
    (0,0.99)
    (1,0.98)
    (2,1.0)
    (3,1.0)
    (4,1.0)
};

\nextgroupplot[x=0.4cm, title style={font=\scriptsize, yshift=-2mm}, title={\texttt{siguard} - src}]
\addplot[ybar, bar width=6pt] coordinates {
    (0,0.95)
    (1,0.91)
    (2,0.94)
    (3,1.0)
    (4,1.0)
};

\nextgroupplot[x=0.4cm, title style={font=\scriptsize, yshift=-2mm}, title={\texttt{slither} - src}]
\addplot[ybar, bar width=6pt] coordinates {
    (0,0.95)
    (1,0.93)
    (2,0.94)
    (3,1.0)
    (4,1.0)
};

\nextgroupplot[x=0.4cm, title style={font=\scriptsize, yshift=-2mm}, title={\texttt{smartcheck} - src}]
\addplot[ybar, bar width=6pt] coordinates {
    (0,1.0)
    (1,1.0)
    (2,1.0)
    (3,1.0)
    (4,1.0)
};

\nextgroupplot[x=0.4cm, title style={font=\scriptsize, yshift=-2mm}, title={\texttt{solhint} - src}]
\addplot[ybar, bar width=6pt] coordinates {
    (0,1.0)
    (1,1.0)
    (2,1.0)
    (3,1.0)
    (4,1.0)
};

\nextgroupplot[x=0.4cm, title style={font=\scriptsize, yshift=-2mm}, title={\texttt{soliditycheck} - src}]
\addplot[ybar, bar width=6pt] coordinates {
    (0,1.0)
    (1,0.99)
    (2,1.0)
    (3,1.0)
    (4,1.0)
};

\nextgroupplot[x=0.4cm, title style={font=\scriptsize, yshift=-2mm}, title={\texttt{vandal} - rb}]
\addplot[ybar, bar width=6pt] coordinates {
    (0,1.0)
    (1,1.0)
    (2,1.0)
    (3,1.0)
    (4,0.98)
};

\nextgroupplot[x=0.4cm, title style={font=\scriptsize, yshift=-2mm}, title={\texttt{vandal} - cb}]
\addplot[ybar, bar width=6pt] coordinates {
    (0,0.96)
    (1,0.99)
    (2,1.0)
    (3,1.0)
    (4,0.98)
};

\nextgroupplot[x=0.4cm, title style={font=\scriptsize, yshift=-2mm}, title={\texttt{ityfuzz} - src}]
\addplot[ybar, bar width=6pt] coordinates {
    (0,1.0)
    (1,1.0)
    (2,1.0)
    (3,1.0)
    (4,1.0)
};

\nextgroupplot[
    hide axis,                
    axis background/.style={fill=white} 
]

\nextgroupplot[
    hide axis,                
    axis background/.style={fill=white} 
]

\end{groupplot}
\end{tikzpicture}}
\caption{Tool compatibility across Solidity versions.}
\label{fig:compatibility-grid}
\end{figure*}

\subsection{\textbf{RQ\textsubscript{2}.} How crash-resilient are detection tools across different versions of Solidity?}

\autoref{tab:rq2_compatibility} shows the crash resilience of each tool across different Solidity versions. The average crash resilience on the 0.4.x version is the highest, confirming that most of the tools were developed for and evaluated on that version. The trend immediately decreases on the 0.5.x version until the 0.7.x version. Then, the crash resilience increases in the latest version. Looking at the tools and their maintenance, we noticed that some tools are still maintained and have been upgraded since their release. These upgrades enhance overall crash resilience, underscoring the trend toward greater crash resilience in the latest version. In particular, Slither, Manticore, Semgrep, Solhint, and Securify have been updated to be compatible with Solidity changes. Some of them explicitly declare their compatibility: Slither can analyze contracts written with Solidity $>=$ 0.4; Securify analyzes contracts written in Solidity $>=$ 0.5.8; Solhint provides two rule sets, one of which is the default rule set, deprecated since Solidity 5.1.0, and the other is the recommended rule set, compatible with the latest Solidity version. In total, 20\% of tools have been maintained, whereas the remaining 80\% do not explicitly report compatibility and have not been maintained since their release; for example, Oyente declares support for version 0.4.19. However, some tools appear fully compatible with all versions. 

\autoref{fig:compatibility-grid} plots tool crash resilience.
A deep analysis of the bar plots highlights an unusual behavior. In particular, ItyFuzz has never failed, but looking at the results, it has never found anything. Indeed, to have a clear understanding of tool crash resilience, we should also consider their performance. RQ$_3$ contributes to increasing awareness of tool crash resilience, since tools may not necessarily fail to demonstrate incompatibility. On the one hand, developers can handle failures and exceptions, masking compatibility issues. On the other hand, detectors implemented in tools should leverage language-specific characteristics and provide more insightful information about their compatibility.

\geribox{\hspace{0.2mm} \textbf{RQ\textsubscript{2} Takeaway.} \major{
Vulnerability detection tools should declare the exact language version they are compatible with.
}
}

\subsection{\textbf{RQ\textsubscript{3}.} What is the performance of the detection tools across different versions of Solidity?}

In the following, we report the results for precision and recall separately.

\begin{table}[ht]
    \rowcolors{2}{gray!10}{white}
    \centering
    \caption{Precision of tools calculated with manual validation.}
    \label{tab:precision}
    \begin{tabular}{lllrrr}
    \toprule
        Analysis Type & Tool & Input Type & 0.4.x & 0.7.x & 0.8.x \\
    \midrule
        Static & Conkas & Sourcecode & 0.49 & - & - \\
        Static & Detecti & Sourcecode & 0.67 & 0.00 & 0.00 \\
        Static & eTainter & Runtime & 0.50 & 1.00 & - \\
        Static & EtherSolve & Creation & 0.08 & 0.64 & 1.00 \\
        Static & Maian & Sourcecode & 1.00 & 0.67 & - \\
        Static & Maian & Creation & 1.00 & 0.67 & - \\
        Static & Osiris & Sourcecode & 0.57 & - & - \\
        Static & Oyente & Sourcecode & 0.43 & - & - \\
        Static & SCAnalyzer & Sourcecode & 0.20 & - & - \\
        Static & Securify & Sourcecode & 1.00 & - & - \\
        Static & Slither & Sourcecode & 0.65 & 0.55 & 0.71 \\
        Static & SmartCheck & Sourcecode & 0.63 & 0.34 & 0.46 \\
        Static & Solhint & Sourcecode & 0.89 & 0.11 & 0.28 \\
        Static & Vandal & Creation & 0.52 & - & 0.96 \\
        Static & Vandal & Runtime & 0.54 & - & 0.95 \\
        Dynamic & ConFuzzius & Sourcecode & 0.38 & 0.45 & 0.00 \\
        Dynamic & ConFuzziusBI & Sourcecode & 0.30 & 0.38 & 0.00 \\
        Dynamic & Mythril & Sourcecode & 0.59 & 0.73 & 0.29 \\
        Dynamic & Mythril & Creation & 0.53 & 0.46 & 0.06 \\
        Dynamic & Mythril & Runtime & 0.31 & 0.10 & 0.00 \\
        Dynamic & sFuzz & Sourcecode & 0.38 & - & - \\
    \midrule
        \multicolumn{3}{l}{Macro Average -- Sourcecode analysis} & 0.58 & 0.40 & 0.25 \\
        \multicolumn{3}{l}{Micro Average -- Sourcecode analysis} & 0.63 & 0.34 & 0.46 \\
        \multicolumn{3}{l}{Macro Average -- Bytecode analysis} & 0.50 & 0.57 & 0.59 \\
        \multicolumn{3}{l}{Micro Average -- Bytecode analysis} & 0.47 & 0.21 & 0.80 \\
        \multicolumn{3}{l}{Macro Average -- Static analysis} & 0.61 & 0.50 & 0.62 \\
        \multicolumn{3}{l}{Micro Average -- Static analysis} & 0.64 & 0.34 & 0.56 \\
        \multicolumn{3}{l}{Macro Average -- Dynamic analysis} & 0.41 & 0.43 & 0.07 \\
        \multicolumn{3}{l}{Micro Average -- Dynamic analysis} & 0.42 & 0.23 & 0.07 \\
    \midrule
        \multicolumn{3}{l}{Macro Average} & 0.56 & 0.47 & 0.39 \\
        \multicolumn{3}{l}{Micro Average} & 0.60 & 0.32 & 0.54 \\
    \bottomrule
    \end{tabular}
\end{table}

\subsubsection{Precision Analysis}
\autoref{tab:precision} shows the precision of tools on different Solidity versions.
Six tools did not report any findings, making it impossible to determine their precision. Among the configurations that produced at least one finding, 21, 13, and 12 tool--input configurations yielded precision values for Solidity 0.4.x, 0.7.x, and 0.8.x, respectively.

\major{
\paragraph{Version-Specific Precision Trends in Macro- and Micro-Average Analysis.}
To provide a consistent view of precision across Solidity versions, we consider both macro- and micro-average precision. The macro-average assigns the same weight to each tool--input configuration and therefore describes the performance of a typical configuration. Conversely, the micro-average pools all true and false positives before computing precision, giving greater weight to configurations that produce more findings. The two measures therefore capture complementary aspects of tool performance.
}

\major{
For Solidity 0.4.x, the macro- and micro-average precision values are 0.56 and 0.60, respectively. The slightly higher micro-average indicates that configurations that produce more findings are, overall, more precise than those that produce fewer. Although some configurations have low precision, such as EtherSolve with creation bytecode, most of the evaluated configurations produce useful findings for this version.
}

\major{
For Solidity 0.7.x, the macro-average decreases from 0.56 to 0.47, while the micro-average decreases more sharply from 0.60 to 0.32. Thus, both the precision of the typical configuration and the overall correctness of the reported findings deteriorate, but the reduction is considerably more pronounced in the micro-average. This discrepancy indicates that configurations that produce more findings yield a higher proportion of false positives. From a practical perspective, auditors may therefore spend substantially more effort validating reports for Solidity 0.7.x.
}

\major{
For Solidity 0.8.x, a different pattern emerges. The macro-average decreases further to 0.39, showing that precision deteriorates across a broad set of tool--input configurations. For example, Detecti, ConFuzzius, ConFuzziusBI, and Mythril with runtime bytecode achieve zero precision, while Solhint and all Mythril configurations perform worse than in Solidity 0.7.x. However, the micro-average increases to 0.54. This difference indicates that, although many configurations perform poorly, a comparatively precise subset accounts for a substantial proportion of the findings. Some configurations, including EtherSolve and Vandal, achieve particularly high precision in Solidity 0.8.x, while Slither also improves compared with Solidity 0.7.x. However, the micro-average recovery should not be interpreted as a general recovery across tools, since the lower macro-average reveals that poor performance remains widespread.
}

\major{
Overall, the macro-average shows a monotonic decline from 0.56 in Solidity 0.4.x to 0.47 in Solidity 0.7.x and 0.39 in Solidity 0.8.x, indicating a progressively broader deterioration across tool--input configurations. In contrast, the micro-average decreases sharply from 0.60 to 0.32, then recovers to 0.54, identifying a large reduction in the correctness of the reported findings in Solidity 0.7.x, followed by a partial recovery in Solidity 0.8.x.
}

\major{
\paragraph{Static vs. Dynamic Analysis.}
When aggregating results by analysis type, static analysis achieves higher precision than dynamic analysis in both macro- and micro-average across all evaluated Solidity versions. For Solidity 0.4.x, static analysis obtains a macro-average precision of 0.61 and a micro-average of 0.64, compared with 0.41 and 0.42 for dynamic analysis, respectively. For Solidity 0.7.x, the difference narrows at the configuration level, with macro-averages of 0.50 for static analysis and 0.43 for dynamic analysis. Nevertheless, the corresponding micro-averages of 0.34 and 0.23 indicate a larger difference in the overall precision of the findings. The gap becomes particularly pronounced for Solidity 0.8.x: static analysis achieves macro- and micro-average precision values of 0.62 and 0.56, whereas both averages decrease to 0.07 for dynamic analysis. The agreement between the two dynamic-analysis averages indicates that the poor precision on Solidity 0.8.x is widespread and is not caused solely by a small number of configurations producing many findings.
}

\major{
These results suggest that dynamic tools are particularly sensitive to changes in execution semantics and environment modeling, potentially increasing false positives when analyzing modern Solidity contracts. Static tools, while not immune to these issues, appear more robust overall, especially when supported by mature rule sets or precise bytecode abstractions.
}

\major{
\paragraph{Source Code vs. Bytecode Inputs.}
The comparison between input types reveals no universally superior representation, but rather version-dependent behavior that also differs between macro- and micro-average precision. For Solidity 0.4.x, source-code-based analyses outperform bytecode-based analyses on both measures, achieving macro- and micro-average precision values of 0.58 and 0.63, compared with 0.50 and 0.47 for bytecode-based analyses.
}

\major{
For Solidity 0.7.x, the two averages provide different conclusions. Bytecode analysis achieves a higher macro-average than source-code analysis, with values of 0.57 and 0.40, respectively, indicating that the typical bytecode configuration is more precise. However, source-code analysis achieves a higher micro-average of 0.34, compared with 0.21 for bytecode analysis. Thus, the bytecode configurations that produce more findings have a higher false-positive rate, despite the stronger performance of the typical bytecode configuration.
}

\major{
For Solidity 0.8.x, bytecode analysis outperforms source-code analysis according to both measures. Bytecode-based configurations achieve macro- and micro-average precision values of 0.59 and 0.80, respectively, whereas source-code-based configurations achieve 0.25 and 0.46. This pattern suggests that bytecode-level analyses may generalize more effectively as the Solidity language evolves because they are less directly affected by changes to the source-language syntax. Nevertheless, the micro-average result for Solidity 0.7.x shows that this greater stability at the configuration level does not necessarily translate into more precise findings overall. In contrast, source-code-based tools require frequent updates to parsers, rules, and vulnerability patterns to remain precise across language versions.
}

\paragraph{Tool-level Observations.}
At the level of individual tools, precision varies widely. Some configurations achieve perfect precision in specific versions, including Maian with source code and creation bytecode and Securify in Solidity 0.4.x, eTainter with runtime bytecode in Solidity 0.7.x, and EtherSolve with creation bytecode in Solidity 0.8.x. Vandal also achieves near-perfect precision with creation and runtime bytecode in Solidity 0.8.x. However, these results may come at the cost of limited applicability or low recall (see \autoref{tab:recall}). Conversely, several configurations report low or zero precision in one or more versions, indicating a high rate of false positives under manual validation.

\begin{table}[ht]
    \centering
    \rowcolors{2}{gray!10}{white}
    \caption{Recall of tools calculated on the injected dataset.}
    \label{tab:recall}
    \begin{tabular}{lllrrr}
    \toprule
        Analysis Type & Tool & Input Type & 0.4.x & 0.6.x & 0.7.x \\
    \midrule
        Static & Conkas & Sourcecode & 0.72 & 0.00 & 0.00 \\
        Static & Detecti & Sourcecode & 0.40 & 0.40 & 0.47 \\
        Static & EtherSolve & Bytecode & 0.69 & 0.59 & 0.75 \\
        Static & EtherTrust & Bytecode & 0.00 & 0.00 & 0.00 \\
        Static & Osiris & Sourcecode & 0.68 & 0.00 & 0.00 \\
        Static & Oyente & Sourcecode & 0.68 & 0.00 & 0.00 \\
        Static & Securify & Sourcecode & 0.60 & 0.00 & 0.00 \\
        Static & Semgrep & Sourcecode & 0.00 & 0.07 & 0.00 \\
        Static & Slither & Sourcecode & 0.52 & 0.92 & 0.98 \\
        Static & SmartCheck & Sourcecode & 0.57 & 0.55 & 0.60 \\
        Static & Solhint & Sourcecode & 0.33 & 0.33 & 0.33 \\
        Static & Vandal & Bytecode & 0.75 & 0.68 & 0.82 \\
        Dynamic & ConFuzzius & Sourcecode & 0.48 & 0.13 & 0.03 \\
        Dynamic & ConFuzziusBI & Sourcecode & 0.48 & 0.12 & 0.03 \\
        Dynamic & ItyFuzz & Sourcecode & 0.00 & 0.00 & 0.00 \\
        Dynamic & Manticore & Sourcecode & 0.00 & 0.00 & 0.00 \\
        Dynamic & Mythril & Sourcecode & 0.09 & 0.12 & 0.03 \\
        Dynamic & Mythril & Bytecode & 0.92 & 0.77 & 0.91 \\
        Dynamic & sFuzz & Sourcecode & 0.73 & 0.00 & 0.00 \\
    \midrule
        \multicolumn{3}{l}{Macro Average -- Sourcecode analysis} & 0.42 & 0.18 & 0.16 \\
        \multicolumn{3}{l}{Micro Average -- Sourcecode analysis} & 0.46 & 0.19 & 0.17 \\
        \multicolumn{3}{l}{Macro Average -- Bytecode analysis} & 0.59 & 0.51 & 0.62 \\
        \multicolumn{3}{l}{Micro Average -- Bytecode analysis} & 0.70 & 0.57 & 0.74 \\
        \multicolumn{3}{l}{Macro Average -- Static analysis} & 0.50 & 0.29 & 0.33 \\
        \multicolumn{3}{l}{Micro Average -- Static analysis} & 0.56 & 0.35 & 0.48 \\
        \multicolumn{3}{l}{Macro Average -- Dynamic analysis} & 0.39 & 0.16 & 0.14 \\
        \multicolumn{3}{l}{Micro Average -- Dynamic analysis} & 0.59 & 0.29 & 0.39 \\
    \midrule
        \multicolumn{3}{l}{Macro Average} & 0.45 & 0.25 & 0.26 \\
        \multicolumn{3}{l}{Micro Average} & 0.57 & 0.32 & 0.45 \\
    \bottomrule
    \end{tabular}
\end{table}

\subsubsection{Recall Analysis}

\major{
\paragraph{Version-Specific Recall Trends in Macro- and Micro-Average Analysis.}
\autoref{tab:recall} reports the recall achieved by each tool on the injected-vulnerability dataset across Solidity versions 0.4.x, 0.6.x, and 0.7.x. To provide a consistent view of the results, we report both macro- and micro-average recall. The macro-average assigns equal weight to each tool--input configuration and therefore represents the recall of a typical configuration. Conversely, the micro-average pools all true positives and false negatives before computing recall, thus reflecting the proportion of vulnerability instances detected across all configurations.
}

\major{
Both averages show a substantial decrease from Solidity 0.4.x to Solidity 0.6.x. The macro-average decreases from 0.45 to 0.25, while the micro-average decreases from 0.57 to 0.32. Thus, the reduction observed in Solidity 0.6.x affects both the typical tool configuration and the overall proportion of detected vulnerability instances.
}

\major{
A different pattern emerges in Solidity 0.7.x. The macro-average remains almost unchanged, increasing only from 0.25 to 0.26, whereas the micro-average increases from 0.32 to 0.45. This difference indicates that the recovery is concentrated in a limited number of high-performing configurations rather than being shared across the evaluated tools. In particular, EtherSolve, bytecode-based Mythril, Slither, and Vandal achieve recall values of 0.75, 0.91, 0.98, and 0.82, respectively, while nine configurations still obtain a recall of 0.00. Consequently, the increase in micro-average recall for Solidity 0.7.x should not be interpreted as a general improvement across tools: although a larger proportion of the evaluated vulnerability instances is detected than in Solidity 0.6.x, the typical tool configuration still achieves low recall. Moreover, the micro-average remains below the 0.57 obtained for Solidity 0.4.x, indicating only a partial recovery in aggregate detection capability.
}

\major{
Overall, Solidity 0.6.x has the lowest recall according to both aggregation strategies. Solidity 0.7.x shows a partial recovery in aggregate detection capability, as reflected by the increase in micro-average recall from 0.32 to 0.45. However, the nearly unchanged macro-average reveals that this improvement depends on a relatively small group of effective tool--input configurations and is not broadly shared across tools.
}

\major{
\paragraph{Static vs. Dynamic Analysis.}
Static analysis achieves a higher macro-average recall than dynamic analysis across all evaluated Solidity versions, indicating that the typical static configuration is consistently more effective. The micro-average results are more nuanced. For Solidity 0.4.x, static analysis obtains a macro-average recall of 0.50, compared with 0.39 for dynamic analysis. However, dynamic analysis achieves a slightly higher micro-average recall than static analysis, with values of 0.59 and 0.56, respectively. This difference indicates that, despite the lower recall of the typical dynamic configuration, a limited number of dynamic configurations detect a comparatively large proportion of the evaluated vulnerability instances.
}

\major{
For Solidity 0.6.x, static analysis achieves macro- and micro-average recall values of 0.29 and 0.35, whereas dynamic analysis achieves 0.16 and 0.29, respectively. In Solidity 0.7.x, static analysis partially recovers, reaching a macro-average of 0.33 and a micro-average of 0.48. Dynamic analysis instead obtains a macro-average of 0.14 and a micro-average of 0.39. Thus, although the micro-average of dynamic analysis increases between Solidity 0.6.x and 0.7.x, its macro-average decreases further, indicating that the improvement is concentrated in a limited number of configurations rather than being shared across dynamic tools.
}

\major{
The persistently low macro-average recall for dynamic analysis suggests that these approaches struggle to trigger injected vulnerabilities in newer compiler versions, possibly due to changes in execution semantics, environment modeling, or insufficient test input generation. Static tools also experience a substantial decline in Solidity 0.6.x, but their recovery in Solidity 0.7.x is stronger: both their macro- and micro-average recall increase, whereas dynamic analysis improves only according to the micro-average.
}

\major{
\paragraph{Source Code vs. Bytecode Input.}
Tools that analyze bytecode consistently outperform those that analyze source code according to both macro- and micro-average recall. For Solidity 0.4.x, bytecode-based analysis achieves macro- and micro-average recall values of 0.59 and 0.70, respectively, compared with 0.42 and 0.46 for source-code-based analysis.
}

\major{
The difference becomes more pronounced for newer Solidity versions. In Solidity 0.6.x, bytecode-based analysis achieves a macro-average of 0.51 and a micro-average of 0.57, whereas source-code-based analysis decreases to 0.18 and 0.19, respectively. In Solidity 0.7.x, bytecode analysis further improves to macro- and micro-average values of 0.62 and 0.74, while source-code analysis decreases to 0.16 and 0.17. The consistently higher micro-average than macro-average for bytecode analysis indicates that the configurations contributing more evaluated vulnerability instances also tend to achieve higher recall. Conversely, the similarly low macro- and micro-average values for source-code analysis in the newer versions indicate that poor recall is widespread across its configurations.
}

\major{
These results suggest that bytecode analysis is less directly sensitive to source-level syntactic and semantic changes introduced across Solidity versions, although EtherTrust fails to detect any injected vulnerability. For example, the transition from optional to mandatory explicit function visibility introduced in Solidity 0.5.0 represents a source-level breaking change that can affect source-code-based analyzers across versions. Bytecode analyzers are less directly affected by this change because visibility-related behavior is handled during compilation. Bytecode-based Mythril and Vandal demonstrate robust performance across all evaluated versions: Mythril achieves recall values between 0.77 and 0.92, while Vandal achieves values between 0.68 and 0.82.
}

\paragraph{Tool-Level Observations}
At the individual tool level, performance varies widely. Slither stands out among static, source-code-based tools, achieving very high recall in Solidity 0.6.x (0.92) and 0.7.x (0.98), indicating strong adaptability to newer language versions. Mythril exhibits a stark contrast between input types: while its source-code analysis yields low recall ($\leq$ 0.12), its bytecode-based variant achieves consistently high recall across all versions.

EtherTrust, ItyFuzz, and Manticore fail to detect any injected vulnerabilities, resulting in zero recall across all versions. This result may indicate limited vulnerability coverage, configuration constraints, or incompatibility with the injected fault models. Older static tools such as Oyente, Osiris, and Securify perform reasonably well on Solidity 0.4.x but completely fail on later versions, suggesting a lack of maintenance or support for newer compiler semantics.

\geribox{\hspace{0.2mm} \textbf{RQ\textsubscript{3} Finding 3.1.}
Static tools generally achieve higher recall than dynamic tools, while bytecode-based analysis remains substantially more stable across Solidity versions than source-code analysis.
}

\subsubsection{Dependence between Solidity Version and Performance} We performed Fisher’s exact test on three sets of tools: (i) all tools, (ii) static analyzers, and (iii) dynamic analyzers. 

\begin{table}[!ht]
\centering
\caption{Pairwise comparison of tools' performance across Solidity versions using Fisher’s Exact Test with Holm correction. Effect size is reported using $\phi$.}
\label{tab:all_tools}
\begin{tabular}{lccc}
\toprule
\multicolumn{4}{c}{\textbf{Precision}}\\
\midrule
\textbf{Comparison} & \textbf{Adjusted $p$-value} & $\boldsymbol{\phi}$ & \textbf{Effect size} \\
\midrule

0.4.x vs 0.7.x & $3.94\times 10^{-51}$ & 0.26 & Small–medium \\
0.4.x vs 0.8.x & $9.21\times 10^{-4}$ & 0.06 & Negligible \\
0.7.x vs 0.8.x & $3.79\times 10^{-25}$ & 0.21  & Small \\

\midrule
\multicolumn{4}{c}{\textbf{Recall}}\\
\midrule
\textbf{Comparison} & \textbf{Adjusted $p$-value} & $\boldsymbol{\phi}$ & \textbf{Effect size} \\
\midrule

0.4.x vs 0.6.x & $4.74\times 10^{-37}$ & 0.24 & Small–medium \\
0.4.x vs 0.7.x & $0.68$ & 0.01 & Negligible \\
0.6.x vs 0.7.x & $1.81\times 10^{-40}$ & 0.25 & Small \\

\bottomrule
\end{tabular}
\end{table}

\paragraph{\textbf{Performance of All Tools.}}
We performed the Fisher–Freeman–Halton Exact test on a 3×2 contingency table, incorporating precision data from all analyzed tools, static and dynamic, to assess True Positives (TP) and False Positives (FP) across three Solidity versions (0.4.x, 0.7.x, and 0.8.x). The global test indicated a highly significant association between Solidity version and detection outcomes (p $<$ 2.2e-16), with a moderate overall effect size measured by Cramér’s V = 0.23.
To identify which version pairs drove this association, we conducted pairwise comparisons using Fisher’s Exact test on 2×2 tables, with Holm correction for multiple testing. As shown in \autoref{tab:all_tools}, all pairs were statistically significant.
However, effect-size analysis indicates that the practical magnitude of these differences is generally small.
The largest practical difference occurs between Solidity 0.4.x and 0.7.x ($\phi$ = 0.26), suggesting that changes introduced in version 0.7 impact tool performance. Between 0.7.x and 0.8.x, the effect is small ($\phi = 0.21$). This indicates that tool detection continues to evolve, but the rate of improvement is more gradual from one version to the next.
Surprisingly, the comparison between 0.4.x and 0.8.x has a negligible effect size ($\phi$ = 0.06). This effect could be explained by the fact that some tools have been updated to be compatible with the latest version. This highlights the importance of considering both statistical significance and effect size, as samples can yield extremely small p-values even when the practical difference is minor.

We conducted a global Fisher–Freeman–Halton Exact test on a 3×2 contingency table, incorporating recall data from all analyzed tools, to assess True Positives (TP) and False Negatives (FN) across Solidity versions 0.4.x, 0.6.x, and 0.7.x. The global association was statistically significant (p $<$ 2.2e-16), with a moderate effect size (Cramér’s V = 0.22), indicating systematic differences in recall across versions.
As shown in \autoref{tab:all_tools}, pairwise Fisher Exact tests with Holm correction revealed that recall differs significantly between Solidity 0.4.x and 0.6.x ($\phi$ = 0.24), and between 0.6.x and 0.7.x ($\phi$ = 0.25).
These results suggest that the global effect is driven primarily by changes in the 0.4.x-0.6.x and 0.6.x-0.7.x transitions.

\geribox{\hspace{0.2mm} \textbf{RQ\textsubscript{3} Finding 3.2.} Most tools perform better on Solidity version 0.4.x, which was used for their evaluations. Tool performance has declined in recent versions of Solidity, except for the maintained ones. Bytecode-based detection tools are less affected by Solidity changes, which, in most cases, are syntactic sugar that disappears after compilation. The micro-averages show a partial recovery after the lowest-performing intermediate version. However, the macro-averages indicate that this recovery is concentrated in a limited subset of tool–input configurations.}




\paragraph{\textbf{Performance of Static Analysis Tools.}}

\begin{table}[!ht]
\centering
\caption{Pairwise comparison of static analysis tools' performance across Solidity versions using Fisher’s Exact Test with Holm correction. Effect size is reported using $\phi$.}
\label{tab:static_pairwise}
\begin{tabular}{lrrr}
\toprule
\multicolumn{4}{c}{\textbf{Precision}}\\
\midrule
\textbf{Comparison} & \textbf{Adjusted $p$-value} & $\boldsymbol{\phi}$ & \textbf{Effect size} \\
\midrule
0.4.x vs 0.7.x & $1.99 \times 10^{-80}$ & 0.37 & Medium--large \\
0.4.x vs 0.8.x & $4.43 \times 10^{-5}$  & 0.08 & Negligible \\
0.7.x vs 0.8.x & $8.56 \times 10^{-45}$ & 0.30 & Medium \\
\midrule
\multicolumn{4}{c}{\textbf{Recall}}\\
\midrule
\textbf{Comparison} & \textbf{Adjusted $p$-value} & $\boldsymbol{\phi}$ & \textbf{Effect size} \\
\midrule
0.4.x vs 0.6.x & $1.57 \times 10^{-28}$ & 0.29 & Medium \\
0.4.x vs 0.7.x & $5.66 \times 10^{-3}$  & 0.07 & Negligible \\
0.6.x vs 0.7.x & $3.23 \times 10^{-47}$ & 0.36 & Medium--large \\
\bottomrule
\end{tabular}
\end{table}

We repeated the previous analysis, focusing only on static analysis tools to verify the extent to which language changes affected their performance.
For both precision and recall, the global test revealed highly significant differences across versions (p-value $<$ 2.2e-16), with moderate-to-large effect sizes (Cramér’s V = 0.32 for precision and 0.31 for recall), indicating a substantial association between Solidity version and tool performance.
The pairwise analyses with Holm correction in \autoref{tab:static_pairwise} further clarified the source of these differences. For precision, all version pairs differ significantly. The largest practical difference emerges between Solidity 0.4 and 0.7 ($\phi$ = 0.37), followed by 0.7 vs 0.8 ($\phi$ = 0.30), while the difference between 0.4 and 0.8, although statistically significant, exhibits a negligible effect size ($\phi$ = 0.08).
A similar pattern is observed for recall. While all pairwise comparisons are statistically significant, the strongest effect is found between Solidity 0.6 and 0.7 ($\phi$ = 0.36), highlighting version 0.6 as a critical transition point. In contrast, the difference between versions 0.4 and 0.7 shows a negligible effect size ($\phi$ = 0.07), indicating a recovery of recall in newer versions.
Overall, these results demonstrate that changes in Solidity versions have a non-negligible, structured impact on static analysis tools, with specific compiler transitions posing significant challenges to both precision and recall.
This suggests that most of the precision degradation occurs during intermediate language transitions rather than accumulating uniformly across versions. Additionally, some tools have been maintained to support the latest version, thereby improving performance to a level comparable to that of version 0.4.

\geribox{\hspace{0.2mm} \textbf{RQ\textsubscript{3} Finding 3.3.} 
There is a dependence between Solidity versions and the detection performance of static analysis tools.
}

\begin{table}[!ht]
\centering
\caption{Pairwise comparison of the dynamic analysis tool's performance across Solidity versions using Fisher’s Exact Test with Holm correction. Effect size is reported using $\phi$.}
\label{tab:dynamic_pairwise}
\begin{tabular}{lrrr}
\toprule
\multicolumn{4}{c}{\textbf{Precision}}\\
\midrule
\textbf{Comparison} & \textbf{Adjusted $p$-value} & $\boldsymbol{\phi}$ & \textbf{Effect size} \\
\midrule
0.4.x vs 0.7.x & $2.55 \times 10^{-4}$ & 0.14 & Small \\
0.4.x vs 0.8.x & $4.71 \times 10^{-6}$ & 0.22 & Small--Medium \\
0.7.x vs 0.8.x & $1.18 \times 10^{-10}$ & 0.32 & Medium \\
\midrule
\multicolumn{4}{c}{\textbf{Recall}}\\
\midrule
\textbf{Comparison} & \textbf{Adjusted $p$-value} & $\boldsymbol{\phi}$ & \textbf{Effect size} \\
\midrule
0.4.x vs 0.6.x & $5.99 \times 10^{-13}$ & 0.21 & Small--Medium \\
0.4.x vs 0.7.x & $5.65 \times 10^{-5}$ & 0.12 & Small \\
0.6.x vs 0.7.x & $1.03 \times 10^{-3}$ & 0.09 & Small \\
\bottomrule
\end{tabular}
\end{table}

\paragraph{\textbf{Performance of Dynamic Analysis Tools.}}
Finally, we focused on dynamic analysis tools to verify the extent to which language changes affected their performance.
The Fisher–Freeman–Halton test reveals a statistically significant association between Solidity versions and the detection performance of dynamic analysis tools, both in terms of precision (p = 2.6e-11) and recall (p = 8.5e-13). The effect size for precision is moderate (Cramér’s V = 0.25), whereas the effect for recall is smaller (Cramér’s V = 0.17), indicating that language evolution impacts precision more strongly than recall.
Pairwise comparisons in \autoref{tab:dynamic_pairwise} show that the greatest differences in precision occur between Solidity versions 0.7 and 0.8 ($\phi$ = 0.32), highlighting the substantial impact of recent language changes on the precision of dynamic analysis tools. Differences between earlier versions (e.g., 0.4 vs. 0.7) exhibit smaller effect sizes.
For recall, although all pairwise comparisons are statistically significant after Holm correction, the corresponding effect sizes remain small to small–medium ($\phi$ $<=$ 0.21). This suggests that dynamic tools exhibit a measurable decline in recall on more recent Solidity versions, yet the practical impact of this degradation remains moderate.

\geribox{\hspace{0.2mm} \textbf{RQ\textsubscript{3} Finding 3.4.} 
There is a dependence between Solidity versions and the detection performance of dynamic analysis tools.
}

{\scriptsize
\begin{longtable}{lcrrrrrrr}
    \caption{Execution time details for each tool across different Solidity versions.}
    \label{tab:rq4_exe_time}    \\
    \toprule
        Tool & Version & Min & Max & Mean & Median & Count & TO & TO+B \\
    \midrule
    \endfirsthead
    \multicolumn{9}{r}{{\footnotesize Continued from previous page}} \\
    \toprule
    Tool & Version & Min & Max & Mean & Median & Count & TO & TO+B \\ 
    \midrule
    \endhead
    \midrule
    \multicolumn{9}{r}{{\footnotesize Continued on next page}} \\
    \midrule
    \endfoot
    \bottomrule
    \endlastfoot
        Conkas & 0.4.x & 3.95 & 1213.36 & 262.74 & 310.88 & 479 & 7 & 0 \\ 
        Conkas & 0.5.x & 4.06 & 1238.19 & 323.73 & 372.31 & 478 & 24 & 0 \\ 
        Conkas & 0.6.x & 3.37 & 27.92 & 8.76 & 7.59 & 477 & 0 & 0 \\ 
        Conkas & 0.7.x & 3.85 & 23.39 & 8.16 & 7.32 & 493 & 0 & 0 \\ 
        Conkas & 0.8.x & 4.23 & 32.33 & 8.36 & 7.38 & 489 & 0 & 0 \\ \midrule
        ItyFuzz & 0.4.x & 1.90 & 1226.07 & 99.71 & 3.28 & 467 & 37 & 0 \\ 
        ItyFuzz & 0.5.x & 1.98 & 1216.99 & 29.28 & 3.37 & 485 & 10 & 0 \\ 
        ItyFuzz & 0.6.x & 1.93 & 1216.00 & 27.04 & 3.31 & 481 & 9 & 0 \\ 
        ItyFuzz & 0.7.x & 2.06 & 1214.53 & 21.91 & 3.37 & 482 & 7 & 0 \\ 
        ItyFuzz & 0.8.x & 2.24 & 1228.92 & 320.05 & 4.15 & 479 & 125 & 0 \\ \midrule
        SCAnalyzer & 0.4.x & 5.40 & 79.25 & 13.70 & 11.92 & 472 & 0 & 0 \\ 
        SCAnalyzer & 0.5.x & 5.91 & 65.02 & 16.05 & 14.18 & 474 & 0 & 0 \\ 
        SCAnalyzer & 0.6.x & 4.12 & 68.08 & 9.81 & 8.09 & 489 & 0 & 0 \\ 
        SCAnalyzer & 0.7.x & 3.83 & 50.56 & 9.31 & 7.76 & 479 & 0 & 0 \\ 
        SCAnalyzer & 0.8.x & 4.18 & 59.52 & 9.27 & 7.90 & 484 & 0 & 0 \\ \midrule
        aChecker & 0.4.x & 4.40 & 1255.08 & 84.58 & 11.73 & 972 & 46 & 0 \\ 
        aChecker & 0.5.x & 4.14 & 1246.21 & 377.37 & 21.32 & 953 & 248 & 0 \\ 
        aChecker & 0.6.x & 4.14 & 1253.22 & 418.77 & 19.07 & 969 & 318 & 0 \\ 
        aChecker & 0.7.x & 4.82 & 1241.41 & 524.22 & 21.82 & 939 & 393 & 0 \\ 
        aChecker & 0.8.x & 4.87 & 1268.43 & 779.53 & 1213.81 & 965 & 611 & 0 \\  \midrule
        Mythril & 0.4.x & 12.37 & 1230.22 & 759.85 & 996.80 & 1433 & 172 & 10 \\ 
        Mythril & 0.5.x & 12.62 & 1230.39 & 688.21 & 992.20 & 1453 & 236 & 4 \\ 
        Mythril & 0.6.x & 10.65 & 1227.48 & 560.36 & 370.18 & 1433 & 189 & 0 \\ 
        Mythril & 0.7.x & 12.60 & 1229.81 & 526.98 & 178.95 & 1447 & 363 & 0 \\ 
        Mythril & 0.8.x & 11.28 & 1224.00 & 615.53 & 986.18 & 1449 & 12 & 0 \\ \midrule
        Maian & 0.4.x & 59.71 & 1229.75 & 428.51 & 214.54 & 1413 & 13 & 224 \\ 
        Maian & 0.5.x & 37.94 & 1235.48 & 310.11 & 171.19 & 1434 & 0 & 77 \\ 
        Maian & 0.6.x & 63.71 & 1217.35 & 273.01 & 162.01 & 1456 & 0 & 31 \\ 
        Maian & 0.7.x & 55.84 & 1231.76 & 393.59 & 170.12 & 1448 & 0 & 115 \\ 
        Maian & 0.8.x & 43.85 & 975.84 & 199.67 & 151.97 & 1440 & 0 & 0 \\ \midrule
        Semgrep & 0.4.x & 5.86 & 42.84 & 11.76 & 10.57 & 474 & 0 & 0 \\ 
        Semgrep & 0.5.x & 6.09 & 44.57 & 12.64 & 11.56 & 472 & 0 & 0 \\ 
        Semgrep & 0.6.x & 6.04 & 41.87 & 12.94 & 11.95 & 487 & 0 & 0 \\ 
        Semgrep & 0.7.x & 5.37 & 45.30 & 12.51 & 11.43 & 474 & 0 & 0 \\ 
        Semgrep & 0.8.x & 6.03 & 62.39 & 12.19 & 10.99 & 483 & 0 & 0 \\ \midrule
        sFuzz & 0.4.x & 2.82 & 1236.14 & 972.25 & 1199.67 & 472 & 114 & 38 \\ 
        sFuzz & 0.5.x & 2.55 & 50.09 & 7.99 & 5.68 & 479 & 0 & 0 \\ 
        sFuzz & 0.6.x & 2.63 & 48.20 & 7.35 & 5.59 & 484 & 0 & 0 \\ 
        sFuzz & 0.7.x & 3.13 & 50.60 & 7.68 & 6.90 & 485 & 0 & 0 \\ 
        sFuzz & 0.8.x & 3.23 & 37.65 & 6.90 & 5.89 & 485 & 0 & 0 \\ \midrule
        Osiris & 0.4.x & 3.43 & 1217.83 & 363.35 & 252.41 & 474 & 2 & 14 \\ 
        Osiris & 0.5.x & 3.35 & 834.59 & 10.32 & 5.32 & 481 & 0 & 0 \\ 
        Osiris & 0.6.x & 3.22 & 33.48 & 6.16 & 5.32 & 490 & 0 & 0 \\ 
        Osiris & 0.7.x & 3.28 & 39.22 & 6.44 & 5.29 & 487 & 0 & 0 \\ 
        Osiris & 0.8.x & 3.78 & 50.37 & 7.04 & 6.02 & 491 & 0 & 0 \\ \midrule
        HoneyBadger & 0.4.x & 3.02 & 1214.32 & 285.34 & 142.34 & 479 & 5 & 12 \\ 
        HoneyBadger & 0.5.x & 3.01 & 1213.40 & 12.57 & 4.98 & 478 & 1 & 0 \\ 
        HoneyBadger & 0.6.x & 2.84 & 61.22 & 5.91 & 4.92 & 482 & 0 & 0 \\ 
        HoneyBadger & 0.7.x & 3.21 & 49.35 & 6.10 & 5.05 & 484 & 0 & 0 \\ 
        HoneyBadger & 0.8.x & 3.53 & 53.81 & 6.66 & 5.79 & 484 & 0 & 0 \\ \midrule
        Siguard & 0.4.x & 9.76 & 1270.80 & 862.14 & 1212.50 & 1428 & 951 & 0 \\ 
        Siguard & 0.5.x & 8.81 & 1249.41 & 792.20 & 1212.31 & 1443 & 859 & 0 \\ 
        Siguard & 0.6.x & 8.77 & 1250.59 & 676.08 & 1108.74 & 1442 & 718 & 0 \\ 
        Siguard & 0.7.x & 10.26 & 1247.57 & 558.65 & 180.34 & 1436 & 582 & 0 \\ 
        Siguard & 0.8.x & 8.12 & 1232.08 & 68.46 & 17.20 & 1436 & 54 & 0 \\ \midrule
        ConFuzziusBI & 0.4.x & 4.86 & 1232.26 & 1002.01 & 1212.29 & 470 & 199 & 137 \\ 
        ConFuzziusBI & 0.5.x & 4.54 & 1252.79 & 516.13 & 10.40 & 476 & 47 & 107 \\ 
        ConFuzziusBI & 0.6.x & 4.41 & 1235.21 & 270.33 & 8.50 & 483 & 39 & 34 \\ 
        ConFuzziusBI & 0.7.x & 4.26 & 1219.48 & 211.37 & 7.69 & 484 & 31 & 22 \\ 
        ConFuzziusBI & 0.8.x & 4.55 & 1243.00 & 803.23 & 1205.96 & 485 & 258 & 16 \\ \midrule 
        EtherSolve & 0.4.x & 1.23 & 134.59 & 33.99 & 33.75 & 1000 & 0 & 0 \\ 
        EtherSolve & 0.5.x & 1.22 & 746.82 & 36.62 & 33.96 & 996 & 0 & 0 \\ 
        EtherSolve & 0.6.x & 1.40 & 623.83 & 35.09 & 34.02 & 997 & 0 & 0 \\ 
        EtherSolve & 0.7.x & 1.42 & 65.21 & 33.96 & 33.57 & 998 & 0 & 0 \\ 
        EtherSolve & 0.8.x & 1.26 & 65.39 & 33.66 & 34.17 & 999 & 0 & 0 \\ \midrule
        EtherTrust & 0.4.x & 4.30 & 1245.78 & 416.29 & 372.18 & 958 & 36 & 0 \\ 
        EtherTrust & 0.5.x & 5.51 & 1238.43 & 506.48 & 392.99 & 961 & 169 & 0 \\ 
        EtherTrust & 0.6.x & 5.78 & 1244.24 & 490.38 & 423.24 & 959 & 100 & 0 \\ 
        EtherTrust & 0.7.x & 5.50 & 1241.70 & 353.08 & 284.09 & 966 & 46 & 0 \\ 
        EtherTrust & 0.8.x & 8.31 & 1217.95 & 54.37 & 28.39 & 955 & 2 & 0 \\ \midrule
        Oyente & 0.4.x & 3.03 & 582.69 & 56.82 & 40.76 & 470 & 0 & 0 \\ 
        Oyente & 0.5.x & 3.57 & 65.42 & 10.51 & 8.52 & 475 & 0 & 0 \\ 
        Oyente & 0.6.x & 4.22 & 60.73 & 9.95 & 8.76 & 477 & 0 & 0 \\ 
        Oyente & 0.7.x & 4.62 & 57.29 & 10.30 & 9.36 & 486 & 0 & 0 \\ 
        Oyente & 0.8.x & 4.11 & 49.64 & 7.60 & 6.69 & 483 & 0 & 0 \\ \midrule
        Securify & 0.4.x & 3.13 & 1249.54 & 805.31 & 965.52 & 466 & 57 & 124 \\ 
        Securify & 0.5.x & 3.16 & 1218.32 & 262.48 & 184.77 & 487 & 11 & 0 \\ 
        Securify & 0.6.x & 3.23 & 1109.08 & 185.92 & 173.31 & 482 & 0 & 0 \\ 
        Securify & 0.7.x & 3.82 & 1214.84 & 305.63 & 321.40 & 483 & 2 & 0 \\ 
        Securify & 0.8.x & 4.84 & 343.39 & 118.45 & 118.32 & 479 & 0 & 0 \\ \midrule
        SmartCheck & 0.4.x & 12.35 & 79.57 & 29.23 & 27.53 & 478 & 0 & 0 \\ 
        SmartCheck & 0.5.x & 11.23 & 92.47 & 33.54 & 31.96 & 485 & 0 & 0 \\ 
        SmartCheck & 0.6.x & 12.73 & 74.85 & 32.05 & 30.35 & 484 & 0 & 0 \\ 
        SmartCheck & 0.7.x & 14.07 & 82.42 & 32.88 & 31.90 & 483 & 0 & 0 \\ 
        SmartCheck & 0.8.x & 12.24 & 78.07 & 29.74 & 28.68 & 480 & 0 & 0 \\ \midrule
        ConFuzzius & 0.4.x & 5.04 & 1220.34 & 805.57 & 897.99 & 471 & 16 & 15 \\ 
        ConFuzzius & 0.5.x & 4.62 & 1225.27 & 297.99 & 12.59 & 480 & 7 & 5 \\ 
        ConFuzzius & 0.6.x & 4.55 & 1209.95 & 151.53 & 8.77 & 475 & 4 & 1 \\ 
        ConFuzzius & 0.7.x & 4.55 & 1213.58 & 147.75 & 8.17 & 485 & 10 & 2 \\ 
        ConFuzzius & 0.8.x & 5.36 & 1220.31 & 601.05 & 796.95 & 488 & 55 & 2 \\ \midrule
        Slither & 0.4.x & 2.83 & 56.78 & 7.02 & 5.57 & 475 & 0 & 0 \\ 
        Slither & 0.5.x & 2.57 & 102.52 & 9.71 & 6.50 & 475 & 0 & 0 \\ 
        Slither & 0.6.x & 2.54 & 594.86 & 10.25 & 6.74 & 484 & 0 & 0 \\ 
        Slither & 0.7.x & 3.64 & 67.95 & 7.56 & 6.49 & 481 & 0 & 0 \\ 
        Slither & 0.8.x & 3.33 & 28.93 & 7.85 & 7.01 & 475 & 0 & 0 \\ \midrule
        Detecti & 0.4.x & 8.52 & 64.95 & 18.70 & 17.13 & 477 & 0 & 0 \\ 
        Detecti & 0.5.x & 8.80 & 1213.71 & 25.93 & 19.27 & 481 & 2 & 0 \\ 
        Detecti & 0.6.x & 7.65 & 1232.35 & 28.26 & 18.55 & 482 & 3 & 0 \\ 
        Detecti & 0.7.x & 8.58 & 70.43 & 20.72 & 19.00 & 484 & 0 & 0 \\ 
        Detecti & 0.8.x & 6.83 & 67.15 & 19.89 & 18.41 & 483 & 0 & 0 \\ \midrule
        Manticore & 0.4.x & 7.94 & 1259.16 & 1110.85 & 1212.94 & 470 & 423 & 0 \\ 
        Manticore & 0.5.x & 13.50 & 1235.53 & 1100.63 & 1212.90 & 483 & 432 & 0 \\ 
        Manticore & 0.6.x & 14.68 & 1264.55 & 991.71 & 1212.68 & 472 & 371 & 0 \\ 
        Manticore & 0.7.x & 11.74 & 1237.05 & 1185.89 & 1213.00 & 477 & 464 & 0 \\ 
        Manticore & 0.8.x & 6.53 & 1241.82 & 385.32 & 37.41 & 477 & 141 & 0 \\ \midrule
        eTainter & 0.4.x & 5.21 & 1236.11 & 58.91 & 11.55 & 968 & 29 & 0 \\ 
        eTainter & 0.5.x & 4.78 & 1235.48 & 346.39 & 17.70 & 971 & 237 & 0 \\ 
        eTainter & 0.6.x & 4.58 & 1250.06 & 399.03 & 17.89 & 953 & 297 & 0 \\ 
        eTainter & 0.7.x & 4.91 & 1265.80 & 519.52 & 23.43 & 958 & 401 & 0 \\ 
        eTainter & 0.8.x & 4.95 & 1232.78 & 71.77 & 15.36 & 962 & 43 & 0 \\ \midrule
        Solhint & 0.4.x & 4.10 & 49.38 & 11.07 & 9.65 & 476 & 0 & 0 \\ 
        Solhint & 0.5.x & 3.99 & 40.13 & 12.11 & 10.92 & 477 & 0 & 0 \\ 
        Solhint & 0.6.x & 3.95 & 30.73 & 11.42 & 10.29 & 482 & 0 & 0 \\ 
        Solhint & 0.7.x & 5.13 & 46.83 & 12.03 & 10.73 & 478 & 0 & 0 \\ 
        Solhint & 0.8.x & 4.33 & 46.41 & 11.17 & 9.75 & 474 & 0 & 0 \\ \midrule
        RNVulDet & 0.4.x & 1.60 & 37.56 & 4.15 & 2.90 & 964 & 0 & 0 \\ 
        RNVulDet & 0.5.x & 1.49 & 55.88 & 4.29 & 2.86 & 970 & 0 & 0 \\ 
        RNVulDet & 0.6.x & 1.60 & 204.46 & 4.57 & 2.93 & 965 & 0 & 0 \\ 
        RNVulDet & 0.7.x & 1.58 & 40.21 & 4.04 & 2.84 & 972 & 0 & 0 \\ 
        RNVulDet & 0.8.x & 1.53 & 41.56 & 4.32 & 2.96 & 966 & 0 & 0 \\ \midrule
        SolidityCheck & 0.4.x & 1.46 & 45.33 & 3.52 & 2.53 & 472 & 0 & 0 \\ 
        SolidityCheck & 0.5.x & 1.52 & 2379.64 & 351.34 & 2.89 & 483 & 97 & 0 \\ 
        SolidityCheck & 0.6.x & 1.46 & 1938.20 & 7.57 & 2.60 & 482 & 1 & 0 \\ 
        SolidityCheck & 0.7.x & 1.49 & 58.23 & 3.78 & 2.62 & 475 & 0 & 0 \\ 
        SolidityCheck & 0.8.x & 1.45 & 25.51 & 3.44 & 2.63 & 481 & 0 & 0 \\ \midrule
        Vandal & 0.4.x & 2.04 & 151.96 & 18.34 & 7.95 & 1000 & 72 & 0 \\
        Vandal & 0.5.x & 1.56 & 146.08 & 38.29 & 11.89 & 1000 & 181 & 25 \\
        Vandal & 0.6.x & 1.54 & 147.49 & 27.07 & 12.88 & 1000 & 102 & 1 \\
        Vandal & 0.7.x & 2.00 & 148.81 & 54.96 & 13.19 & 1000 & 346 & 1 \\
        Vandal & 0.8.x & 2.10 & 150.49 & 43.05 & 11.01 & 1000 & 248 & 1 \\ \midrule
        Overall & 0.4.x & 1.23 & 1270.80 & 364.50 & 53.06 & 17178 & 2107 & 574 \\ 
        Overall & 0.5.x & 1.22 & 2379.64 & 304.12 & 27.88 & 17330 & 2380 & 193 \\ 
        Overall & 0.6.x & 1.40 & 1938.20 & 248.58 & 19.30 & 17367 & 2049 & 66 \\ 
        Overall & 0.7.x & 1.42 & 1265.80 & 257.29 & 21.33 & 17364 & 2299 & 139 \\ 
        Overall & 0.8.x & 1.26 & 1268.43 & 191.58 & 18.83 & 17372 & 1301 & 18 \\
\end{longtable}
}

\subsection{\textbf{RQ\textsubscript{4}.} What is the impact of Solidity versions on the execution time of
smart contract detection tools?}

\autoref{tab:rq4_exe_time} reports the execution time details for each tool across different Solidity versions. For execution-time analysis, we rely on the median, as it allows us to reason about results that are less affected by outliers introduced by timeouts. In addition to the central tendency of the execution time distribution, \autoref{tab:rq4_exe_time} also provides information about execution termination. Specifically, the column TO denotes executions that were terminated due to a timeout without producing any findings or errors, whereas TO+B denotes executions that were terminated due to a timeout while still producing findings or errors.
\major{
The results show that execution-time analysis must be interpreted in light of tool crash resilience. Overall, the results show longer execution times for Solidity version 0.4.x. This behavior can be attributed to the fact that tools tend to crash less frequently on this version, as they were originally designed for and validated against it, as also discussed in RQ$_2$. Consequently, executions are more likely to complete rather than terminate prematurely.
However, some tools exhibit different behavior: instead of failing and stopping execution, they enter non-terminating loops and are eventually terminated by the imposed timeout. On one hand, this phenomenon significantly impacts execution times, artificially inflating them and introducing additional variability in the results. On the other hand, it means that tools are not version-aware and do not handle changes introduced in Solidity upgrades.
}

\begin{table}[ht]
\centering
\caption{Statistically significant differences of static analyzers across Solidity versions (Kruskal--Wallis, Dunn post hoc with Bonferroni correction, and Cliff's $\delta$ for effect size. $\alpha = 0.05$)}
\label{tab:time_static}
\begin{tabular}{c|ccccc}
\toprule
 & 0.4.x & 0.5.x & 0.6.x & 0.7.x & 0.8.x \\
\midrule
0.4.x & -- &  Negligible  & Negligible & Negligible & Small \\
0.5.x &  & -- & Negligible & Negligible & Small \\
0.6.x &  &  & -- & Negligible & Negligible \\
0.7.x &  &  &  & -- & Negligible \\
0.8.x &  &  &  &  & -- \\
\bottomrule
\end{tabular}
\end{table}

The statistical analysis of static analysis tools indicates that, overall, effect sizes across Solidity versions are predominantly negligible, indicating that most version-to-version differences have a very limited practical impact on tool behavior. Small effect sizes are observed only in comparisons involving Solidity 0.8.x (e.g., 0.4.x vs. 0.8.x and 0.5.x vs. 0.8.x). These results come from major semantic changes introduced in Solidity 0.8.x, such as built-in arithmetic checks and stricter typing rules.
Changes to the Solidity version have a minor impact on the performance of static analysis tools.

\geribox{\hspace{0.2mm} \textbf{RQ\textsubscript{4} Finding 4.1.} There is a statistically significant difference in the execution time of static analysis tools across different Solidity versions. However, its negligible or small effect size makes it challenging to distinguish the compatibility of static analysis tools based on execution time.}

\begin{table}[ht]
\centering
\caption{Statistically significant differences of dynamic analyzers across Solidity versions (Kruskal--Wallis, Dunn post hoc with Bonferroni correction, $\alpha = 0.05$)}
\label{tab:time_dynamic}
\begin{tabular}{c|ccccc}
\toprule
 & 0.4.x & 0.5.x & 0.6.x & 0.7.x & 0.8.x \\
\midrule
0.4.x & -- & Small & Small & Small & Small \\
0.5.x &  & -- & Negligible & Negligible & -- \\
0.6.x &  &  & -- & -- & Negligible \\
0.7.x &  &  &  & -- & Negligible \\
0.8.x &  &  &  &  & -- \\
\bottomrule
\end{tabular}
\end{table}

Also, for dynamic analysis tools, the effect sizes across Solidity versions are predominantly negligible, indicating minimal differences in tool execution time.
Small effect sizes are observed only when comparing Solidity 0.4.x with later versions (0.5.x to 0.8.x), suggesting a gradual deviation from the earliest compiler version rather than abrupt changes across consecutive releases.
Notably, comparisons among more recent versions ($>=$ 0.5) consistently yield negligible effects, indicating that tool behavior stabilizes after early language revisions. These results suggest that major practical differences are limited to the transition from early Solidity versions, while subsequent evolution introduces only a marginal impact on execution time.

\geribox{\hspace{0.2mm} \textbf{RQ\textsubscript{4} Finding 4.2.} The findings reported for static analysis tools also apply to dynamic analysis tools, with a few cases when execution times are impacted by the Solidity language in a statistically significant manner.}

\section{Discussion}
\label{discussion}
This section discusses the results of our study in light of the research questions, particularly those related to the effects of the Solidity code changes on the vulnerability detectors and their crash resilience, as well as the trade-offs between the tools' crash resilience and performance.

\major{
\subsection{Solidity Language Changes Affecting Vulnerability Detection}
We discuss the most relevant changes to the Solidity language and compiler that may affect both vulnerability occurrence and the effectiveness of detection tools. While the main results of this study are reported at the aggregate level, this section examines selected changes in more detail to discuss and explain some of the version-specific performance differences observed in our evaluation. For each change, we first describe the modification introduced across Solidity versions and provide code snippets illustrating how the same operation is expressed before and after the change. We then examine the precision or recall of the tools for the vulnerability classes potentially affected by that change. These vulnerability-level metrics are used to identify empirical patterns consistent with the corresponding language change, rather than to establish it as the sole cause of the observed performance differences. Additional vulnerability-level results and detailed tables are available in the online appendix.}

\major{
Before discussing the Solidity-specific causes of tool failures, we highlight the broader empirical lesson emerging from our results. The effectiveness of vulnerability detection tools should not be interpreted as an intrinsic and stable property of the tool alone. Instead, precision, recall, crash resilience, and execution time characterize a tool under a specific technological configuration, including the language version, compiler behavior, input representation, analysis type, and execution environment. This finding complements previous empirical evaluations by showing that technology evolution can change both whether a tool executes and whether its internal analysis assumptions remain valid. Therefore, evaluations of software analysis tools should report and analyze technological versions as first-class experimental factors, rather than treating them as incidental metadata.
}

\begin{description}
\item[\textbf{Now and block.timestamp keywords}.]
Two of the most widely used keywords affected by Solidity changes are \textsc{now} and \textsc{block.timestamp}. 
Before Solidity version 0.7.0, these keywords were used interchangeably, with \textsc{now} being widely adopted in examples and real-world contracts. To avoid semantic ambiguity and to clarify that the returned value refers to the block timestamp rather than to real, continuously flowing time during execution, the Solidity development team removed the alias \textsc{now}. From version 0.7.0 onward, any use of \textsc{now} results in a compilation error and must be replaced with \textsc{block.timestamp}.
Using these two keywords is essential for detecting SWC-116 Timestamp Dependency vulnerabilities. Most static analysis approaches identify this vulnerability by observing the presence of timestamp-related keywords within conditional statements, whose evaluation can be indirectly influenced by miners. By manipulating the block timestamp within its permitted range, a miner can affect the outcome of time-dependent conditions, potentially altering the contract's behavior.
Since the \textsc{now} keyword has been deprecated, static analysis tools must handle its absence and correctly interpret \textsc{block.timestamp} as its semantic equivalent. However, most existing tools were released before July 28, 2020 — the release date of Solidity 0.7.0 — and were primarily evaluated on legacy benchmarks composed almost exclusively of Solidity 0.4.x contracts. It is therefore reasonable to assume that many detectors rely heavily on the presence of the keyword \textsc{now} to identify timestamp dependencies.
Consequently, the absence of \textsc{now} in Solidity versions $>=$0.7.0 adversely affects detectors that rely exclusively on this keyword. Static analyses based on Abstract Syntax Tree (AST) parsing must be updated to support language versions beyond 0.7.0. Moreover, \textsc{now} is no longer considered a global variable in these newer versions, further affecting detection logic that depends on legacy language constructs.

\begin{figure}[htbp]
    \centering
    \begin{subfigure}[b]{0.49\textwidth}
        \centering
        \includegraphics[width=\textwidth]{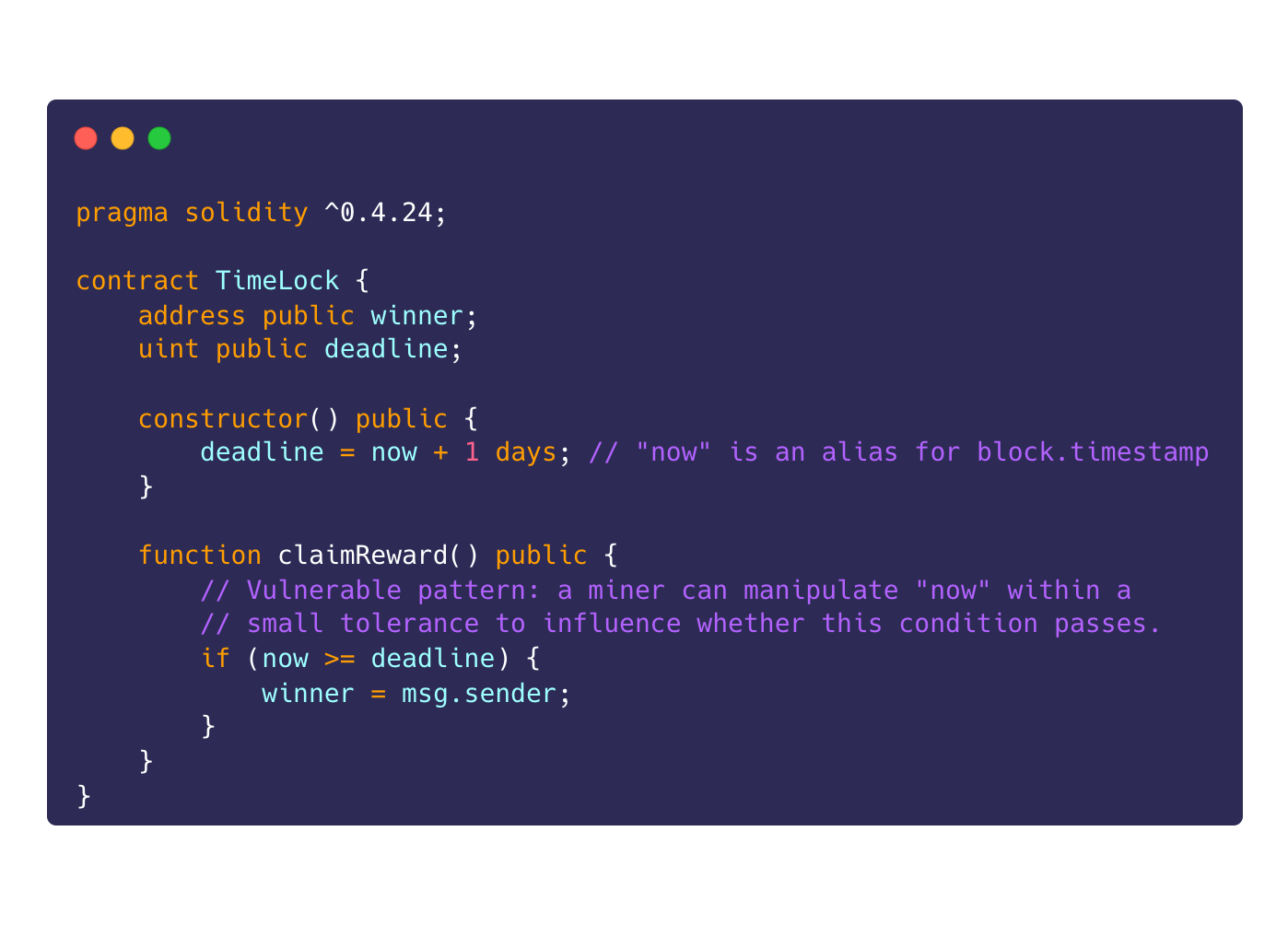} 
        \caption{Use of \textsc{now}}
        \label{fig:comparison-left-1}
    \end{subfigure}
    \hfill 
    \begin{subfigure}[b]{0.49\textwidth}
        \centering
        \includegraphics[width=\textwidth]{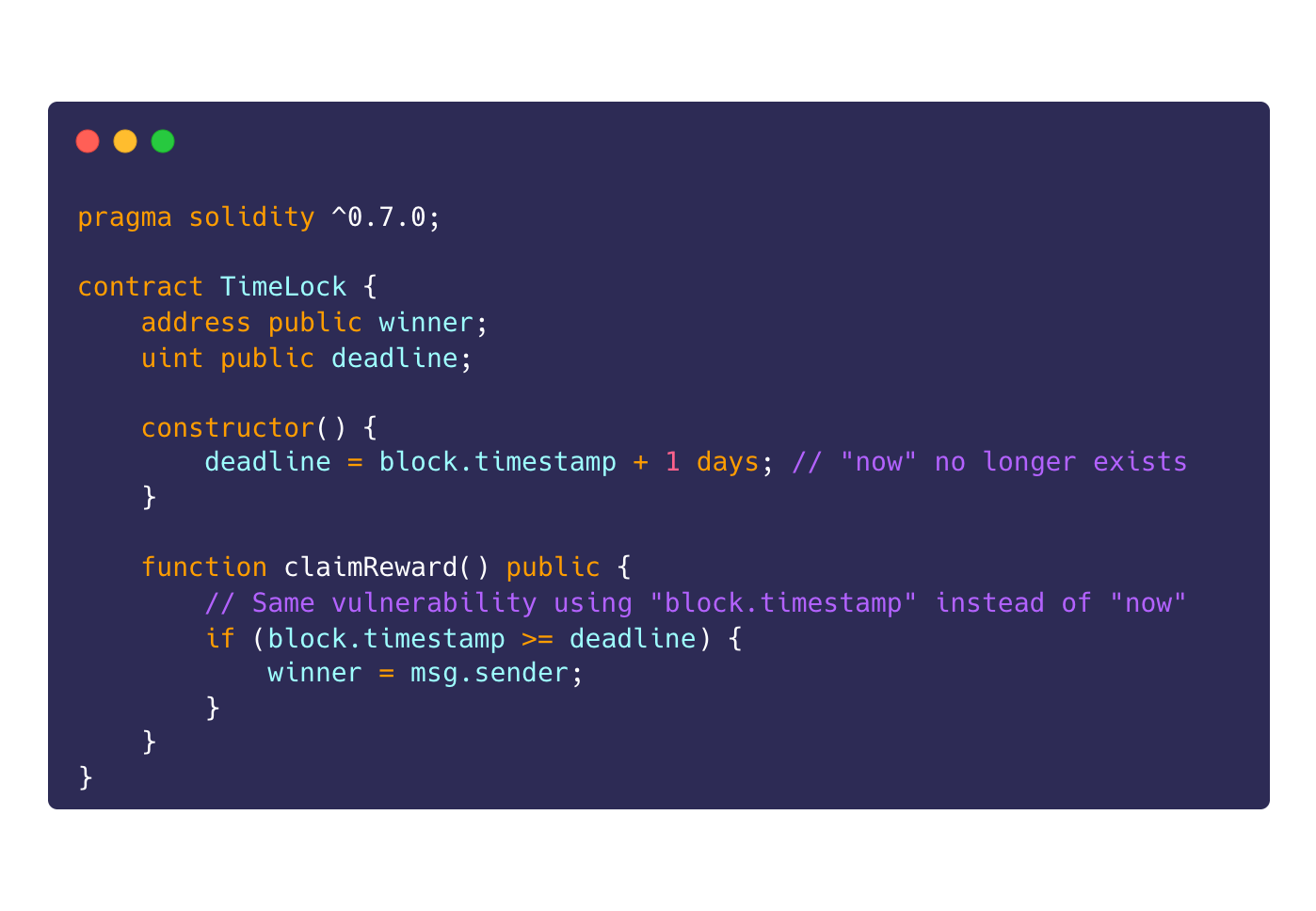} 
        \caption{Use of \textsc{block.timestamp}}
        \label{fig:comparison-right-1}
    \end{subfigure}
    
    \caption{SWC-116 in two different versions.}
    \label{fig:main-comparison-1}
\end{figure}

\major{
The removal of \textsc{now} and the related change in the AST representation of timestamp expressions appear to have affected detectors developed around legacy Solidity constructs. The precision results show that Conkas, Detecti, Osiris, Oyente, and sFuzz correctly detect SWC-116 in Solidity 0.4.x, achieving a precision of 1.00. However, none of these tools reports an SWC-116 finding in the newer versions considered, which may indicate that their detection logic does not recognize the timestamp representation used by newer Solidity versions, although precision alone cannot confirm this explanation. The recall results provide stronger evidence: Conkas and sFuzz achieve a recall of 0.93 in Solidity 0.4.x but fail to detect any injected SWC-116 instance in 0.7.x. Mythril shows a more gradual degradation. Its recall decreases from 0.59 in Solidity 0.4.x to 0.52 in 0.6.x and 0.45 in 0.7.x, while its precision decreases from 1.00 to 0.58 between Solidity 0.4.x and 0.7.x. Slither exhibits a different pattern: its precision decreases from 1.00 in Solidity 0.4.x to 0.62 in 0.7.x, where \textsc{now} was removed, but returns to 1.00 in 0.8.x. This recovery suggests that subsequent tool maintenance adapted its detection logic to the updated language and AST representation. 
}

\item[\textbf{Payable Keyword}.]
\major{
The introduction of address payable represents another change in the Solidity language that directly affects the compilation and analysis of smart contracts across versions. Before Solidity 0.5.0, the address type implicitly allowed Ether transfers, and functions such as transfer and send could be invoked on any address. Starting with Solidity 0.5.0, the language introduced a strict separation between \textit{address} and \textit{payable address}, requiring explicit annotation to indicate that an address is payable. As a consequence, contracts written for Solidity versions earlier than 0.5.0 will fail to compile if payable is used as a type modifier, resulting in errors such as ``Expected identifier but got \textsc{payable}''. 
An address payable can be implicitly converted to an address, but the reverse conversion is not allowed. Converting an address to an address payable is only possible via an explicit cast through uint160. When c is a contract, address(c) yields an address payable only if the contract defines a payable fallback or receive function. Otherwise, the resulting type is a non-payable address. In practice, code that follows the withdraw pattern often requires little or no modification, since transfer is typically invoked on \textsc{msg.sender} rather than on stored addresses. Because \textsc{msg.sender} is of type address payable, such usage remains valid without changes.
This change enforces safer programming practices but also affects static analysis and benchmarking, as tools must correctly account for version-specific typing rules when parsing and analyzing contracts. In particular, analyses based on AST parsing or type inference must account for the Solidity version to correctly interpret Ether transfer semantics and avoid misclassifications or compilation failures during vulnerability detection.
Manual analysis of the results revealed the use of the \textsc{payable} keyword in vulnerable code snippets. Specifically, vulnerabilities such as SWC-107 Reentrancy, SWC-104 Unchecked Call Return Value, SWC-114 Transaction Ordering Dependency, and SWC-115 Authorization via tx.origin can involve payable because they relate to transactions and Ether transfers, and therefore often involve address payable variables.
These vulnerabilities can occur across all Solidity versions. However, in version 0.4.x — the most commonly used version for our evaluation — the payable keyword is not required to indicate that an address can receive or send Ether. Since version 0.5.x, it is mandatory to specify the payable address. Given that version 0.4.x was predominantly used in the tool evaluations, we infer that the tools evaluated on this version do not account for the payable keyword associated with the address type. Consequently, for versions beyond 0.4.0, the behavior of these tools is heavily affected by the \textsc{payable} keyword, which adds extra information to the parsing output that must be accounted for.}

\begin{figure}[htbp]
    \centering
    \begin{subfigure}[b]{0.49\textwidth}
        \centering
        \includegraphics[width=\textwidth]{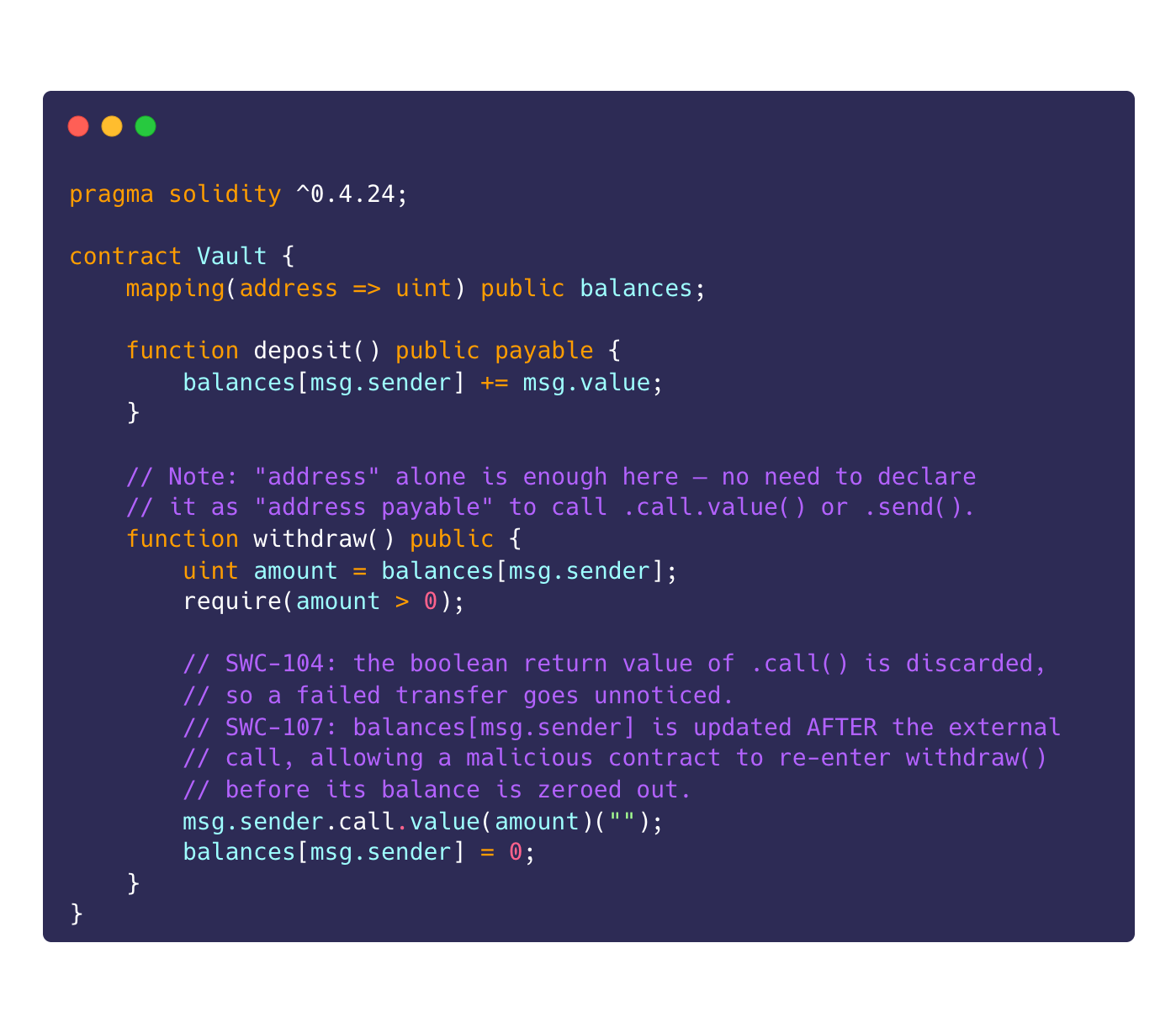} 
        \caption{\textsc{Payable} does not exist}
        \label{fig:comparison-left-2}
    \end{subfigure}
    \hfill 
    \begin{subfigure}[b]{0.49\textwidth}
        \centering
        \includegraphics[width=\textwidth]{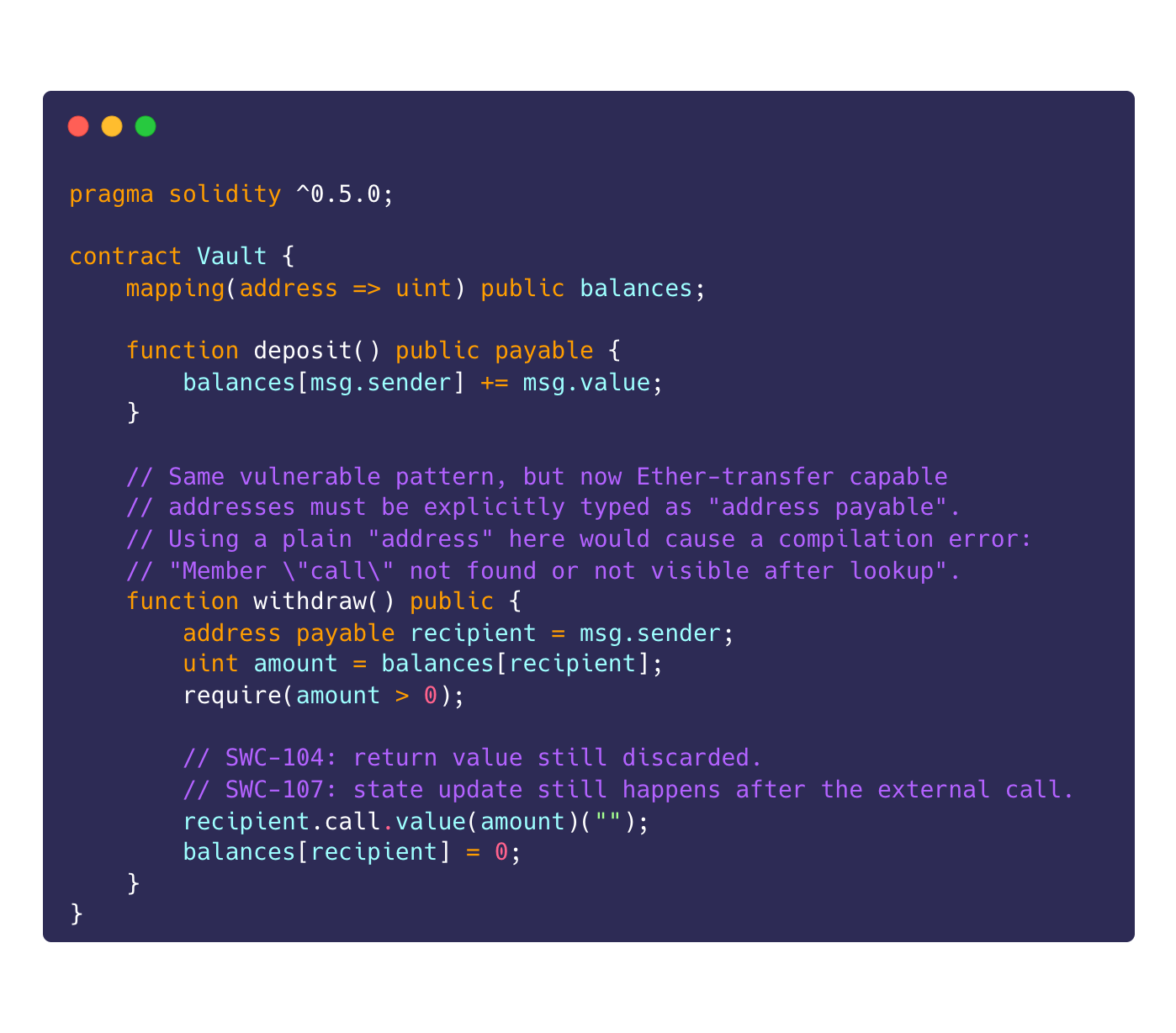} 
        \caption{\textsc{Payable} explicitly needed}
        \label{fig:comparison-right-2}
    \end{subfigure}
    
    \caption{SWC-104 and SWC-107 in two different versions.}
    \label{fig:main-comparison-2}
\end{figure}

\major{
The \textsc{address payable} in Solidity 0.5.x appears to separate the evaluated tools into two groups. Legacy tools show a sharp loss of recall after Solidity 0.4.x: ConFuzzius, ConFuzziusBI, Conkas, Osiris, Oyente, and sFuzz detect SWC-107 with a recall of 0.80 in Solidity 0.4.x, but achieve either 0.00 or only 0.07 in Solidity 0.6.x and 0.7.x. Similar reductions occur for SWC-104 and SWC-114: Oyente's recall for SWC-104 decreases from 0.93 to 0.00, while Conkas and Osiris no longer detect SWC-114 after reaching 0.73 and 0.93, respectively, in Solidity 0.4.x. These results may be produced by parsers and detection rules designed around the former \textsc{address} type and not updated to handle the explicit \textsc{address payable} representation used in Ether-transfer operations. Slither shows the opposite pattern, with SWC-107 recall increasing from 0.00 in Solidity 0.4.x to 0.93 in 0.6.x and 1.00 in 0.7.x, suggesting that its maintenance better supports the newer type system. However, the observed degradation cannot be attributed exclusively to \textsc{payable}, since the syntax used for value-transferring calls also changed in later Solidity versions. 
}

\item[\textbf{Disallow dot syntax for value}.]
In Solidity, the syntax for transferring Ether during a function call changed significantly starting from version 0.7.0. In earlier versions, it was common to use \textsc{.value(amount)} to specify the amount of Ether to send along with a call, for example, \textsc{address.call.\-value(amt)(data)}. From version 0.7.0 onward, this syntax was deprecated and replaced by the new \textsc{\{value: amount\}} form, which is clearer and more flexible, and also allows the specification of additional options, such as gas. Consequently, all calls that transfer Ether must now use the \textsc{\{value: ...\}} syntax, whereas \textsc{.value(...)} was not supported in recent versions and results in a compilation error. 
This change has improved code readability and standardized the handling of Ether transfers in modern Solidity versions. Also, it has significant implications for static analysis tools used to detect vulnerabilities in Solidity contracts.
Tools that were originally developed for versions before 0.7.0 may rely on the \textsc{.value(...)} syntax to identify Ether transfers associated with potentially unsafe calls. With the introduction of \textsc{\{value: ...\}}, these tools may fail to recognize equivalent operations in newer contracts unless they are updated to parse the modern syntax. As a result, the accuracy of vulnerability detection, such as SWC-107 Reentrancy, SWC-104 Unchecked Call Return Value, or SWC-115 Authorization via tx.origin, can be strongly influenced by the tool’s ability to interpret \textsc{\{value: ...\}} expressions correctly.

\begin{figure}[htbp]
    \centering
    \begin{subfigure}[b]{0.49\textwidth}
        \centering
        \includegraphics[width=\textwidth]{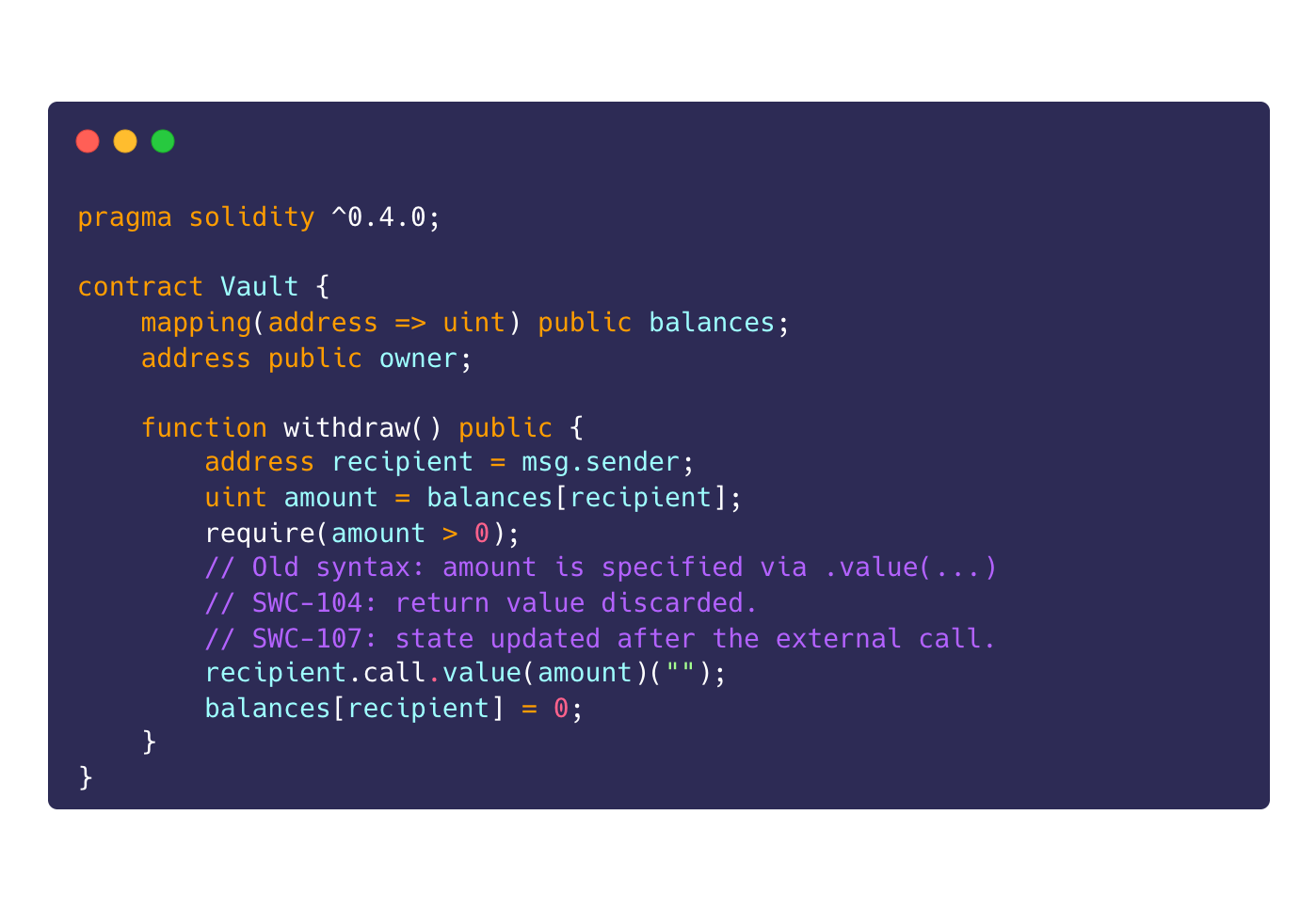} 
        \caption{\textsc{.value(...)}}
        \label{fig:comparison-left-3}
    \end{subfigure}
    \hfill 
    \begin{subfigure}[b]{0.49\textwidth}
        \centering
        \includegraphics[width=\textwidth]{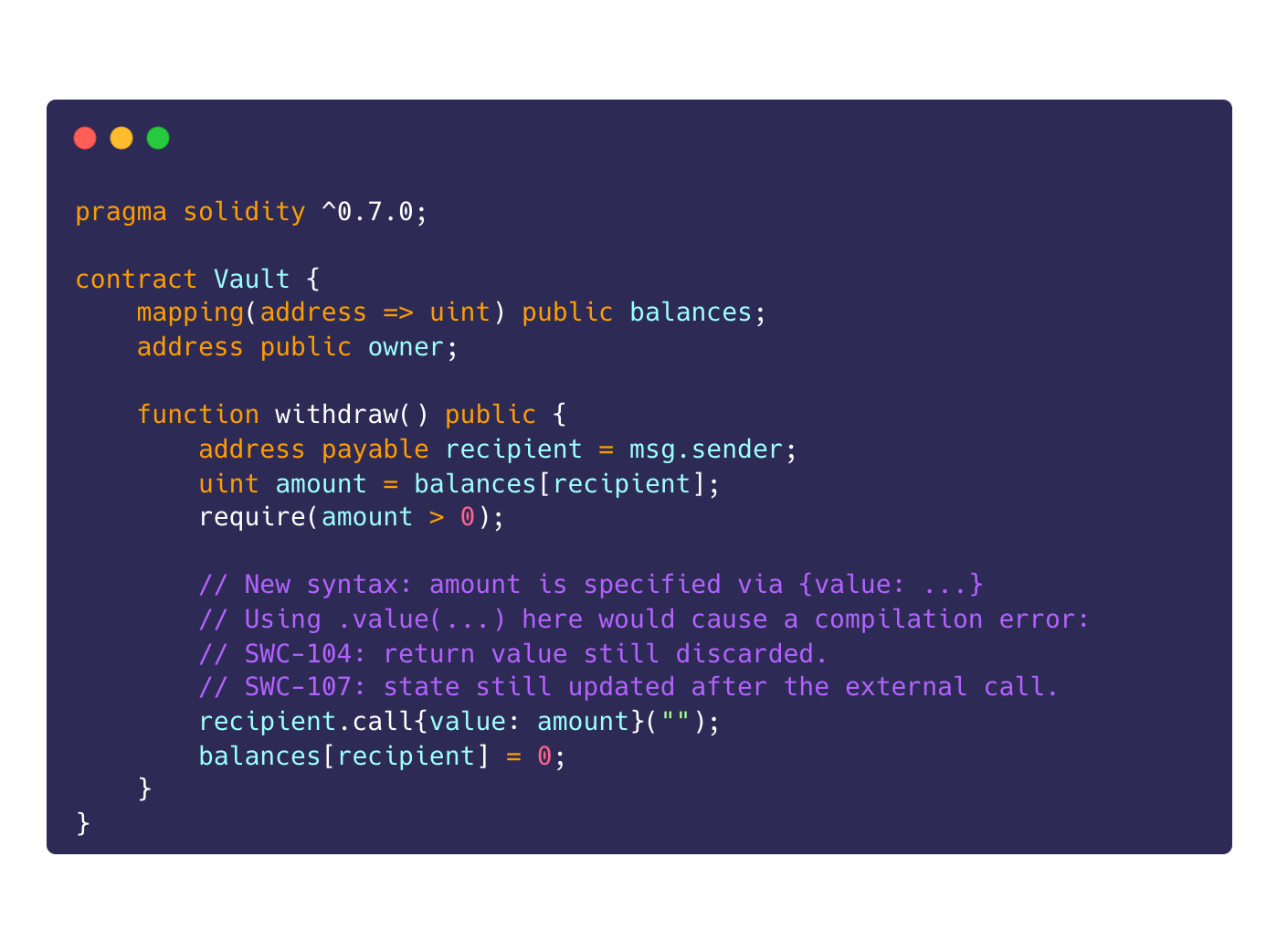} 
        \caption{\textsc{\{value: ...\}}}
        \label{fig:comparison-right-3}
    \end{subfigure}
    
    \caption{SWC-104 in two different shapes.}
    \label{fig:main-comparison-3}
\end{figure}

\major{
The transition from the legacy \textsc{.value(...)} syntax to the \textsc{{value: ...}} form appears to have temporarily affected the accuracy of detectors for vulnerabilities involving Ether-transferring calls. The clearest evidence concerns SWC-104. Slither's precision decreases from 1.00 in Solidity 0.4.x to 0.00 in 0.7.x, where the new syntax was introduced, before recovering to 0.92 in 0.8.x. This V-shaped pattern is consistent with a temporary mismatch between the new call representation and the tool's detection logic, followed by an update that restored support in the subsequent version. ConFuzziusBI shows a similar loss, with SWC-104 precision decreasing from 1.00 in 0.4.x to 0.00 in 0.7.x and becoming undefined in 0.8.x because no finding was reported. However, ConFuzziusBI has never been maintained.
By contrast, maintained tools show improved precision for SWC-107: Slither increases from 0.29 in 0.4.x to 0.71 in 0.7.x and 0.98 in 0.8.x, while Solhint increases from 0.67 to 1.00 and retains this value in 0.8.x. Although these changes cannot be attributed exclusively to the call syntax, the results indicate that support for the new \textsc{{value: ...}} representation and subsequent tool maintenance influenced the correctness of detections involving Ether transfers.}

\item[\textbf{Default arithmetic operations check}.]
Starting with Solidity 0.8.0, arithmetic operations on integer types are checked by default for overflow and underflow, introducing a major breaking change from previous versions. In earlier releases, arithmetic operations would silently wrap on overflow, requiring developers to explicitly use external libraries such as SafeMath to prevent unintended behavior. With Solidity 0.8.0, any overflow or underflow in arithmetic expressions triggers a runtime exception and causes the transaction to revert, unless the operation is explicitly enclosed within an unchecked block. This change significantly improves safety by eliminating SWC-101 Integer Overflow and Underflow by default; however, it also affects backward compatibility and the behavior of existing contracts. In this context, static analysis tools that previously classified contracts as safe based on the presence of the SafeMath library may produce false positives when analyzing contracts written for newer Solidity versions, where such protections are implicit and the library is no longer used. This observation highlights the importance of analysis tools that are aware of the Solidity version under analysis, as the security semantics of arithmetic operations vary substantially across versions of the language.

\begin{figure}[htbp]
    \centering
    \begin{subfigure}[b]{0.49\textwidth}
        \centering
        \includegraphics[width=\textwidth]{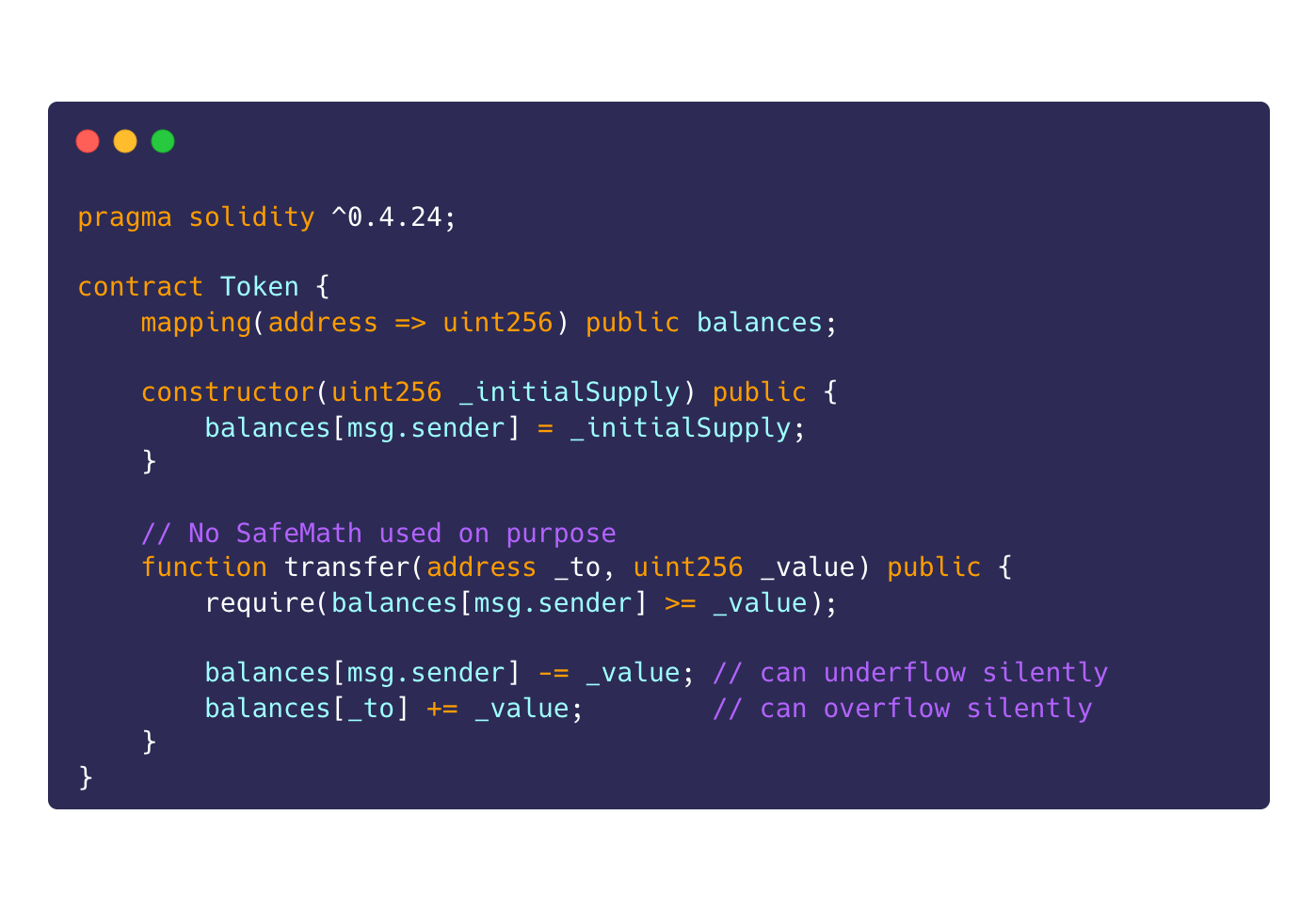} 
        \caption{\textsc{Silent overflow/underflow.}}
        \label{fig:comparison-left-4}
    \end{subfigure}
    \hfill 
    \begin{subfigure}[b]{0.49\textwidth}
        \centering
        \includegraphics[width=\textwidth]{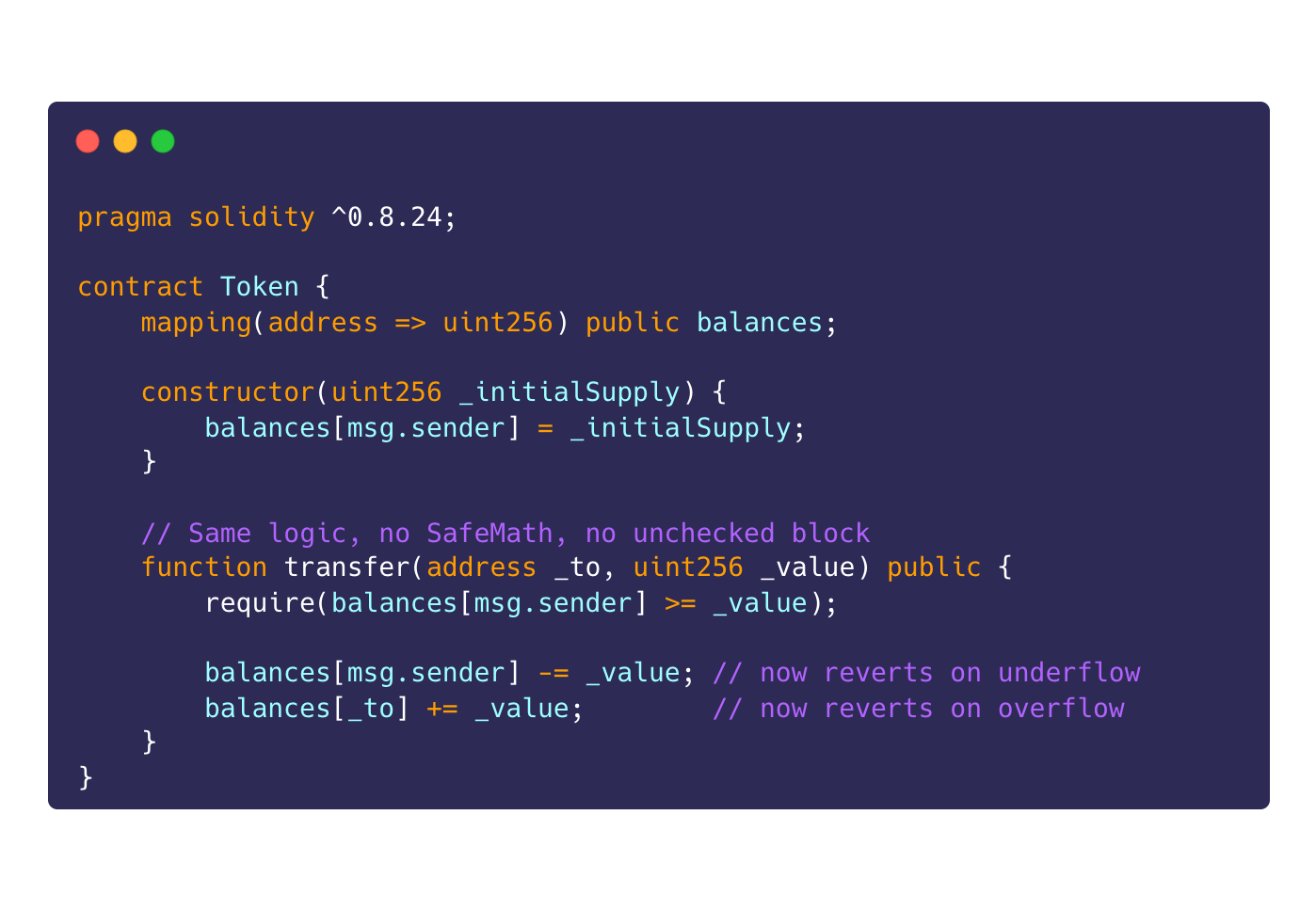} 
        \caption{\textsc{Arithmetic checks by default.}}
        \label{fig:comparison-right-4}
    \end{subfigure}
    
    \caption{SWC-101 checked by default in Solidity 0.8.x.}
    \label{fig:main-comparison-4}
\end{figure}

\major{
The introduction of default arithmetic checks in Solidity 0.8.x had a clear effect on SWC-101 detection. ConFuzzius, ConFuzziusBI, and Mythril all obtain a precision of 0.00 in Solidity 0.8.x, meaning that none of their reported integer overflow or underflow findings correspond to a true vulnerability. This result is consistent with detection logic that does not account for the automatic runtime checks introduced in Solidity 0.8.0 and therefore continues to flag arithmetic operations that now revert by default. The degradation cannot be attributed exclusively to the 0.8.0 change. Nevertheless, the complete absence of correct SWC-101 findings in 0.8.x provides direct evidence that these tools do not correctly adapt their reports to the new arithmetic semantics. The results therefore show that vulnerability definitions must be interpreted according to the compiler version, particularly because SWC-101 remains applicable in Solidity 0.8.x only when arithmetic operations are placed inside an \textsc{unchecked} block.
}

\item[\textbf{Shadowing State Variables}.] Variable shadowing occurs when a variable declared in a child contract or in a narrower scope has the same name as a variable declared in a parent contract or in an outer scope. In such cases, the inner declaration shadows the outer one, potentially leading to unintended behavior.
In Solidity, variable shadowing is particularly dangerous in the presence of inheritance, as developers may inadvertently read from or write to the wrong variable, leading to logic errors or broken access-control mechanisms.
This vulnerability (SWC-119) is exploitable only in contracts compiled with Solidity versions before 0.6.0, where the compiler allowed state-variable shadowing without issuing warnings or errors. As a result, ambiguous and error-prone code could compile successfully.
Starting with Solidity 0.6.0, the compiler issued warnings for state-variable shadowing; since Solidity 0.7.0, such shadowing is disallowed and results in a compilation error. These changes significantly reduce the likelihood of this vulnerability by allowing developers to detect and fix shadowing issues at compile time.
The tool affected by the change is Slither.

\begin{figure}[htbp]
    \centering
    \begin{subfigure}[b]{0.49\textwidth}
        \centering
        \includegraphics[width=\textwidth]{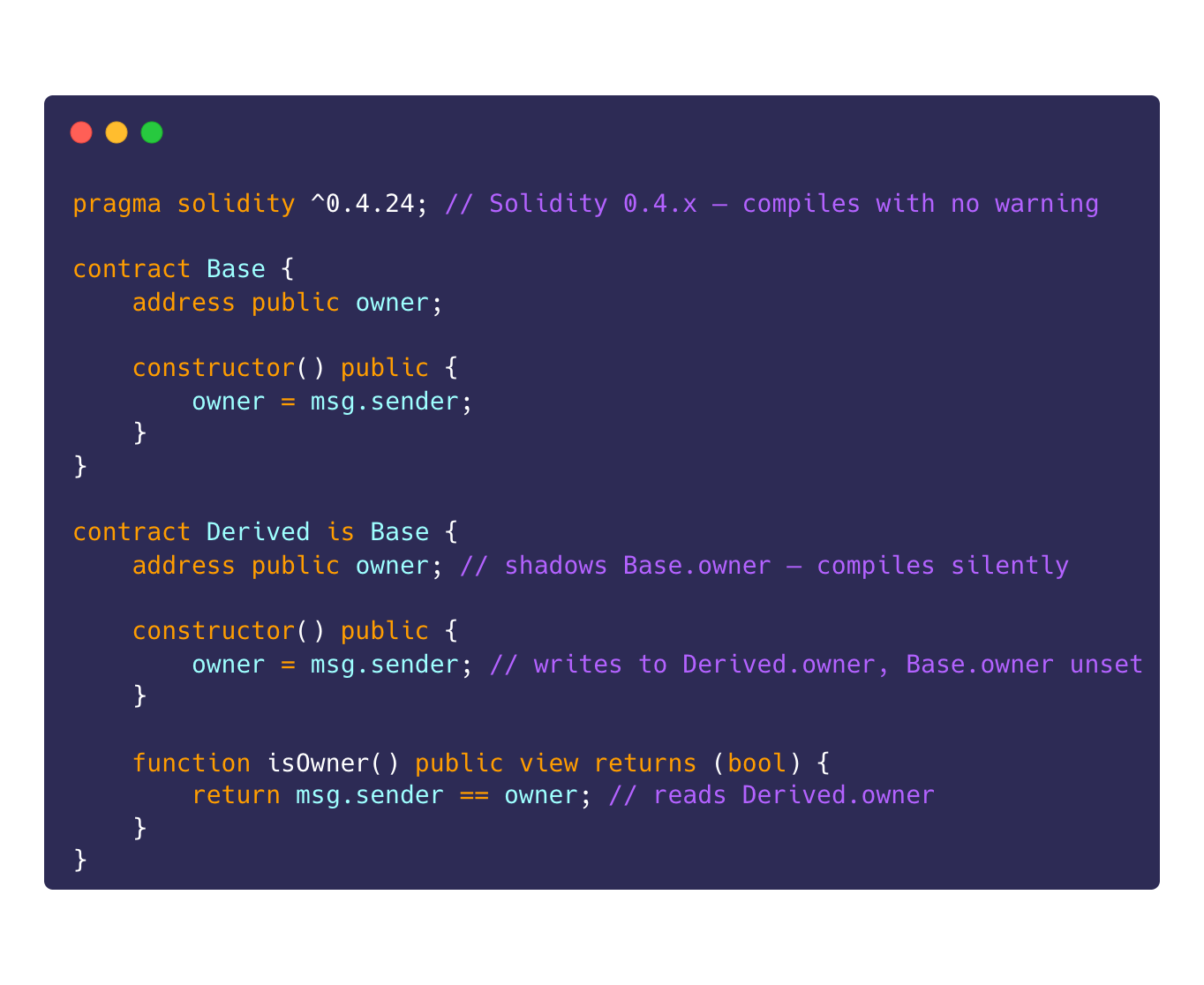} 
        \caption{\textsc{SWC-119 compiles.}}
        \label{fig:comparison-left-5}
    \end{subfigure}
    \hfill 
    \begin{subfigure}[b]{0.49\textwidth}
        \centering
        \includegraphics[width=\textwidth]{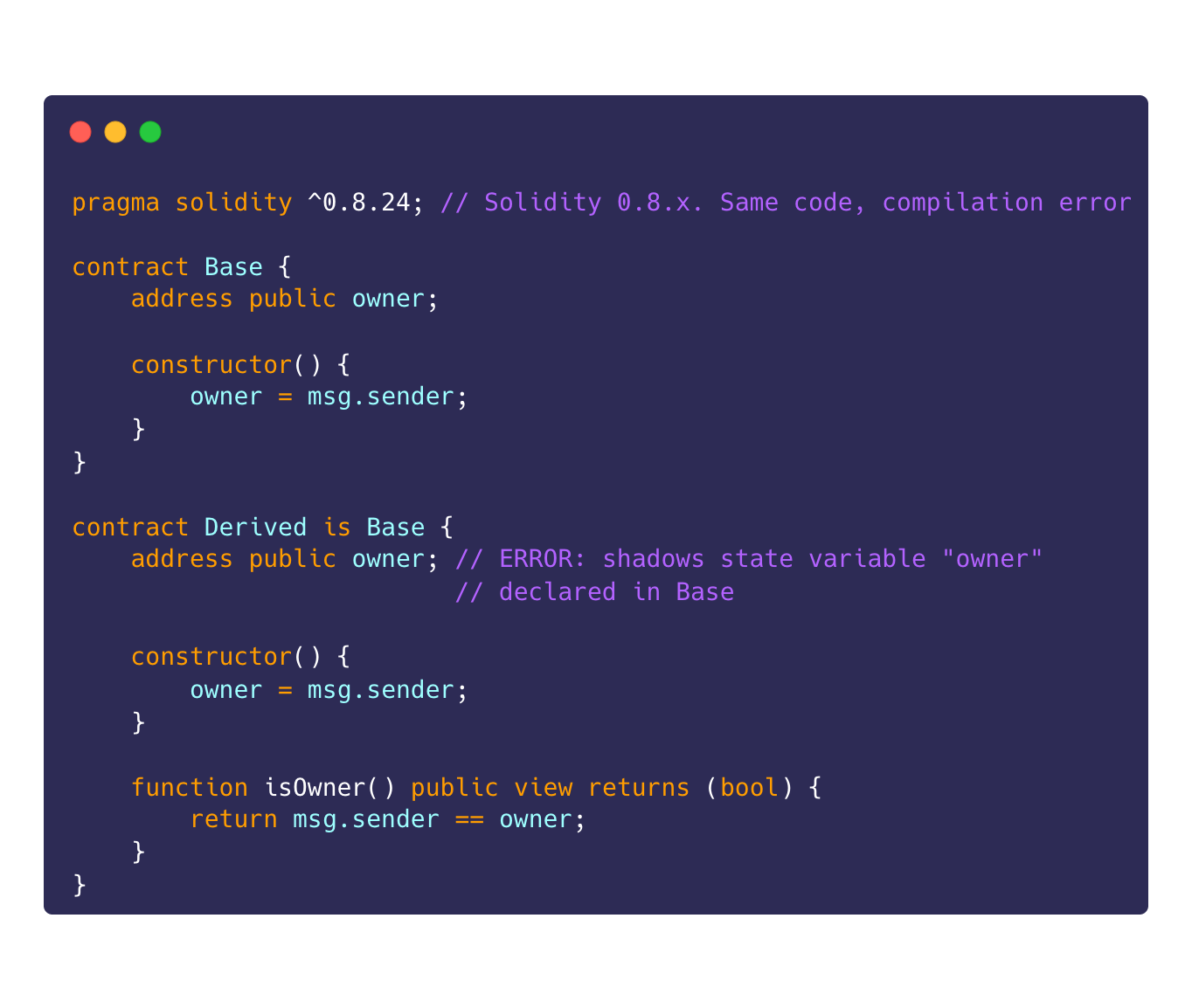} 
        \caption{\textsc{SWC-119 does not compile.}}
        \label{fig:comparison-right-5}
    \end{subfigure}
    
    \caption{SWC-119 absent from Solidity 0.7.0}
    \label{fig:main-comparison-5}
\end{figure}

\major{
Slither achieves a precision of 0.94 in Solidity 0.4.x, where the compiler accepts the vulnerable pattern, but its precision decreases to 0.00 in both Solidity 0.7.x and 0.8.x. This result shows that the rules are not version-aware and a detector may continue to report a vulnerability after the language has eliminated its underlying cause.}

\item[\textbf{Explicitness Requirements}.]
Starting from Solidity 0.5.0, explicit function visibility is mandatory. Developers must add \textsc{public} to every function and constructor, and \textsc{external} to every fallback or interface function that does not already specify its visibility. 
Starting from Solidity 0.7.0, the visibility (\textsc{public}/\textsc{internal}) is no longer needed for constructors. To prevent a contract from being created, it can be marked \textsc{abstract}, making the visibility concept for constructors obsolete. 

\begin{figure}[htbp]
    \centering
    \begin{subfigure}[b]{0.49\textwidth}
        \centering
        \includegraphics[width=\textwidth]{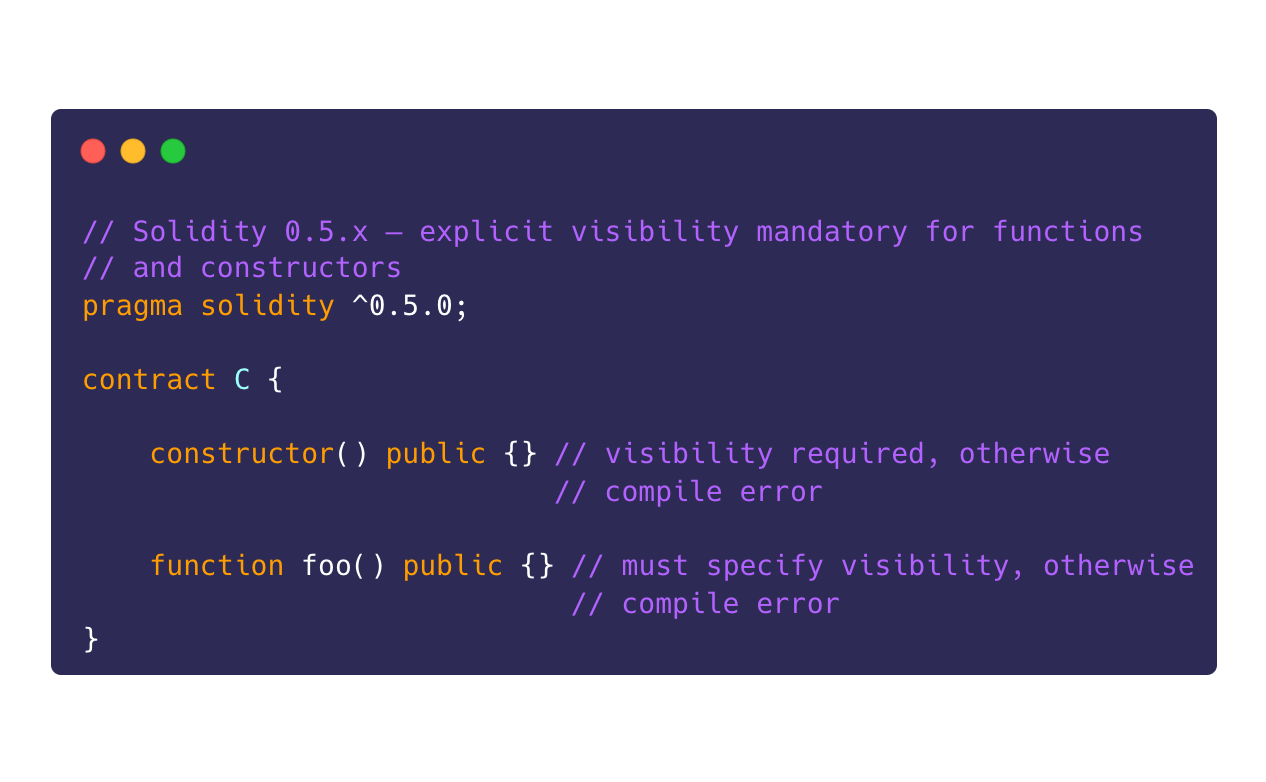} 
        \caption{\textsc{Explicit visibility mandatory}}
        \label{fig:comparison-left-6}
    \end{subfigure}
    \hfill 
    \begin{subfigure}[b]{0.49\textwidth}
        \centering
        \includegraphics[width=\textwidth]{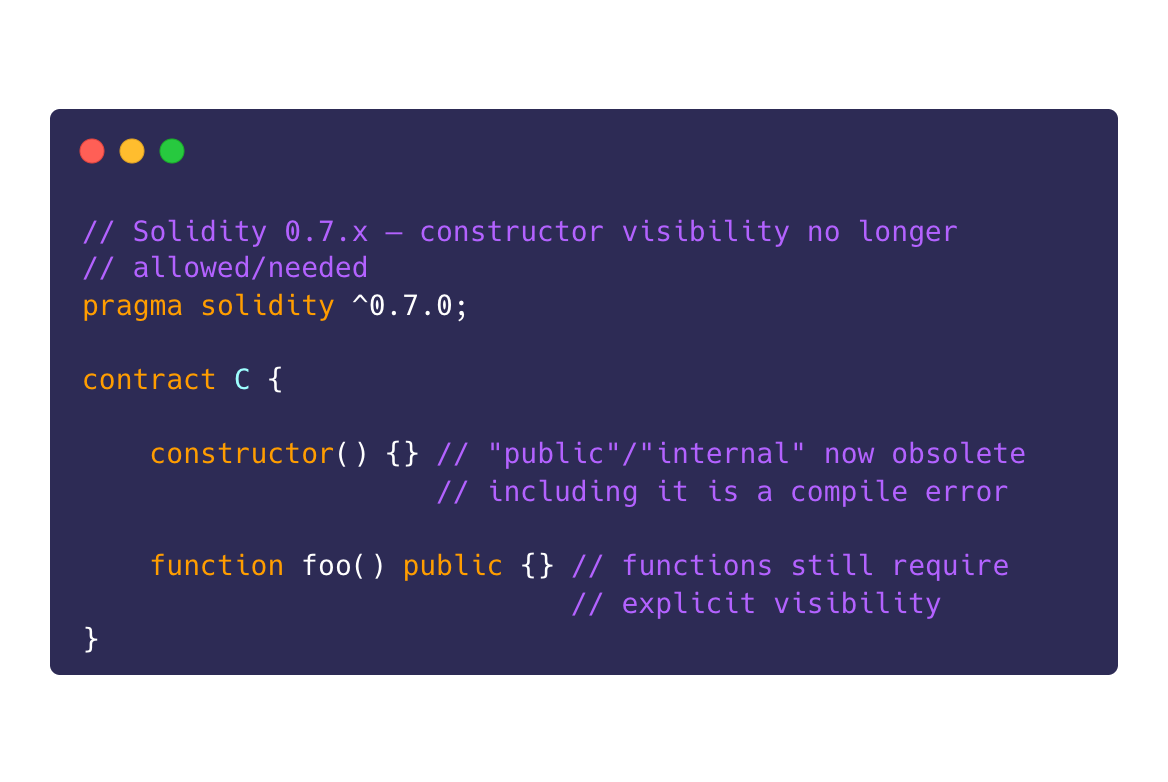} 
        \caption{\textsc{Obsolete visibility for constructor}}
        \label{fig:comparison-right-6}
    \end{subfigure}
    
    \caption{Visibility requirements in different Solidity versions.}
    \label{fig:main-comparison-6}
\end{figure}


\end{description}

\subsection{Practical Examples of the Effect of Language Changes on Vulnerability Detection Performance}
The section discusses how detector implementations are affected by Solidity changes, highlighting detector compatibility across language versions.

\paragraph{\textbf{SmartCheck}.} 
The tool reports the \textsc{constant} keyword as deprecated (SOLIDITY\_DEPRECATED\_CONSTRUCTIONS), which is incorrect for Solidity 0.4.x, where \textit{constant} is still valid. This behavior indicates that the tool is not version-aware and indiscriminately applies rules from newer Solidity versions.
The tool also reports state variable declarations without an explicit visibility specifier. In Solidity 0.4.x, state variables without an explicit visibility are assigned the default visibility internal; therefore, this behavior is correct and does not represent a vulnerability. These findings should be considered false positives.
Additionally, the tool flags the fallback function as having incorrect or missing visibility. In Solidity 0.4.x, the fallback function must be explicitly or implicitly declared as public; otherwise, the contract does not compile. Consequently, reporting this pattern as a visibility error is incorrect and highlights another instance of version-unaware analysis.
Finally, the tool exhibits incorrect behavior when analyzing function declarations spanning multiple lines. When function parameters or return values are declared on separate lines, the tool reports a separate SOLIDITY\_VISIBILITY issue for each line, resulting in duplicated and spurious warnings. This behavior indicates a parsing or pattern-matching limitation, rather than a genuine visibility-related issue.
The tool also reports gas-limit issues (SOLIDITY\_GAS\_LIMIT\_IN\_LOOPS) in constructors that contain loops with a fixed, non-manipulable size. However, such loops do not pose a gas-related risk, as their iteration count is statically bounded and cannot be influenced by user input. Moreover, because the code is executed within a constructor, it is invoked only once at deployment time and therefore does not introduce the risk of repeated or unbounded execution. As a result, these findings should be considered false positives.
Some detectors remain compatible across all Solidity versions; for example, the SOLIDITY\_PRAGMAS\_VERSION detector works well because it relies on a syntax feature that is not affected by Solidity upgrades.

\paragraph{\textbf{ConFuzzius, ConFuzzius-BI, and Mythril}.} 
ConFuzzius, ConFuzziusBI, and Mythril frequently report integer overflows even when the code is protected by explicit assert statements that prevent such overflows. A typical example is the use of SafeMath functions, which guard arithmetic operations with pre- and post-condition assertions to prevent overflow.
Despite these safeguards, the tools still flag the arithmetic operations as vulnerable. Moreover, since ConFuzzius is a fuzzing-based tool, when it generates inputs that trigger an overflow, the corresponding assert condition is violated, causing the transaction to revert and effectively preventing the overflow from being exploited.
As a consequence, the tool reports not only an Integer Overflow but also an Assertion Failure, which is directly triggered by the former. In this context, both findings are false positives, as the assertions correctly enforce safety properties and the overflow is never observable in a successful execution.

\paragraph{\textbf{Oyente}.} 
In the analysis logs, Oyente reports: \textit{``You are using solc version 0.4.26. The latest supported version is 0.4.19.''}
This message indicates that the tool is being executed with a Solidity compiler version newer than the officially supported ones, which may affect the correctness of the analysis results.
Oyente relies on symbolic execution, and in some cases, we observed that integer overflows are reported for variables represented as strings in the analysis output. This behavior suggests that, under certain conditions, Oyente incorrectly handles variable types during symbolic execution, treating a string token as an integer. As a result, arithmetic operations are symbolically evaluated on an incorrect type, leading to spurious overflow reports.
These issues highlight limitations in Oyente’s robustness to type inference during symbolic execution, leading to false positives in overflow detection.

\paragraph{\textbf{Solhint 3.3.8}.} 
Solhint reports a compiler version issue (compiler-version) whenever a contract does not use the latest Solidity version known to the tool. In the version used in our study (3.3.8), the latest recommended Solidity version is 0.5.8. Because the tool has not been updated to handle contracts written in newer Solidity versions, it does not handle them properly.
As a consequence, when analyzing contracts compiled with Solidity 0.6.x, 0.7.x, or 0.8.x, Solhint incorrectly flags the compiler version as outdated. These reports are therefore false positives, as the contracts actually rely on more recent Solidity versions than the tool supports.
Similarly, Solhint reports a function visibility issue (func-visibility) when a function’s access modifier is omitted. However, starting from Solidity 0.5.0, function visibility specifiers are mandatory, and contracts that omit them do not compile. As a result, for all contracts written in Solidity versions $>=$ 0.5.0, this warning cannot correspond to a real issue and should be considered a false positive.
Overall, these behaviors indicate that Solhint lacks adequate version awareness, leading to systematic false positives when analyzing contracts written in Solidity versions newer than those officially supported by the tool.

\major{
\subsection{Crash Resilience in Relation to Precision and Recall}
The section provides further discussion about how crash resilience, along with detection performance in terms of precision and recall, clarifies the actual compatibility of vulnerability detection.}

\begin{table}[htbp]
\rowcolors{2}{gray!10}{white}
\myColor{
\centering
\small
\caption{Crash Resilience (CR), Precision, and Recall by version}
\label{tab:compatibility-performance-triple}
\resizebox{\textwidth}{!}{
\begin{tabular}{ll|rrr|rrr|rrr|rrr}
\toprule
 \multirow{2}{*}{Tool} &  \multirow{2}{*}{Input} & \multicolumn{3}{c}{0.4.x} & \multicolumn{3}{c}{0.6.x} & \multicolumn{3}{c}{0.7.x} & \multicolumn{3}{c}{0.8.x} \\
\cmidrule(lr){3-5} \cmidrule(lr){6-8} \cmidrule(lr){9-11} \cmidrule(lr){12-14}
 &  & CR & Prec & Rec & CR & Prec & Rec & CR & Prec & Rec & CR & Prec & Rec \\
\midrule
Confuzzius & Sourcecode & 0.72 & 0.38 & 0.48 & 0.16 & - & 0.13 & 0.15 & 0.45 & 0.03 & 0.62 & 0.00 & - \\
Confuzziusbi & Sourcecode & 0.73 & 0.30 & 0.48 & 0.19 & - & 0.12 & 0.16 & 0.38 & 0.03 & 0.62 & 0.00 & - \\
Conkas & Sourcecode & 0.87 & 0.49 & 0.72 & 0.00 & - & 0.00 & 0.00 & - & 0.00 & 0.00 & - & - \\
Detecti & Sourcecode & 0.96 & 0.67 & 0.40 & 0.97 & - & 0.40 & 0.97 & 0.00 & 0.47 & 1.00 & 0.00 & - \\
eTainter & Runtime & 1.00 & 0.50 & - & 0.98 & - & - & 1.00 & 1.00 & - & 1.00 & - & - \\
EtherSolve & Creation & 0.88 & 0.08 & 0.69 & 0.93 & - & 0.59 & 0.90 & 0.64 & 0.75 & 0.92 & 1.00 & - \\
EtherTrust & Runtime & 0.96 & - & 0.00 & 0.87 & - & 0.00 & 0.97 & - & 0.00 & 0.99 & - & - \\
HoneyBadger & Sourcecode & 0.87 & 0.00 & - & 0.00 & - & - & 0.00 & - & - & 0.00 & - & - \\
ItyFuzz & Sourcecode & 1.00 & - & 0.00 & 1.00 & - & 0.00 & 1.00 & - & 0.00 & 1.00 & - & - \\
Maian & Runtime & 0.91 & 1.00 & - & 0.68 & - & - & 0.81 & 0.67 & - & 0.95 & - & - \\
Maian & Sourcecode & 0.83 & 1.00 & - & 0.64 & - & - & 0.17 & 0.67 & - & 0.93 & - & - \\
Manticore & Sourcecode & 0.93 & - & 0.00 & 0.93 & - & 0.00 & 0.99 & - & 0.00 & 1.00 & - & - \\
Mythril & Creation & 0.99 & 0.53 & - & 0.99 & - & - & 1.00 & 0.46 & - & 0.99 & 0.06 & - \\
Mythril & Runtime & 1.00 & 0.31 & 0.92 & 1.00 & - & 0.77 & 1.00 & 0.10 & 0.91 & 0.99 & 0.00 & - \\
Mythril & Sourcecode & 0.91 & 0.59 & 0.09 & 0.72 & - & 0.12 & 0.97 & 0.73 & 0.03 & 0.96 & 0.29 & - \\
Osiris & Sourcecode & 0.95 & 0.57 & 0.68 & 0.00 & - & 0.00 & 0.00 & - & 0.00 & 0.00 & - & - \\
Oyente & Sourcecode & 0.93 & 0.43 & 0.68 & 0.00 & - & 0.00 & 0.00 & - & 0.00 & 0.00 & - & - \\
SCAnalyzer & Sourcecode & 0.51 & 0.20 & - & 0.03 & - & - & 0.02 & - & - & 0.22 & - & - \\
Securify & Sourcecode & 0.94 & 1.00 & 0.60 & 0.93 & - & 0.00 & 1.00 & - & 0.00 & 1.00 & - & - \\
Semgrep & Sourcecode & 1.00 & - & 0.00 & 1.00 & - & 0.07 & 1.00 & - & 0.00 & 1.00 & - & - \\
sFuzz & Sourcecode & 0.92 & 0.38 & 0.73 & 0.00 & - & 0.00 & 0.00 & - & 0.00 & 0.00 & - & - \\
Slither & Sourcecode & 0.95 & 0.65 & 0.52 & 0.94 & - & 0.92 & 1.00 & 0.55 & 0.98 & 1.00 & 0.71 & - \\
SmartCheck & Sourcecode & 1.00 & 0.63 & 0.57 & 1.00 & - & 0.55 & 1.00 & 0.34 & 0.60 & 1.00 & 0.46 & - \\
Solhint & Sourcecode & 1.00 & 0.89 & 0.33 & 1.00 & - & 0.33 & 1.00 & 0.11 & 0.33 & 1.00 & 0.28 & - \\
Vandal & Runtime & 1.00 & 0.54 & 0.75 & 1.00 & - & 0.68 & 1.00 & - & 0.82 & 0.98 & 0.95 & - \\
Vandal & Creation & 0.96 & 0.52 & - & 1.00 & - & - & 1.00 & - & - & 0.98 & 0.96 & - \\
\bottomrule
\end{tabular}}}
\end{table}

\major{A closer look at Table~\ref{tab:compatibility-performance-triple} shows that crash resilience alone is not a reliable measure of tool compatibility. Several tools remain stable across Solidity versions, while their vulnerability detection performance is poor. EtherTrust is a clear example: its runtime crash resilience remains high across versions (0.96, 0.87, 0.97, and 0.99), yet its recall is zero in all versions for which it is reported. Thus, the tool does not detect any true positives despite rarely crashing. A similar pattern is observed for ItyFuzz and Manticore, which achieve near-perfect or perfect crash resilience across the four versions while their recall remains at 0.00. Semgrep also shows perfect crash resilience, but its recall remains close to zero, reaching only 0.07 in Solidity 0.6.x.
}

\major{
A similar pattern can be observed for precision. Maian with runtime bytecode maintains high and stable crash resilience, while its precision on version 0.7.x indicates that its reported findings are mainly false positives. A third group includes tools that complete their execution successfully but whose precision decreases as the Solidity version becomes newer. Detecti and Solhint follow this pattern, with precision decreasing from relatively high values in Solidity 0.4.x (0.67 and 0.89, respectively) to near-zero values in Solidity 0.7.x, while their recall remains almost unchanged.
}

\major{
Mythril with runtime bytecode shows stable recall across versions 0.4 and 0.7, respectively, at 0.92 and 0.91, while its precision decreases from 0.31 to 0.00, suggesting that it detects more true vulnerabilities in newer versions at the cost of false positives. Mythril with source code, instead, shows consistently low recall, ranging from 0.03 to 0.12 across the four versions, even though its crash resilience remains relatively high.
}

\major{
Finally, EtherSolve with creation bytecode shows a different issue. Its precision changes considerably across versions, with values of 0.08, 0.64, and 1.00, without a clear trend. At the same time, its crash resilience and recall remain relatively stable, which may indicate that the tool handles or classifies certain Solidity constructs differently depending on the language version, rather than reflecting a general limitation.
}

\subsection{Connection with Core Solidity Initiative}
The results of our study are consistent with recent statements from the Solidity development team~\cite{solidity2025coresolidity}, which acknowledge structural limitations in the current language design and the increasing difficulty of safely evolving the compiler and type system. In particular, the ad hoc nature of language upgrades and the introduction of features with inconsistent or evolving semantics have been recognized as major obstacles to compiler correctness and to tool support.
Our empirical findings provide quantitative evidence of these challenges. We observe statistically significant and practically relevant variations in tool performance across Solidity versions, with more pronounced degradation in precision and recall on recent compiler versions. This trend suggests that existing analysis tools, many of which were designed for earlier versions of the language, struggle to model its evolving behavior accurately.
Notably, the Solidity team explicitly identifies the expansion of verification and analysis tooling as a key motivation for introducing Core Solidity, a redesigned compiler front-end and middle-end aimed at improving language consistency and extensibility. Our results empirically support this motivation, highlighting how current language evolution directly impacts the reliability and effectiveness of analysis tools.
Overall, our study suggests that the observed tool incompatibilities are not isolated implementation issues, but rather symptoms of deeper language-level challenges. The transition to Core Solidity represents a promising direction for restoring semantic stability and improving long-term tool support, thereby enabling more robust analysis and verification of smart contracts.

\section{Implications}
\label{sec:implication}

\major{
\subsection{Implications for Practitioners}
We provide practical guidance to allocate analysis budget during vulnerability detection. By relating execution time to detection performance across tools and Solidity versions, we show how practitioners can select tools according to their time constraints and analysis goals. In particular, we consider the average time each tool takes to produce a true positive, a false positive, or a false negative, thereby enabling practitioners to balance analysis speed, detection capability, and the effort required to inspect incorrect findings.}

\begin{table}[!ht]
    \rowcolors{2}{gray!10}{white}
    \myColor{
    \centering
    \caption{Computational time in seconds for vulnerability detection outcomes (TP, FP, FN) by tool and Solidity version.}
    \label{tab:time_x_performance}
    \resizebox{\textwidth}{!}{
    \begin{tabular}{r|rrrr|rrrr|rrr|rr|rrr|rr}
        Tool  &  \multicolumn{4}{c}{\# Contracts per version}  &  \multicolumn{4}{c}{Total time per version}  &  \multicolumn{3}{c}{Avg time  0.4}  &  \multicolumn{2}{c}{Avg time  0.6}  &  \multicolumn{3}{c}{Avg time  0.7}  &  \multicolumn{2}{c}{Avg time  0.8} \\ 
    \toprule
        tool  &  0.4  &  0.6  &  0.7  &  0.8  &  0.4  &  0.6  &  0.7  &  0.8  &  TP  &  FP  &  FN  &  TP  &  FN  &  TP  &  FP  &  FN  &  TP  &  FP \\ 
    \midrule
        ConFuzzius & 105 & 55 & 63 & 5 & 90,996 & 14,830 & 7,023 & 3,025 & 1,625 & 2,022 & 2,935 & 1,854 & 285 & 1,003 & 1,171 & 121 & - & 605\\
        ConFuzziusBI & 90 & 55 & 61 & 5 & 98,521 & 24,395 & 8,875 & 6,040 & 2,189 & 2,663 & 3,178 & 3,485 & 460 & 1,775 & 1,775 & 153 & - & 1,208\\
        Conkas & 139 & 65 & 68 &  -  & 38,491 & 507 & 556 &  -  & 267 & 414 & 1,833 & - & 7 & - & - & 7 &  -  &   - \\
        Detecti & 49 & 42 & 43 & 3 & 906 & 889 & 831 & 66 & 41 & 453 & 34 & 49 & 33 & 40 & 831 & 35 & - & 22\\
        eTainter & 14 &  -  & 1 &  -  & 552 &  -  & 46 &  -  & 79 & 79 & - &  -  &  -  & 46 & - & - &  -  &   - \\
        EtherSolve & 40 & 30 & 41 & 3 & 2,907 & 2,171 & 2,775 & 270 & 97 & 242 & - & 75 & 2,171 & 75 & 694 & - & 68 & -\\
        EtherTrust & 15 & 15 & 15 &  -  & 13,669 & 20,432 & 11,454 &  -  & - & - & 911 & - & 1,362 & - & - & 764 &  -  &   - \\
        ItyFuzz & 29 & 28 & 30 &  -  & 1,314 & 102 & 113 &  -  & - & - & 44 & - & 3 & - & - & 4 &  -  &   - \\
        Maian & 4 &  -  & 3 & - & 2,372 &  -  & 2,445 & - & 474 & - & - &  -  &  -  & 611 & 1,222 & - & - & -\\
        Manticore & 30 & 28 & 30 &  -  & 31,831 & 28,248 & 35,832 &  -  & - & - & 1,061 & - & 942 & - & - & 1,194 &  -  &   - \\
        Mythril & 143 & 64 & 173 & 37 & 347,924 & 132,436 & 257,534 & 77,431 & 2,096 & 3,192 & 4,518 & 2,499 & 1,540 & 2,575 & 1,922 & 3,179 & 19,358 & 1,721\\
        Osiris & 128 & 53 & 58 &  -  & 53,231 & 313 & 375 &  -  & 532 & 1,183 & 2,802 & - & 5 & - & - & 6 &  -  &   - \\
        Oyente & 147 & 53 & 58 &  -  & 8,907 & 481 & 574 &  -  & 83 & 104 & 469 & - & 8 & - & - & 10 &  -  &   - \\
        SCAnalyzer & 15 &  -  &  -  &  -  & 188 &  -  &  -  &  -  & 63 & 16 & - &  -  &  -  &  -  &  -  &  -  &  -  &   - \\
        Securify & 94 & 41 & 40 &  -  & 93,972 & 9,023 & 13,537 &  -  & 1,093 & - & 5,528 & - & 210 & - & - & 315 &  -  &   - \\
        Semgrep & 30 & 28 & 30 &  -  & 321 & 405 & 374 &  -  & - & - & 11 & 202 & 14 & - & - & 12 &  -  &   - \\
        sFuzz & 79 & 52 & 56 &  -  & 81,132 & 375 & 472 &  -  & 1,531 & 5,409 & 5,071 & - & 6 & - & - & 8 &  -  &   - \\
        Slither & 120 & 53 & 170 & 100 & 869 & 486 & 1,285 & 790 & 7 & 17 & 30 & 9 & 97 & 5 & 9 & 1,285 & 4 & 9\\
        SmartCheck & 169 & 53 & 180 & 135 & 4,786 & 1,705 & 5,900 & 4,089 & 17 & 34 & 184 & 52 & 63 & 34 & 22 & 246 & 35 & 30\\
        Solhint & 162 & 43 & 174 & 127 & 1,724 & 476 & 1,967 & 1,349 & 6 & 54 & 57 & 32 & 16 & 37 & 7 & 66 & 14 & 5\\
        Vandal & 110 & 41 & 41 & 75 & 2,932 & 3,131 & 5,407 & 4,781 & 26 & 43 & 489 & 87 & 626 & 139 & - & 2,703 & 24 & 478\\
    \bottomrule
    \end{tabular}}}
\end{table}

\major{
\autoref{tab:time_x_performance} relates each tool's computational time to its detection outcomes. For each Solidity version, we report the total execution time and the amortized time per true positive (TP), false positive (FP), and false negative (FN), computed by dividing the total execution time by the number of outcomes of the corresponding type. A low time per TP is desirable because it indicates that the tool produces useful findings frequently. In contrast, a low time per FP or FN indicates that incorrect reports or missed vulnerabilities occur frequently. Total execution times should also be interpreted alongside the number of analyzed contracts, which varies across tools due to differing analysis methods and input types. FP- and FN-related values are available for the Solidity versions considered in the respective precision and recall analyses.
}

\major{
When the available analysis time is limited, lightweight static tools provide the most useful first analysis step. Slither shows the most stable computational efficiency, requiring between 4 and 9 seconds per TP across the four Solidity versions. Solhint and SmartCheck also produce TPs at a relatively low computational cost, ranging from 6 to 37 seconds and from 17 to 52 seconds per TP, respectively. However, execution speed must be considered together with the cost of incorrect findings. For example, Solhint requires 54 seconds per FP in Solidity 0.4.x, but only 7 and 5 seconds in Solidity 0.7.x and 0.8.x. Since a lower value indicates that FPs are generated more frequently, using it in recent versions may increase the manual effort required to inspect the reports. Detecti provides another version-dependent example: it requires between 40 and 49 seconds per TP from Solidity 0.4.x to 0.7.x, but produces no TP in Solidity 0.8.x. Therefore, a tool suitable on a limited budget for one Solidity version may not remain suitable for another.
Some tools provide an intermediate trade-off between execution cost and detection performance. For example, EtherSolve requires between 68 and 97 seconds per TP, while Vandal requires between 24 and 139 seconds. These costs remain much lower than those of most dynamic and symbolic tools. Nevertheless, their FP and FN costs vary across versions, showing that their inclusion in an audit workflow should depend on the Solidity version and on whether the main goal is to reduce false reports or missed vulnerabilities.
}

\major{
A larger time budget does not necessarily lead to better detection outcomes. Mythril requires approximately 2,096--2,575 seconds per TP from Solidity 0.4.x to 0.7.x, and this cost increases to 19,358 seconds per TP in Solidity 0.8.x. Similarly, ConFuzzius and ConFuzziusBI require between 1,003 and 3,485 seconds per TP on versions that detect at least one vulnerability, whereas neither produces a TP in Solidity 0.8.x. More importantly, Manticore, EtherTrust, and ItyFuzz produce no TP in the evaluated versions despite consuming non-negligible execution time. Thus, computationally expensive analyses should not be selected only because they perform deeper program exploration. Their additional cost should be justified by a measurable improvement in detection performance for the target Solidity version.
}

\major{
These results support a staged allocation of the available audit time. Practitioners with a limited budget can first execute fast tools with a low time per TP on all compatible contracts. Additional tools can then be applied selectively to high-risk contracts or vulnerability classes not adequately covered by the initial analysis. When manual inspection capacity is limited, practitioners should also consider the frequency of FPs and avoid configurations that generate incorrect findings at a high rate. Conversely, when missing a vulnerability has a higher cost than inspecting additional reports, more time can be assigned to tools that complement the first-stage analysis and reduce the number of FNs.
The table does not identify a single tool that is optimal under every condition. Instead, it allows practitioners to construct a version-aware tool portfolio subject to three constraints: the available computational budget, the desired detection coverage, and the effort available for manual validation of the findings. In particular, the results show that execution time should never be considered in isolation: fast execution is useful only when the tool produces relevant findings, while a high computational cost is justified only when it provides additional detection value.}

\major{
Practitioners should also treat published tool performance as conditional on the technological configurations considered in the original evaluation. Before including a tool in an audit workflow, they should verify that its supported Solidity and compiler versions, input representation, and execution environment match the target contracts. In particular, an execution that terminates without reporting findings should not be interpreted as evidence that the contract is secure.
}

\major{
\subsection{Implications for Researchers and Developers}
Our findings have implications that extend beyond the detection of Solidity vulnerabilities. They show that the behavior of software analysis tools is conditional on the technological environment in which they operate. Language versions, compiler versions, input representations, optimization settings, and runtime platforms should therefore be treated as first-class experimental factors rather than as incidental metadata.
}

\major{
For empirical researchers, this means that tool evaluations should clearly define the technological and temporal scope of their conclusions. Reporting a single aggregate performance value may hide substantial differences across language or compiler versions. Benchmarks should therefore record the versions under which each artifact was produced and the configurations for which its ground truth remains valid. When possible, evaluations should be conducted across different versions or generations and designed to incorporate new versions or updates over time.
}

\major{
For tool developers, our results highlight the importance of explicitly documenting supported language versions, compiler configurations, and input representations. A tool should distinguish an unsupported configuration from a successfully analyzed artifact with no detected vulnerabilities. Otherwise, compatibility failures may be interpreted as poor detection performance, while silent execution may be incorrectly interpreted as successful support.
The choice of input representation may also affect the tool's resilience to technological evolution. Source-based tools are directly exposed to changes in syntax, semantics, and compiler requirements. Bytecode-based tools may be less affected by source-level changes, but they remain sensitive to compiler transformations, optimization strategies, metadata, and virtual-machine revisions. Bytecode should therefore be regarded as a potentially more stable target for analysis, rather than as a version-independent representation.
}

\major{
More generally, evaluations of tools targeting evolving technologies should be designed as maintainable and longitudinal studies. Experimental infrastructures should separate artifact collection, version identification, input preparation, tool execution, failure classification, and metric computation. This separation allows future technology versions and tools to be integrated without redesigning the entire evaluation. The lasting implication of our study is therefore not a fixed ranking of Solidity analyzers, but the need for version-aware tools, benchmarks, and empirical evaluations.
}

\section{Threats to Validity}
\label{threats}
This section examines the threats to validity that could affect our results and the mitigation strategies we adopted.

\paragraph{\textbf{Threats to Construct Validity.}}
A possible threat concerns the creation of Docker images. Implementing the image could affect its performance. Consequently, SmartBugs results on tool execution time may be affected by Docker image performance. We followed the official best practices\footnote{\url{https://docs.docker.com/build/building/best-practices/}} to build these images. 

Another threat comes from vulnerability injection. Vulnerabilities were injected using SolidiFi, which directly introduces vulnerable code snippets into smart contracts. These snippets are selected from a large and diverse set of vulnerability examples. Nevertheless, we acknowledge that relying on a single injection tool may compromise the validity of our study, as the injected vulnerabilities reflect the patterns and assumptions embodied in SolidiFi and may not fully capture the diversity of vulnerabilities observed in real-world smart contracts.

Crash resilience is defined as the proportion of successful executions among all executions. Tools may exhibit high crash resilience either because they correctly handle errors and exceptions or because they do not rely on a strict parsing or compilation phase. For this reason, we cross-analyzed crash resilience data with performance metrics. Although a tool may appear crash-resilient, this does not necessarily imply that it operates correctly. Performance results allowed us to identify several tools that, despite exhibiting high crash resilience, fail to produce meaningful outputs. In such cases, robust error handling effectively hides the tool’s inability to adapt to different Solidity versions, leading to an overestimation of its practical applicability.

Finally, our study relies on a dataset of Solidity smart contracts, which may not fully capture the diversity of constructs across all contract versions. While we prioritized mining the most recent contracts to reflect current development practices, certain patterns or older constructs may be underrepresented, potentially affecting the representativeness of the dataset for some Solidity versions.

\paragraph{\textbf{Threats to Internal Validity.}}
Compilation issues arise when smart contracts import external libraries with different Solidity versions, potentially causing errors even when the contract's version is supported by the tool. Identifying the root cause of compilation errors can be challenging, so we excluded contracts that import external libraries. Our preliminary analysis further supports this decision, revealing that datasets typically do not include the external libraries referenced by smart contracts.

Another significant threat is determining whether a tool is crash-resilient with a specific Solidity version. The SmartBugs framework injects the appropriate compiler into the Docker image based on the pragma directive in the contract. However, compilation errors may still occur, as documented in the SmartBugs repository. For instance, a contract containing commented pragma solidity directives can lead to misinterpretation by the framework. We mitigated this risk by removing commented-out pragma directives.

Additionally, a threat arises from pragma directives that specify a range of compatible compiler versions or offer multiple version options, such as $<, \leq, >, \geq$, and \textasciicircum. Since there is no standard for selecting the appropriate compiler version within a range, the result depends on the specific environment used. To address this, we replaced range-based or multi-version pragma directives with the exact version declared in the contract, ensuring consistency in the analysis process.

A potential threat to internal validity concerns the compatibility of the analyzed tools with different versions of the Solidity language. Our results show that several tools are highly crash-resilient with Solidity 0.4.x. This behavior can be largely attributed to the fact that many of these tools were originally developed to analyze contracts written in this version and were primarily validated on it.
As the Solidity language evolved, multiple syntactic and semantic changes were introduced, negatively affecting tool performance. In particular, tool crash resilience decreases sharply beyond version 0.4.x, suggesting that language evolution poses significant challenges for existing analysis techniques.
Interestingly, for the latest version of Solidity, some tools partially restore compatibility. This trend cannot be fully explained by the language evolution alone and may be influenced by additional factors not captured in our analysis.

After injecting vulnerabilities into the smart contracts, the source files were compiled to generate the corresponding bytecode, which was then analyzed by bytecode-based detection tools. The compilation process produces a separate bytecode file for each contract defined within the same source file. As a result, a vulnerability injected at the source level may manifest in one or more of the bytecode files generated during compilation.
To ensure a fair and accurate evaluation of bytecode-based detection tools, we considered all bytecode files generated during compilation. This choice is necessary because the injected vulnerability may reside in an auxiliary contract (e.g., a utility contract, library, interface, or standard token implementation) rather than the main contract.
This situation does not occur for bytecode deployed on the blockchain, where only the main contract is deployed. In such cases, auxiliary contracts and libraries are linked to the main contract but are not deployed as standalone bytecode.

Finally, precision analysis distinguishes between creation bytecode and runtime bytecode, assuming that the two representations differ exclusively in the presence of deployment-specific instructions, such as constructor logic and initialization code. While this assumption generally holds, it may oversimplify the impact of compilation and deployment on the resulting bytecode. In particular, constructor-related control-flow paths present in creation bytecode may interact with vulnerability detection heuristics in tool-specific ways, potentially inflating false positives and biasing precision measurements. Conversely, the higher precision observed for runtime bytecode may partly stem from the reduced complexity of the analyzed code or tool implementations rather than from a more accurate modeling of contract behavior. As a result, differences in precision between creation and runtime bytecode may reflect representational artifacts rather than intrinsic tool effectiveness. This threat is partially mitigated by evaluating both representations across multiple tools and Solidity versions; however, caution is required when generalizing conclusions about tool precision based solely on the choice of bytecode representation.

\paragraph{\textbf{Threats to External Validity.}} 
A potential threat concerns the frequency of the 0.7.x and 0.8.x pragmas. The preliminary analysis shows that these versions are less used in the literature. We mitigated the lack of recent versions by extracting smart contracts from the Ethereum blockchain using Etherscan, starting with the latest blocks that contain smart contracts compiled with a recent version of Solidity.

Our findings are based on contracts from different Solidity versions. However, the results may not generalize to all versions, especially older or less commonly used ones. We selected these versions to capture the evolution of Solidity, from the most widely used (0.4.x) to the latest (0.8.x), providing a representative overview of the language’s evolution.

\major{
A threat to external validity concerns the generalizability of the precision results to the entire Ethereum blockchain. The population of deployed smart contracts is continuously evolving and cannot be completely enumerated in advance; therefore, it is not possible to construct a statistically representative random sample. Consequently, the reported precision values should not be interpreted as estimates of the tool's precision across the entire blockchain.
Furthermore, the balanced dataset should not be interpreted as reflecting the natural distribution of Solidity versions on Ethereum. This design choice supports fair cross-version comparisons, but it limits conclusions about tool performance under the naturally skewed distribution of deployed contracts.
}

\major{
The selection of tools may affect the external validity of our findings. We identified candidate tools from the collections compiled by Iuliano \etal~\cite{iuliano_dinucci} and Rameder \etal~\cite{rameder2022review}, which represented the most recent and comprehensive lists available when we began the study. Nevertheless, tools not included in these collections or released after our data-collection cutoff may have been unintentionally excluded. Moreover, reducing the initial set of 267 tools to the 26 tools integrated into SmartBugs may limit the representativeness of the evaluated sample. To mitigate selection bias, we followed a systematic procedure for each candidate tool and documented the number of tools and the reasons for exclusion at each stage. However, our conclusions should be interpreted with respect to the tools discoverable and reusable under these criteria at the time of the study, and they may not generalize to tools excluded from the considered lists or released subsequently.
}

\paragraph{\textbf{Threats to Conclusion Validity.}}

The main threat to the validity of our conclusions may arise from the interpretation of statistical significance. To address the limitations of p-values, we primarily relied on effect-size magnitude in our analysis~\cite{kampenes2007systematic}. This choice ensures that the differences discussed are not just statistically significant but also practically meaningful.

\section{Conclusion}
\label{conclusion}
We conducted an exploratory study that combined empirical analysis with structured evaluation to assess tool crash resilience, performance, and execution efficiency across Solidity versions. We developed two distinct datasets: one comprising real-world smart contracts and the other featuring injected vulnerabilities. We leveraged the SmartBugs framework for tool execution and integration. Our initial survey identified 267 tools, of which 150 had links to artifacts. Following a rigorous artifact evaluation process, we determined that 103 tools were available, 53 were functional, and 38 were reusable. Out of 38 reusable tools, 26 were integrated into SmartBugs. Crash resilience revealed significant limitations: many tools are restricted to the Solidity version in use at the time of their original development, typically 0.4.x. While tools that do not explicitly declare compatibility may still execute on newer versions, their behavior is often compromised by changes in the Solidity compiler. Aggregate precision and recall are highest on Solidity 0.4.x. Execution times are also generally longer for this version, partly because fewer executions terminate prematurely. Notably, our analysis highlights that bytecode-based configurations exhibit greater cross-version stability in recall, whereas source-code-based tools are more directly exposed to language evolution. However, bytecode does not consistently provide higher precision; its behavior remains version- and representation-dependent. We presented and discussed the key changes to the Solidity language that affect tool performance, offering actionable insights for both researchers and practitioners.

\major{Beyond the specific Solidity versions and tools analyzed in this study, our results show that empirical evaluations of software analysis tools should explicitly account for technological evolution. Tool effectiveness is not a time-invariant property, but depends on language versions, compiler behavior, input representations, and execution environments. Future Solidity releases may change some of the numerical results reported here, but they reinforce rather than invalidate the central lesson of this work: benchmarks, tools, and empirical studies should be version-aware and designed for longitudinal re-evaluation.}

\subsection*{\textbf{Author Contributions}} 
Gerardo Iuliano: Formal analysis, Investigation, Data Curation, Validation, Writing - Original Draft, Visualization.
Davide Corradini: Investigation, Data Curation, Writing - Review \& Editing.
Michele Pasqua: Investigation, Data Curation, Validation, Writing - Original Draft.
Matteo Cicalese: Investigation, Data Curation.
Mariano Ceccato: Supervision, Resources, Writing - Review \& Editing.
Dario Di Nucci: Supervision, Resources, Writing - Review \& Editing.

\subsection*{\textbf{Funding}}
Funded by the European Union – Next Generation EU, Mission 4 Component 1, CUP D53D23008400006, and by the Italian Government (MUR, PRIN 2022) code 202233YFCJ. Gerardo Iuliano is funded by the European Union – Next Generation EU, Mission 4 Component 2, CUP D42B24002220004, by the Italian Government (MUR), and AstraKode S.r.l.

\subsection*{\textbf{Data Availability}}
The datasets and results of this paper are publicly available at \url{https://figshare.com/s/ec72394075625a122300}. The extended SmartBugs is open-source on GitHub: \url{https://github.com/GerardoIuliano/SmartBugs-3.0}.

\section*{\textbf{Statements and Declarations}}

\paragraph{\textbf{Conflict of Interest.}} 
Gerardo Iuliano reports financial support provided by the European Union, the Italian Government (MUR), and AstraKode S.r.l. Dario Di Nucci reports a relationship with AstraKode S.r.l. for funding grants. The other authors report that financial support was provided by the European Union and the Italian Government (MUR). 
\paragraph{\textbf{Ethical Approval.}} Not applicable.
\paragraph{\textbf{Informed Consent.}} Not applicable.
\paragraph{\textbf{Clinical Trial Number.}} Not applicable.

\bibliographystyle{spmpsci}
\bibliography{main}

\end{document}